\documentclass[aps,prx,twocolumn,superscriptaddress,longbibliography,nofootinbib]{revtex4-2}
\usepackage{graphicx,amssymb,amsmath,amsfonts,empheq,latexsym,braket,bm,bbm,dsfont,upgreek}
\usepackage[normalem]{ulem}
\usepackage{subfig}
\usepackage[dvipsnames]{xcolor}
\usepackage{comment}
\usepackage{tabularx}
\usepackage{color}
\usepackage[dvipsnames]{xcolor}
\usepackage{makecell}
\usepackage{multirow}
\usepackage{url}
\usepackage{soul}
\usepackage{ulem}
\usepackage{algorithm}
\usepackage{enumitem}
\usepackage{algpseudocode}
\usepackage[colorlinks=true,linkcolor=blue,citecolor=blue,urlcolor=blue,breaklinks=true]{hyperref}

\usepackage{tikz}
\usetikzlibrary{arrows.meta,positioning,calc}

\makeatletter
\long\def\@makecaption#1#2{%
    \par\vskip\abovecaptionskip
    {\small\rmfamily\flushing\noindent
    \@make@capt@title{#1}{#2}\par}%
    \vskip\belowcaptionskip}
\makeatother

\begin{document}

\title{Surface codes with coherent errors: optimal decoding and finite-size crossovers}

\author{Xu-Dong Dai}
\email{daixudong@ust.hk}
\affiliation{New Cornerstone Science Laboratory, Department of Physics,\\ \text{The Hong Kong University of Science and Technology,
Clear Water Bay, Kowloon 999077, Hong Kong, China}}
\author{Zhou-Quan Wan}
\affiliation{Center for Computational Quantum Physics, Flatiron Institute, New York, NY 10010, USA}
\author{Xi Dai}
\affiliation{New Cornerstone Science Laboratory, Department of Physics,\\ \text{The Hong Kong University of Science and Technology,
Clear Water Bay, Kowloon 999077, Hong Kong, China}}
\author{Guo-Yi Zhu}
\affiliation{\text{The Hong Kong University of Science and Technology (Guangzhou), Nansha, Guangzhou, 511453, Guangdong, China}}
\begin{abstract}  
We study the surface code under uniform coherent \(Z\) rotations \(\exp{(i\theta Z)}\) followed by perfect syndrome measurements, a setting known to exhibit strong finite-size effects that complicate the extraction of the asymptotic decoding threshold.
We identify finite-size behavior of both Pauli decoding measures and the coherent information with two characteristic lengths: an algebraic ballistic-crossover scale \(d_{\mathrm{ba}}\sim 1/\beta_R^{\alpha}\) ($\alpha=1$ for odd $d$ and $\alpha=2$ for even $d$) and a much larger renormalization-group scale \(d_\mathrm{RG}\sim \exp{(C/\beta^{4}_R)}\), where \(\beta_R=-\tfrac{1}{2}\ln\tan\theta\) and $d$ is the system size.
As \(\theta\) approaches the ballistic SWAP point \(\pi/4\), the two length scales diverge at very different rates, leaving a broad intermediate regime \(d_{\mathrm{ba}}\ll d\ll d_\mathrm{RG}\) in which these quantities show substantial finite-size drift and approximate plateaus.
To establish the microscopic basis of this crossover picture, we derive a leading inter-flavor quartic interaction of order \(\beta_R^2\) in the replicated Majorana theory, which provides a route to the class-D nonlinear sigma model.
The resulting marginal RG flow motivates the scaling variable \(X_4=\beta_R^4\ln d\), in which free-fermion simulations show approximate collapse of decoding measures.
These results are consistent with a Pauli-decoding threshold at \(\theta_c=\pi/4\) in the thermodynamic limit.
For any even $d$, we prove that maximum-likelihood Pauli decoding maximizes entanglement fidelity over all deterministic syndrome-conditioned quantum recoveries, and that its decoding threshold attains the quantum-optimal limit, coinciding with the one defined by coherent information.
\end{abstract}

\maketitle
\tableofcontents
\section{Introduction}
Quantum error correction protects quantum information through encoding it across multiple physical qubits and using syndrome measurement to guide recovery~\cite{Shor1995, Steane1996}. 
Surface codes provide a framework for topological quantum memory, combining local stabilizers with nonlocal logical operators~\cite{Kitaev2003, DennisEtAl2002}. 
Below a decoding threshold, logical failure can be suppressed arbitrarily by increasing the code distance~\cite{GoogleQuantumAI2023, GoogleQuantumAI2025}.  
For independent bit- or phase-flip errors with perfect syndrome measurements, maximum-likelihood decoding (MLD) maps to the random-bond Ising model (RBIM) and the decoding threshold corresponds to the phase transition along the Nishimori line~\cite{Nishimori1981, DennisEtAl2002, WANGPreskill2003, KatzgraberPRL2009, BombinPRX2012, Terhal2015, Chubb_2021}.

While stochastic Pauli channels mentioned above describe an incoherent mixture of error strings, systematic control errors instead generate coherent rotations on the physical qubits~\cite{SandersWallmanSanders2016, WallmanEtAl2015, KuengEtAl2016, WallmanEmerson2016, CaiXuBenjamin2020, DebroyEtAl2018}. 
Error strings with the same syndrome then contribute amplitudes that interfere, making their relative phases relevant to logical recovery~\cite{GreenbaumDutton2018, BealeEtAl2018, ChamberlandEtAl2017, HuangDohertyFlammia2019, IversonPreskill2020}. 
For the spatially uniform coherent $Z$ rotations followed by the perfect syndrome measurement,
\begin{equation}
    U(\theta)=\prod_j e^{i\theta Z_j},\quad
    \rho\to\Pi_s U(\theta)\rho U^\dagger(\theta)\Pi_s^\dagger,
    \label{eq:coherent_channel}
\end{equation}
where \(0\leq\theta\leq\pi/4\), \(j\) labels the physical qubits, and \(\Pi_s\) projects onto the stabilizer eigenspace with syndrome \(s\); the rotation $U(\theta)$ alone is a finite-depth local unitary that leaves the pure state in the same universality class without any transition. 
Syndrome measurement, however, introduces a nonunitary projection and can reveal logical information when the outcome probabilities depend on the encoded input.
The relevant decoding transition concerns whether the encoded logical information remains recoverable after this measurement~\cite{BravyiEtAl2018, VennBeri2020, behrends2025surface}.

Early studies developed tractable noise approximations and simulated small codes under realistic or coherent noise~\cite{MagesanEtAl2013, PuzzuoliEtAl2014, TomitaSvore2014, BarnesEtAl2017}. 
Majorana simulations and mappings to complex RBIM and scattering networks have enabled larger-scale studies~\cite{SuzukiFujiiKoashi2017, BravyiEtAl2018, VennBehrendsBeri2023, PatoStaplesBrown2025, YanBaoVijay2026SWSSB, NegariBeri2026Longrange}.
A network analysis estimated a coherent-error threshold at $\theta\simeq0.14\pi$~\cite{VennBehrendsBeri2023}, while the boundary-entanglement scaling in a related measurement formulation suggested a transition at $\theta\simeq0.125\pi$~\cite{EcksteinEtAl2025}.
However, optimal-decoding simulations exhibit finite-size drift that complicates extrapolation~\cite{PatoMiaoBrown2024}. 
Recent class-D field theory and monitored Majorana dynamics suggest that optimal Pauli decoding remains asymptotically successful for every fixed $\theta<\pi/4$, with strong finite-size effects obscuring this behavior~\cite{YanBaoVijay2026, YangLudwigJian2026}.
Moreover, an established thermodynamic phase diagram alone does not determine recovery performance at finite code distance. 
A related distinction arises in the two-dimensional toric code at thermal equilibrium: although quantum topological order is absent at any nonzero temperature in the thermodynamic limit, finite systems retain topological signatures below size-dependent crossover temperatures that approach zero as the system grows \cite{CastelnovoChamon2007, Hastings2011, LuEtAl2020}. 
For coherent-error decoding, this motivates two connected objectives: 
\begin{itemize}
    \item Determining the Pauli-decoding threshold.
    \item Investigating the finite-size crossover scales.
\end{itemize}

\begin{figure*}[t]
    \noindent
    \includegraphics[width=1.0\linewidth]{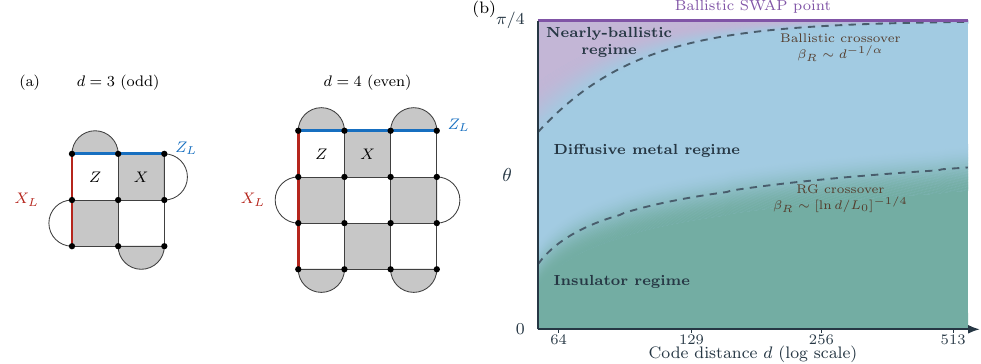}
    \caption{\textbf{(a) Surface code geometry.} Rotated planar surface codes with odd distance \(d=3\) (left) and even distance \(d=4\) (right), encoding one logical qubit. Black dots represent physical qubits. Shaded and unshaded plaquettes and boundary caps denote \(X\)- and \(Z\)-type stabilizers, respectively, acting on four qubits in the bulk or two qubits along a boundary cap. Physical representatives of the logical operators \(X_L\) (red) and \(Z_L\) (blue) lie along the left and top boundaries, respectively. The depicted logical strings have weight \(d\), while all stabilizer generators have even weight. 
    \textbf{(b) Phase diagram with two crossover scales under coherent error.}
Schematic crossover diagram as a function of coherent rotation angle $\theta$ and code distance $d$, showing three regimes: an insulator regime, an intermediate diffusive metal regime, and a nearly-ballistic regime. 
Dashed guides indicate the size-parity-dependent ballistic crossover \(\beta_R\sim d^{-1/\alpha}\), with \(\alpha=1\) for odd \(d\) and \(\alpha=2\) for even \(d\), and the RG crossover \(\beta_R\sim[\ln(d/L_0)]^{-1/4}\), where \(\beta_R=-\tfrac12\ln\tan\theta\).
The upper solid boundary marks the ballistic SWAP point \(\theta=\pi/4\). 
For fixed $d$, in terms of the effective temperature $T_{\rm eff} \equiv 1/\beta_R$ (diverging as $\theta \to \pi/4$), the two crossovers translate into the lower scale $T_{\rm RG} \sim [\ln(d/L_0)]^{1/4}$ and the upper scale $T_{\rm ba} \sim d^{1/\alpha}$, separating three distinct regimes. 
Table~\ref{tab:bulk_logical_regimes} summarizes the corresponding logical-recovery behavior.
For fixed $\theta$, the corresponding crossover scales are $d_{\mathrm{ba}} \sim \beta_R^{-\alpha}$ (ballistic crossover) and $d_\mathrm{RG} \sim L_0 \exp(C/\beta_R^4)$ (RG crossover), where $L_0, C > 0$ are nonuniversal matching parameters. 
Because $d_{\mathrm{RG}}$ diverges exponentially faster than $d_{\mathrm{ba}}$ as $\theta \to \pi/4$, the intermediate diffusive window ($d_{\mathrm{ba}} \ll d \ll d_{\mathrm{RG}}$) broadens across decades in system size, giving rise to pronounced decoding plateaus and substantial finite-size drift (Fig.~\ref{fig:measure_parity}).}
    \label{fig:setup_phase_diagram}
\end{figure*}

Understanding these finite-size crossovers requires a microscopic connection between uniform coherent errors and their continuum description. 
Ref.~\cite{YanBaoVijay2026} derives a nonlinear sigma model (NL\(\sigma\)M) by averaging Gaussian-distributed rotation angles. 
Here, we derive the model directly from the spatially uniform errors, providing a microscopic basis for the resulting finite-size crossover scales.
For the uniform coherent errors, we derive a leading interaction quadratic in effective inverse temperature \(\beta_R=-1/2\ln\tan\theta\) in the replicated Majorana theory, which provides a microscopic route to the class-D NL\(\sigma\)M with bare inverse stiffness proportional to \(\beta_R^2\). 
The resulting picture contains two parametrically separated scales: a ballistic-crossover scale \(d_\mathrm{ba}\sim\beta_R^{-\alpha}\) and a much larger RG scale \(d_{\mathrm{RG}}\sim L_0\exp(C/\beta_R^4)\), which organize the finite-size regimes illustrated in Fig.~\ref{fig:setup_phase_diagram}(b). 
The RG analysis suggests the scaling variable \(X_4=\beta_R^4\ln d\), in which the Pauli-decoding and coherent information data from free-fermion simulations at different sizes show approximate collapse.
Both the theoretical and numerical results are consistent with a Pauli-decoding threshold at \(\theta_c=\pi/4\) in the thermodynamic limit.
Closer to $\theta=\pi/4$ at fixed \(d\), the near-ballistic crossover grows beyond the system size.
In this regime, \(d< d_{\mathrm{ba}}\), the logical response shows a pronounced dependence on the parity of $d$.

Interpreting these finite-size decoding results requires distinguishing limitations of Pauli recovery from information lost through syndrome measurement.
Coherent information \(I_c\) provides a benchmark of quantum recoverability \cite{SchumacherNielsen1996, Lloyd1997, SchumacherWestmoreland2002}. 
For surface codes, it is known that any finite odd code distance gives conditional logical rotations, permitting perfect syndrome-dependent quantum recovery and \(I_c=\ln2\) \cite{BravyiEtAl2018, VennBeri2020, behrends2025surface}. 
This raises the question of whether and when Pauli decoding can attain the quantum recovery limit, analogous to the incoherent error where the thresholds of $I_c$ and MLD coincide \cite{Hauserbaosang2024, NiwaLee2025, WanDaiZhu2026}. 
Thus, our third objective is
\begin{itemize}
    \item Establishing MLD optimality in quantum recovery.
\end{itemize}
At every even code distance, we prove that MLD maximizes entanglement fidelity for maximally mixed logical input over all syndrome-conditioned quantum recoveries after syndrome measurement. 
We further prove that MLD fidelity, posterior purity, and coherent information define the same asymptotically correctable set, and thus share the same threshold for even distances.

The rest of the paper is organized as follows. 
Secs.~\ref{sec:logical_channel}--\ref{sec:measure_parity} establish the exact logical-channel results, including the common-threshold equality and the MLD optimality.
Secs.~\ref{sec:rbim_maj_net}--\ref{sec:numerical_nlsm} derive the NL$\sigma$M for the effective-field description and provide the numerical test.
Finally, Sec.~\ref{sec:near_swap} analyzes the size-parity effect near $\theta\simeq \pi/4$ and gives statistical benchmarks for the fidelity plateaus.

\begin{table*}[t]
\caption{\textbf{Bulk terminology and logical information.} The table summarizes the schematic finite-size correspondence in Fig.~\ref{fig:setup_phase_diagram}(b). 
For odd \(d\), \(I_c=\ln2\) throughout. The qualitative descriptions in the “Logical information” column therefore refer to Pauli-recovery performance for both parities and to coherent information only for even \(d\); reduced Pauli fidelity at odd distance does not imply irreversible loss of logical quantum information.
Arrows denote $\theta\to\pi/4$ at fixed $d$.}
\label{tab:bulk_logical_regimes}
\begingroup
\small
\renewcommand{\arraystretch}{1.15}
\begin{ruledtabular}
\begin{tabular}{cccc}
Phase regime & Logical information & $P_{\mathrm{succ}}$ & $I_c/\ln2$ (even $d$) \\
\hline
Insulator & High-fidelity  & $\simeq1$ & $\simeq1$ \\
Diffusive metal & Intermediate& fractional & fractional \\
Nearly ballistic & Parity-dependent loss& $\to1/2$ & $\to0$ \\
Ballistic SWAP& Totally lost & $1/2$ & $0$ \\
\end{tabular}
\end{ruledtabular}
\endgroup
\end{table*}

\section{Syndrome-resolved logical channel}\label{sec:logical_channel}
Uniform coherent \(Z\) rotations followed by perfect syndrome measurement induce a syndrome-resolved logical channel. 
Measurement removes coherence between distinct syndromes while retaining interference among error strings with the same syndrome, so each conditional logical operation is determined by coherent sector amplitudes.
Following Refs.~\cite{BravyiEtAl2018, VennBeri2020}, we formulate the syndrome measurement and reference correction as a logical quantum instrument in Sec.~\ref{subsec:channel_setup}. 
Then, Sec.~\ref{subsec:logical_kraus} expresses each conditional Kraus operator through two logical-sector amplitudes and their interference.
To characterize the logical recovery, Sec.~\ref{subsec:measures} introduces three measures: MLD fidelity \(P_{\mathrm{succ}}\), posterior-sampling Pauli decoding fidelity \(P_{\mathrm{Bayes}}\), and coherent information \(I_c\).

\subsection{Uniform coherent error and syndrome measurement}\label{subsec:channel_setup}
We consider a surface code encoding one logical qubit, subject to uniform coherent \(Z\) rotations and perfect syndrome measurement (Eq.~\ref{eq:coherent_channel}).
In this setting, physical errors are detected exclusively by \(X\)-type stabilizers, and only a single logical operator is relevant (see Fig.~\ref{fig:setup_phase_diagram}). 

Let \(V: \mathcal{H}_L \rightarrow \mathcal{H}_{\mathrm{phys}}\) denote the linear encoding isometry mapping the logical space onto the codespace \(\mathcal{H}_{\mathrm{c}} = \mathrm{Im}(V) \subset \mathcal{H}_{\mathrm{phys}}\), defined as the simultaneous \(+1\) eigenspace of all stabilizer generators. 
The relevant logical Pauli operator \(Z_L\) is represented on physical qubits by a nontrivial physical \(Z\) string supported on \(\Gamma\), \(Z_\Gamma = \prod_{l\in\Gamma}Z_l\) satisfying \(V Z_L = Z_\Gamma V\).

The encoded state then undergoes uniform coherent rotations, where all \(N\) physical qubits experience an identical single-qubit unitary rotation generated by
\begin{equation}
    U(\theta) = \prod_{j=1}^{N} e^{i\theta Z_j} = (\cos\theta)^N \sum_{E \in \Lambda } (i\tan\theta)^{|E|} Z(E). \label{eq:U_expansion}
\end{equation}
Here, \(E\) denotes the support of the Pauli error string \(Z(E) \equiv \prod_{j\in E} Z_j\), and $\Lambda$ gives the set of all possible error string configurations. \(|E|\) is its Hamming weight. The noise channel thus produces a coherent superposition over all \(2^N\) error configurations, each weighted by a complex amplitude \((i\tan\theta)^{|E|}\). 

Let \(A_v\) denote the \(X\)-type stabilizer generators that detect \(Z\) errors, and denote the measured syndrome configuration by \(s = \{s_v = \pm 1\}\). The orthogonal projector onto the corresponding syndrome sector is
\begin{equation}
    \Pi_s = \prod_v \frac{I + s_v A_v}{2}, \quad \sum_s \Pi_s = I.
\end{equation}
Projective measurement of \(s\) decoheres superpositions across distinct syndrome configurations while preserving interference among error strings sharing the same syndrome \(\partial E = s\). In chain notation, we identify \(s\) with the binary syndrome \((1-s_v)/2\) whose nonzero entries mark violated \(X\) stabilizers.

For each syndrome configuration \(s\), we fix a reference error chain \(C_s\) satisfying the boundary condition \(\partial C_s = s\), and define the reference Pauli correction as \(R_s = Z(C_s)=\prod_{j\in C_s}Z_j\). 
Applying \(R_s\) returns the state conditioned on syndrome \(s\) to the code space. 
Relative to the reference correction \(R_s\), we label the two logical sectors by \(\mu=0,1\), corresponding to the logical actions \(I_L\) and \(Z_L\), respectively.
The residual operation generally contains coherent contributions from both logical sectors, whose recoverability depends on the subsequent recovery \cite{Terhal2015}. The sequence mapping the logical input $\rho_L$ to the unnormalized reference-corrected logical output is summarized by
\begin{equation}
    \rho_L \xrightarrow{V} \rho_{\mathrm{c}} \xrightarrow{U(\theta)} U \rho_{\mathrm{c}} U^\dagger \xrightarrow{\Pi_s} \rho_s \xrightarrow{R_s} R_s \rho_s R_s^\dagger \xrightarrow{V^\dagger} K_s\rho_LK_s^\dagger,
    \label{eq:workflow}
\end{equation}
where \(\rho_s = \Pi_s U(\theta) \rho_{\mathrm{c}} U^\dagger(\theta) \Pi_s\) is the unnormalized state conditioned on measuring syndrome \(s\) and $V^\dagger$ is the isometry that maps the physical space back to the logical space. 
With the syndrome probability \(p_s(\rho_L) = \mathrm{Tr}(K_s \rho_L K_s^\dagger)\), the normalized logical state after the reference Pauli correction is
\begin{equation}
   \rho'_L(s) = \frac{K_s \rho_L K_s^\dagger}{p_s(\rho_L)}, \quad K_s = V^\dagger R_s \Pi_s U(\theta) V,
   \label{eq:conditional_state}
\end{equation}
where \(K_s\) is the conditional logical Kraus operator. Retaining the syndrome outcome as a classical register yields the quantum instrument (channel)
\begin{equation}
    \mathcal{M}_\theta(\rho_L) = \sum_s |s\rangle\langle s| \otimes K_s \rho_L K_s^\dagger, \quad \sum_s K_s^\dagger K_s = I_L.
    \label{eq:instrument}
\end{equation}
The completeness relation, shown in Appendix~\ref{app:kraus_completeness}, ensures that $\mathcal M_\theta$ is trace-preserving. 

\subsection{Logical-sector amplitudes and interference}\label{subsec:logical_kraus}
Every error configuration \(E\) compatible with syndrome \(s\) satisfies \(\partial E = s\). Upon applying the reference correction \(R_s = Z(C_s)\), the residual string \(E\oplus C_s\) has zero measured syndrome. Modulo \(Z\)-type stabilizers,  the residual string belongs to one of two homological classes, $[E \oplus C_s] = \mu \in \{0, 1\}$, corresponding to the logical operations $I_L$ ($\mu = 0$) and $Z_L$ ($\mu = 1$).
Projecting onto the logical space via \(K_s = V^\dagger R_s \Pi_s U(\theta) V\) yields the two-sector decomposition \cite{BravyiEtAl2018,behrends2025surface,VennBeri2020}
\begin{equation}
    K_s = A_0^s I_L + A_1^s Z_L.
    \label{eq:Ks_decomposition}
\end{equation}
The sector amplitudes are coherent sums over configurations within each homological class:
\begin{equation}
    A_\mu^s = (\cos\theta)^N \sum_{E \mid [E \oplus C_s] = \mu} (i \tan\theta)^{|E|}, \quad \mu \in \{0, 1\}.
    \label{eq:A_mu_def}
\end{equation}

The action on a logical input separates into contributions diagonal in the logical Pauli basis and interference between the two logical sectors \cite{BravyiEtAl2018, behrends2025surface, VennBeri2020}:
\begin{equation}
\begin{aligned}
K_s\rho_LK_s^\dagger
={}&\underbrace{
|A_0^s|^2\rho_L+|A_1^s|^2Z_L\rho_LZ_L
}_{\text{Pauli-diagonal contribution}}
\\
&+\underbrace{
(A_0^s)^*A_1^s Z_L\rho_L
+A_0^s(A_1^s)^*\rho_LZ_L
}_{\text{interference between logical sectors}}.\label{eq:kraus_logical_output}
\end{aligned}
\end{equation}
The first two terms have the form of an incoherent Pauli mixture, while the cross terms retain the relative phase between the \(I_L\) and \(Z_L\) amplitudes and distinguish unitary rotations from weak measurements in the logical space (Sec.~\ref{subsec:logical_parity}). 

For each syndrome \(s\), we define the total weight 
\begin{equation}
    D_s = |A_0^s|^2 + |A_1^s|^2,
    \label{eq:syndrome_weight}
\end{equation}
which corresponds to the outcome probability $p_s$ for initial logical state $\ket{+}$ or unpolarized logical state as shown in Eq.~\eqref{eq:syndrome_prob}.
The competition and quantum interference between the two logical sectors are characterized by the magnitude ratio \(r_s \ge 0\) and the normalized interference parameter \(m_s\):
\begin{equation}
    \frac{A_1^s}{A_0^s} = r_s e^{i\Delta\phi_s},\quad 
    m_s \equiv \frac{2\operatorname{Re}\!\left[(A_0^s)^* A_1^s\right]}{D_s} = \frac{2r_s \cos\Delta\phi_s}{1 + r_s^2},
    \label{eq:m_s_def}
\end{equation}
where \(\Delta\phi_s \in (-\pi, \pi]\) is the relative phase, and \(m_s \in [-1, 1]\). 
While the polar coordinates \((r_s, \Delta\phi_s)\) assume both amplitudes to be nonzero, the definition of \(m_s\) in terms of amplitudes remains smooth and non-singular whenever \(D_s > 0\).
Thus, we have $K_s^\dagger K_s=D_s(I_L+m_sZ_L)$, where \(D_s\) sets the overall scale of \(K_s^\dagger K_s\), and \(m_s\) determines whether \(K_s\) is proportional to a unitary.
For \(m_s=0\), the conditional logical evolution is unitary after normalization.
For \(0<|m_s|<1\), \(K_s\) weights the two logical \(Z_L\) eigenstates unequally; at \(|m_s|=1\), it is proportional to a rank-one projector.

For an arbitrary logical input \(\rho_L\), the syndrome probability depends on \(\rho_L\) only through its \(Z_L\) polarization:
\begin{equation}
    p_s(\rho_L) = \mathrm{Tr}\!\left(K_s \rho_L K_s^\dagger\right) = D_s \left[1 + m_s \langle Z_L \rangle_{\rho_L}\right],
    \label{eq:syndrome_prob}
\end{equation}
where \(\langle Z_L \rangle_{\rho_L} \equiv \mathrm{Tr}(Z_L \rho_L)\). Thus, the same parameter \(m_s\) that governs the conditional logical operation also measures the syndrome's sensitivity to the input state. For a branch with \(m_s = 0\), its probability is independent of the input: \(p_s(\rho_L) = D_s\).
More generally, any state with zero logical \(Z_L\) polarization, \(\langle Z_L \rangle = 0\)—including the maximally mixed state \(I_L/2\) and the logical $X_L$ eigenstates $|\pm\rangle_L$—yields the same syndrome distribution \cite{BravyiEtAl2018}:
\begin{equation}
    p_s \left(|\pm\rangle_L\right) = p_s \left(I_L/2\right) = D_s, \quad \sum_s D_s = 1.
    \label{eq:unpolarized_prob}
\end{equation}
The shared syndrome distribution \(p_s=D_s\) for these zero-polarization inputs does not imply unitary conditional evolution: branches with \(m_s\neq0\) still change the logical state by weak measurements. 
The full syndrome distribution is independent of the logical input if and only if \(m_s=0\) on every nonzero branch, in which case each normalized branch is unitary. 
Details on input dependence are provided in Appendix~\ref{app:logical_input}.
 
\subsection{Pauli decoding fidelity and coherent information}\label{subsec:measures}
In this part, we introduce logical recoverability using three complementary measures.
We quantify Pauli decoding performance using the entanglement fidelities of two recovery procedures: maximum-likelihood Pauli decoding, denoted by \(P_{\mathrm{succ}}\), and a randomized Pauli recovery defined below, denoted by \(P_{\mathrm{Bayes}}\).
We also evaluate the coherent information \(I_c\) for a maximally mixed logical input purified by a reference qubit, with the syndrome record retained~\cite{SchumacherNielsen1996, Lloyd1997, SchumacherWestmoreland2002, Devetak2005}.
Syndrome averages use the Born weights \(p_s=D_s\) (Eq.~\eqref{eq:syndrome_prob}).

\paragraph{Pauli decoding fidelity}
We quantify Pauli decoding performance by its entanglement fidelity: the overlap of the recovered logical state with its initial maximally entangled state~\cite{Schumacher1996, Nielsen1996EntanglementFidelity}. 
Applying \(Z_L^\mu\) on a syndrome gives conditional fidelity $|\langle+|Z^\mu_LK_s|+\rangle|^2/D_s=|A_\mu^s|^2/D_s$ \cite{BravyiEtAl2018, IversonPreskill2020, VennBehrendsBeri2023}.
These fidelities depend only on sector magnitudes and are reproduced exactly by the following logical dephasing representation, where the decoding fidelities admit classical decoding-success interpretations: the MLD recovery fidelity and randomized recovery fidelity equal their entanglement fidelities (Sec.~\ref{subsubsec:even_decoder_optimality}); we use this correspondence below to introduce \(P_{\mathrm{succ}}\) and \(P_{\mathrm{Bayes}}\).

For each syndrome with \(D_s>0\), normalizing the Pauli-diagonal contribution in Eq.~\eqref{eq:kraus_logical_output} acts like a logical dephasing channel with effective posterior weights $q^s_{\mu}$,
\begin{equation}
    \begin{gathered}
    \mathcal D_s(\rho_L)=q^s_{0}\rho_L+q^s_{1}Z_L\rho_LZ_L,\\
    q^s_{\mu} = |A_\mu^s|^2/D_s, \quad q^s_{0} + q^s_{1} = 1.
    \label{eq:q_mu_def}
        \end{gathered}
\end{equation}
For this effective logical dephasing channel, Pauli recovery amounts to choosing between \(I_L\) and \(Z_L\). We consider two natural decoding procedures and their fidelities.
The first one is MLD, which chooses the correction corresponding to the larger posterior weight \cite{DennisEtAl2002, Terhal2015, VennBehrendsBeri2023}. Its conditional fidelity and Born average are
\begin{equation}
   \begin{gathered}
    P_{\mathrm{succ}}(s) = \max\left(q^s_{0}, q^s_{1}\right), \\
    P_{\mathrm{succ}} = \sum_s D_s P_{\mathrm{succ}}(s) = \sum_s \max\!\left(|A_0^s|^2, |A_1^s|^2\right).
    \label{eq:P_succ_def}
    \end{gathered}
\end{equation}
The second is the randomized Pauli decoding fidelity $P_{\mathrm{Bayes}}$ \cite{YanBaoVijay2026, NiwaLee2025, WanDaiZhu2026}.
Randomized recovery instead chooses correction \(Z_L^\mu\) with probability \(q^s_{\mu}\). 
Since this correction succeeds with fidelity \(q^s_{\mu}\), its conditional fidelity and Born average are
\begin{equation}
\begin{gathered}
    P_{\mathrm{Bayes}}(s) = (q^s_{0})^2+(q^s_{1})^2, \\
    P_{\mathrm{Bayes}} = \sum_s D_s P_{\mathrm{Bayes}}(s) = \sum_s \frac{|A_0^s|^4 + |A_1^s|^4}{D_s}.
    \label{eq:P_bayes_def}
    \end{gathered}
\end{equation}
This measure quantifies posterior concentration without selecting a dominant sector. 

For either procedure, the recovery fidelity for an arbitrary pure logical input \(|\phi\rangle_L\), averaged over syndrome outcomes and any random choice of correction, satisfies
\begin{equation}
        F_\phi
    =P+(1-P)
    |\langle\phi|_L Z_L|\phi\rangle_L|^2\geq P,
\end{equation}
where \(P=P_{\mathrm{succ}}\) for MLD and \(P=P_{\mathrm{Bayes}}\) for the randomized recovery.
This follows from the \(I_L/Z_L\) form of the corrected logical operators and trace preservation of the full recovery operation.
Each entanglement fidelity equals the minimum recovery fidelity over all pure logical inputs, which is attained by states with zero logical $Z_L$ polarization $\langle\phi|_L Z_L|\phi\rangle_L=0$.
Thus, we choose $\ket+$ as the representative to compute both $P_\mathrm{succ}$ and $P_\mathrm{Bayes}$. 

For each syndrome branch, the two Pauli measures satisfy $\tfrac12\leq P_{\mathrm{Bayes}}(s)
\leq P_{\mathrm{succ}}(s)\leq1$.
Both measures equal \(1/2\) for equal sector weights (\(|A_0^s| = |A_1^s|\)) and reach unity only when a single sector carries the entire weight. For this effective logical dephasing part, both \(P_{\mathrm{succ}}\) and \(P_{\mathrm{Bayes}}\) only depend on the sector magnitudes \(|A_\mu^s|\) and do not concern the interference between the two logical sectors, which is encoded in the relative phase \(\Delta\phi_s\).

\paragraph{Logical channel and coherent information}
Coherent information is evaluated from the full syndrome-resolved channel \(\mathcal M_\theta\) (Eq.~\eqref{eq:instrument}), retaining the interference terms in Eq.~\eqref{eq:kraus_logical_output} \cite{FanEtAl2024,Lee2025CoherentInformation,ColmenarezEtAl2024,EcksteinEtAl2024}.
We purify the maximally mixed logical input \(I_L/2\) with a reference qubit \(R\), forming a maximally entangled state $|\Phi\rangle_{LR}$. 
With the classical syndrome register denoted by \(S\), the joint output and its reduction are $\rho_{SLR}=(\mathcal M_\theta\otimes\mathrm{id}_R)(|\Phi\rangle\langle\Phi|)$ and $\rho_{SL}=\operatorname{tr}_R\rho_{SLR}$.
Then the coherent information is
\begin{equation}
    I_c=S(\rho_{SL})-S(\rho_{SLR}),\quad
    S(\rho)=-\operatorname{tr}(\rho\ln\rho).\label{eq:Ic_def}
\end{equation}
For each outcome with \(D_s>0\), the conditional logical-reference state $\tfrac{1}{\sqrt{D_s}}(K_s\otimes I_R)|\Phi\rangle_{LR}$ is pure.
Its reduced logical state is
\begin{equation}
    \rho_{L|s}=\frac{K_sK_s^\dagger}{2D_s}
    =\frac12\left(I_L+m_sZ_L\right).
\end{equation}
Because both full output states retain the syndrome register, their classical syndrome entropies cancel. We have
\begin{equation}
    I_c(s)=h_2\!\left(\frac{1+m_s}{2}\right),\qquad
    I_c=\sum_sD_sI_c(s),\label{eq:Ic_ms_def}
\end{equation}
where \(h_2(x)=-x\ln x-(1-x)\ln(1-x)\).
Thus \(I_c\) is the Born-averaged entanglement entropy of the conditional logical-reference states. 
Unlike the two Pauli measures, it depends on the relative sector phase through \(m_s\) (Eq.~\eqref{eq:m_s_def}). 
In Sec.~\ref{subsec:logical_parity}, we will see that a branch proportional to a unitary preserves the maximal value \(\ln2\), whereas unequal singular values of \(K_s\) reduce this entanglement through logical measurement backaction.
Accessible syndrome information is also essential to Eq.~\eqref{eq:Ic_ms_def}; its derivation and the effect of discarding the syndrome record are in Appendix~\ref{app:syndrome_record}. 

We choose the three measures to address complementary aspects of logical recovery. \(I_c\) benchmarks quantum recoverability beyond Pauli recovery. \(P_{\mathrm{succ}}\) gives the optimal entanglement fidelity achievable by Pauli recovery; under even code distance, it attains the optimum over all deterministic syndrome-conditioned quantum recoveries, as proved in Sec.~\ref{subsec:common_threshold}. \(P_{\mathrm{Bayes}}\) measures posterior concentration smoothly without selecting a dominant sector, providing complementary sensitivity to finite-size behavior. All three measures are independent of the reference \(Z\)-type Pauli correction. Their dependence on sector magnitudes and relative phase underlies the code distance parity-dependent relations established next.
Other standard measures include the syndrome-conditioned diamond distance~\cite{BravyiEtAl2018, SandersWallmanSanders2016, KuengEtAl2016} and the averaged logical gate infidelity~\cite{PatoStaplesBrown2025, IversonPreskill2020}, which for single-qubit channels is fixed by the entanglement fidelity~\cite{Nielsen2002AverageFidelity}.

\section{Odd-even effects and the common-threshold equality}\label{sec:measure_parity}
In this section, we give exact results about the logical channel.
The code-distance parity dependence of the syndrome-resolved logical operation is established for coherent \(Z\) errors when all \(Z\)-type stabilizers have even weight. 
Odd distance yields a conditional logical rotation, whereas even distance yields a logical weak measurement \cite{BravyiEtAl2018, VennBeri2020}. 
In Sec.~\ref{subsec:logical_parity}, we reconstruct these established conditional logical operations in terms of the interference parameter \(m_s\), providing the basis for the common-threshold equality and the MLD optimality in the even code distance.
In Sec.~\ref{subsec:common_threshold}, we derive size-independent bounds showing that \(P_{\mathrm{succ}}\), \(P_{\mathrm{Bayes}}\), and \(I_c\) define the same asymptotically correctable set for even code distance, and give the same threshold $\theta_c$.
We further prove that MLD is optimal among all CPTP recoveries for even code distance.
Numerical results in Fig.~\ref{fig:measure_parity} compare the three measures across code distances, showing an odd-distance bump and contrasting even- and odd-distance behavior near \(\theta=\pi/4\).
Sec.~\ref{subsec:swap_endpoint_recovery} gives the exact values of fidelities at the maximal error angle $\theta=\pi/4$.

\subsection{The conditional logical channels}\label{subsec:logical_parity}
From the expansion in Eq.~\eqref{eq:U_expansion}, an error string $E$ contributes $(\cos\theta)^N(i\tan\theta)^{|E|}$, with a complex phase $i^{|E|}$. Fix a syndrome $s$ and one logical sector. Two representatives \(E\) and \(E'\) in the same sector differ by the support \(B\) of a product of \(Z\)-type stabilizers: \(E'=E\oplus B\).
Their weights satisfy
\begin{equation}
    \begin{aligned}
        |E\oplus B|&=|E|+|B|-2|E\cap B|,\\
|E\oplus B|&\equiv|E|+|B|\pmod 2 .\label{eq:error_string_deform}
    \end{aligned}
\end{equation}
We consider all the $Z$ stabilizer generators to have even weight, so $|B|$ is even and the two representatives have the same weight parity.
It follows that all strings in a fixed logical sector have the same weight parity. Their phases may still differ by a sign because their weights can differ modulo four, but they lie on the same axis of the complex plane. The corresponding sector amplitude is therefore either real or purely imaginary:
\begin{equation}
    |E|\ \text{even}\ \Longrightarrow\ A_\mu^s\in\mathbb R,\quad
    |E|\ \text{odd}\ \Longrightarrow\ A_\mu^s\in i\mathbb R .
\end{equation}

For the square-code family considered here, choose a weight-\(d\) representative \(Z_\Gamma\) of the logical \(Z\) operator, acting as \(Z_L\) on the encoded qubit.
Representatives of the two logical sectors differ by $\Gamma$, up to an even-weight stabilizer deformation, so
\begin{equation}
    |E\oplus\Gamma|
\equiv
|E|+d
\pmod 2. 
\end{equation}
Thus the parity of \(|\Gamma|=d\) fixes the relative phase of the two sector amplitudes \cite{BravyiEtAl2018, VennBeri2020}. For nonzero $A_0^s$ and $A_1^s$, their relative phase is
\begin{equation}
\arg\left(\frac{A_1^s}{A_0^s}\right)=
\begin{cases}
\pm\pi/2, & d\ \text{odd},\\[2mm]
0\ \text{or}\ \pi, & d\ \text{even}.\label{eq:phase_case}
\end{cases}
\end{equation}

\subsubsection{Odd distance: effective logical rotation}\label{subsubsec:odd_parity}
\begin{figure}[t]
    \centering
    \includegraphics[width=1.0\linewidth]{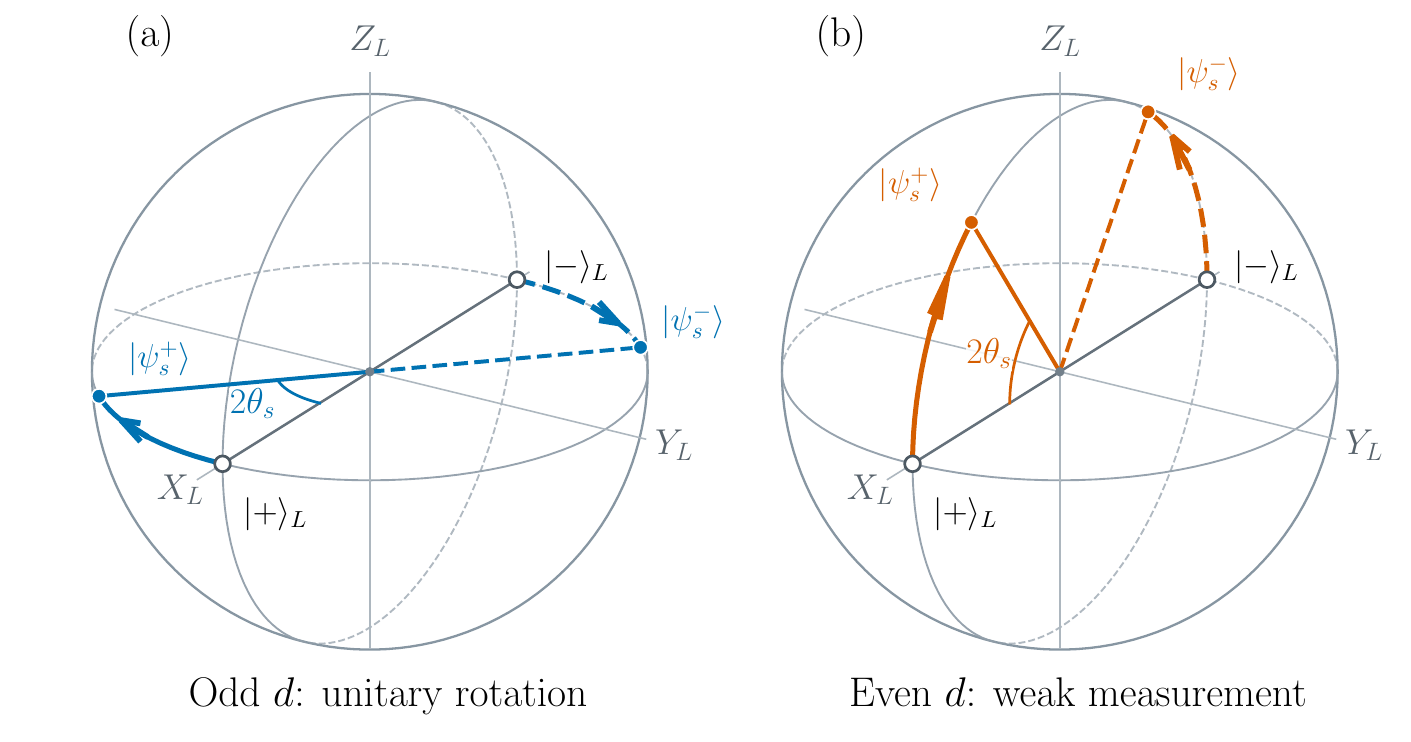}
    \caption{\textbf{Conditional logical operations on the Bloch sphere.}
    Two illustrative syndrome branches acting separately on the logical inputs $|\pm\rangle_L$, after reference correction and before logical recovery. Open markers indicate the inputs; filled markers indicate the normalized conditional outputs $|\psi_s^\pm\rangle$. Solid and dashed colored arrows correspond to the $+$ and $-$ inputs, respectively.
    (a) For odd $d$, the unitary operation in Eq.~\eqref{eq:odd_logical_rotation}, illustrated with $\theta_s=\pi/8$, rotates both Bloch vectors about $Z_L$ by $-2\theta_s=-\pi/4$ (blue), preserving their antipodal relation and orthogonality $\langle\psi_s^+|\psi_s^-\rangle=0$.
    (b) For even $d$, the conditional weak measurement in Eq.~\eqref{eq:even_logical_filter}, illustrated with $\sin(2\theta_s)=\tanh(2\alpha_s)=0.8$, tilts both outputs toward $+Z_L$ (orange), making them nonorthogonal $\langle\psi_s^+|\psi_s^-\rangle=0.8$. All illustrated conditional outputs remain pure. The distributions of both logical rotation angles and logical tilt angles $\theta_s$ are shown in Fig.~\ref{fig:angle_d128_129}.
    }
    \label{fig:logical_bloch_vector}
\end{figure}

For odd $d$, phase pinning in Eq.~\eqref{eq:phase_case} implies that the amplitude ratio is purely imaginary. The logical rotation angle $\theta_s$ can be defined as
\begin{equation}
\frac{A_1^s}{A_0^s}=i\tan\theta_s,
\qquad
-\frac{\pi}{2}\leq\theta_s<\frac{\pi}{2}.
\end{equation}
The conditional logical Kraus operator $K_s=A_0^sI_L+A_1^sZ_L$ is a unitary rotation about the logical $Z_L$ axis:
\begin{equation}
 \frac{K_s}{\sqrt{D_s}}=\left(\cos\theta_sI_L+i\sin\theta_sZ_L\right)=e^{i\theta_sZ_L},
 \label{eq:odd_logical_rotation}
\end{equation}
up to a global phase. Thus, we have $K_s^\dagger K_s=D_s I_L$ and $m_s=0$: after removal of its scalar factor and global phase, every nonzero syndrome branch is a unitary logical rotation \cite{BravyiEtAl2018}. See Fig.~\ref{fig:logical_bloch_vector}(a).

\subsubsection{Even distance: effective logical measurement}\label{subsubsec:even_parity}
For even $d$, phase pinning makes the amplitude ratio real. First consider a nonzero branch with $|A^s_1/A^s_0|\equiv r_s<1$. We define the real non-compact parameter $\alpha_s$ by
\begin{equation}
\frac{A_1^s}{A_0^s}=\tanh\alpha_s,
\qquad \alpha_s\in\mathbb R.
\end{equation}
Up to a global phase, the conditional logical Kraus operator takes the form
\begin{equation}
\frac{K_s}{\sqrt{D_s}}=\frac{\cosh\alpha_s I_L+\sinh\alpha_s Z_L}{\sqrt{\cosh(2\alpha_s)}}
=\frac{e^{\alpha_sZ_L}}{\sqrt{\cosh(2\alpha_s)}}.
\label{eq:even_logical_filter}
\end{equation}
The exponential weights the two logical $Z_L$ components by $e^{\pm\alpha_s}$, producing a weak measurement that is nonunitary unless $\alpha_s=0$. 
Thus $\alpha_s$ is the weak-measurement strength that controls relative amplitudes \cite{VennBeri2020}.

To visualize the measurement backaction on the Bloch sphere, we introduce the parameter \(\theta_s\in(-\pi/4,\pi/4)\) through 
\begin{equation}
\tan\theta_s=\tanh\alpha_s,\label{eq:even_logical_tilt}
\end{equation}
such that \(\sin(2\theta_s)=\tanh(2\alpha_s)=m_s\).
For the logical inputs \(|\pm\rangle_L\), the normalized conditional outputs are tilted by \(2|\theta_s|\) from the equator toward the north pole for \(\alpha_s>0\), or the south pole for \(\alpha_s<0\), as illustrated in Fig.~\ref{fig:logical_bloch_vector}(b). 
The outputs remain pure but become nonorthogonal, with overlap \(\langle\psi_s^+|\psi_s^-\rangle=\sin(2\theta_s)\). 
Thus, \(\alpha_s\) controls the logical weak-measurement strength, while \(\theta_s\) gives the logical tilt angle on these two inputs.

For $r_s>1$, exchanging $A_0^s$ and $A_1^s$, equivalently replacing $K_s$ by $Z_LK_s$, gives the same parameterization with the inverted ratio $A_0^s/A_1^s=\tanh\alpha_s$. 
The original Kraus operator then contains an additional logical $Z_L$ factor. 
At $r_s=1$, the ratios $A_1^s/A_0^s=\pm1$ correspond to the limits $\alpha_s\to\pm\infty~(\theta_s\to\pm \pi/4)$, in which the normalized operator becomes a rank-one projector.

\begin{figure*}[t]
    \noindent
    \includegraphics[width=1.0\linewidth]{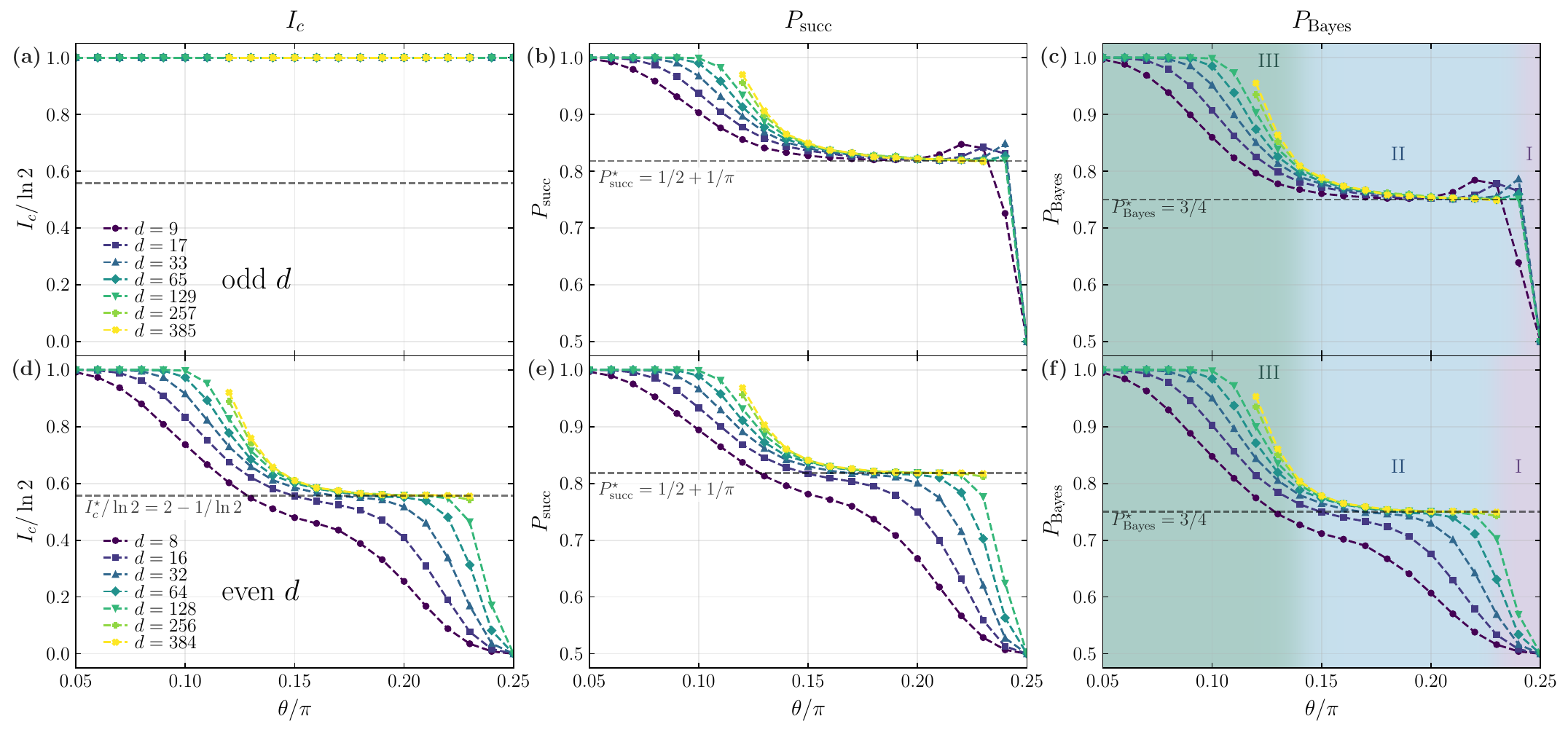}
    \caption{\textbf{Code distance dependence of measures under uniform coherent $Z$ rotations.}
    Columns show $I_c$, $P_{\mathrm{succ}}$, and $P_{\mathrm{Bayes}}$ versus $\theta/\pi\in[0.05,0.25]$; the upper and lower rows show odd and even distances, respectively. Averages use the Born weights $p_s=D_s$ for maximally mixed logical input. Each point uses $50{,}000$ syndrome samples, except $d=384,385$, which use $10{,}000$; the four largest distances cover only $0.12\leq\theta/\pi\leq0.23$. One-standard-error bars are smaller than the markers and are omitted. The odd coherent-information curves show the exact phase-pinning result $I_c/\ln2=1$. Odd-distance Pauli curves show a bump before the SWAP point $\theta=\pi/4$. Horizontal dashed lines indicate the reference values $P_{\mathrm{succ}}^*=1/2+1/\pi$, $P_{\mathrm{Bayes}}^*=3/4$, and $I_c^{*,\mathrm{even}}/\ln2=2-1/\ln2$, obtained by assuming a uniform Born-weighted distribution of the folded logical angle. Panels (c) and (f) schematically mark the finite-size crossovers for the largest simulated size in the odd and even cases, respectively.  Regimes I, II, and III correspond to the nearly-ballistic, intermediate diffusive, and insulator regimes of Fig.~\ref{fig:setup_phase_diagram} (see also Table~\ref{tab:bulk_logical_regimes}). The numerical simulations are based on the Majorana simulator developed in Ref.~\cite{PatoStaplesMsim}.}
    \label{fig:measure_parity}
\end{figure*}

\subsection{Optimal decoding and common-threshold equality}\label{subsec:common_threshold}
Under the phase-pinning condition of Sec.~\ref{subsec:logical_parity}, we first examine exact quantum recovery at odd distance. For even distance, we establish that the deficits of $P_{\rm succ}$, $P_{\rm Bayes}$, and $I_c$ obey size-independent bounds that define the same asymptotically correctable set and enforce a common threshold. Furthermore, we prove the exact finite-size optimality of Pauli MLD, showing that it achieves the maximum entanglement fidelity over all deterministic syndrome-conditioned CPTP recoveries.

\subsubsection{Odd distance: exact recoverability}
For odd $d$, every nonzero syndrome branch acts as the logical unitary $e^{i\theta_sZ_L}$ after normalization. 
A recovery with access to the syndrome can apply the inverse rotation $U_s=e^{-i\theta_sZ_L}$ to restore any logical input, although this generally requires a non-Clifford logical rotation~\cite{GottessmannKnill} that needs magic~\cite{BravyiKitaev2005}. 
Equivalently, the logical information remains intact, thus
\begin{equation}
I_c(s)=\ln 2,
\quad
I_c=\sum_sD_s I_c(s)=\ln 2,
\quad d\ \text{odd}.
\label{eq:odd_exact_coherent_information}
\end{equation}
This also follows directly by substituting $m_s=0$ into Eq.~\eqref{eq:Ic_ms_def}.
Remarkably, Eq.~\eqref{eq:odd_exact_coherent_information} holds for every odd size and any physical error angle~\cite{BravyiEtAl2018,behrends2025surface}, proving that the full quantum recoverability is guaranteed whenever the syndrome-dependent logical rotation $\theta_s$ is known.
Perfect coherent recovery does not imply perfect Pauli decoding, which restricts the action to \(I_L\) or \(Z_L\). The logical-sector weights are $q^s_{0}=\cos^2\theta_s$ and $q^s_{1}=\sin^2\theta_s$, so the Pauli measures can be expressed in logical rotation angle $\theta_s$:
\begin{equation}
    \begin{gathered}
P_{\mathrm{succ}}(s)=\max\{\cos^2\theta_s,\sin^2\theta_s\},\\
P_{\mathrm{Bayes}}(s)=\cos^4\theta_s+\sin^4\theta_s.\label{eq:pauli_meas_angle}
    \end{gathered}
\end{equation}
At $|\theta_s|=\pi/4$, both Pauli measures equal $1/2$ while $I_c(s)=\ln 2$. 
The upper row of Fig.~\ref{fig:measure_parity} illustrates this distinction: \(I_c/\ln2=1\) throughout all error angles, while both Pauli measures fall substantially below unity. 
Their variation reflects limitations of Pauli recovery despite exact syndrome-conditioned coherent recoverability.

\subsubsection{Even distance: common-threshold equality}
For even code distance, phase pinning makes the relative amplitude ratio real, giving $m_s^2=4q^s_{0}q^s_{1}$. All three measures can be expressed in terms of the interference parameter $m_s$:
\begin{equation}
1-P_{\mathrm{succ}}(s)=\frac{1-\sqrt{1-m_s^2}}{2},\
1-P_{\mathrm{Bayes}}(s)=\frac{m_s^2}{2}.\label{eq:p_succ_ms}
\end{equation}
Together with $I_c(s)=h_2(\tfrac{1+m_s}{2})$ (Eq.~\eqref{eq:Ic_ms_def}), the deficits of three measures are even functions of $m_s$ and are bounded by positive constants times $m_s^2$. Averaging with the Born weights $p_s=D_s$, and writing $\langle f_s\rangle_s=\sum_sp_sf_s$, gives
\begin{equation}
\begin{gathered}
\frac14\langle m_s^2\rangle_s
\leq1-P_{\mathrm{succ}}\leq\frac12\langle m_s^2\rangle_s,\\
1-P_{\mathrm{Bayes}}=\frac12\langle m_s^2\rangle_s,\\
\frac12\langle m_s^2\rangle_s
\leq\ln2-I_c\leq(\ln2)\langle m_s^2\rangle_s.
\end{gathered}
\label{eq:even_averaged_bounds}
\end{equation}
Details of the proof are given in Appendix~\ref{app:measure_inequalities}. Since the comparison constants are independent of code size, at fixed physical error angle $\theta$ as $d\to\infty$, all three deficits vanish if and only if $\langle m_s^2\rangle_s\to0$. 
\begin{equation}
    P_{\mathrm{succ}}\to1
\Leftrightarrow
P_{\mathrm{Bayes}}\to1
\Leftrightarrow
I_c\to\ln2
\Leftrightarrow
\langle m_s^2\rangle_s\to0.
\end{equation}
The three measures therefore define the same asymptotically correctable set. Whenever this set has a well-defined threshold,
\begin{equation}
\theta_c(P_{\mathrm{succ}})
=\theta_c(P_{\mathrm{Bayes}})
=\theta_c(I_c),\quad d\ \text{even}.
\label{eq:even_common_threshold}
\end{equation}
The bounds constrain the deficits within constant factors but do not determine their asymptotic ratios, finite-size crossing positions, or detailed scaling functions. Odd-weight stabilizer deformations can change the relative phase, so the common-threshold argument does not generally extend to those geometries; see Appendix~\ref{app:odd_weight_limits}.

Independently of phase pinning, the two Pauli measures satisfy bounds
\begin{equation}
\frac12\left(1-P_{\mathrm{Bayes}}\right)\leq 1-P_{\mathrm{succ}}\leq \left(1-P_{\mathrm{Bayes}}\right),
\label{eq:even_pauli_bounds}
\end{equation}
which sharpens the lower bound from $1/(4\ln2)$ given in Ref.~\cite{YanBaoVijay2026} to $1/2$ in our result (Appendix~\ref{app:measure_inequalities}).

\subsubsection{Even distance: optimality of maximum-likelihood recovery}
\label{subsubsec:even_decoder_optimality}
Under the even phase-pinning condition,  we prove that maximum-likelihood Pauli decoding maximizes entanglement fidelity over \emph{all} deterministic, syndrome-conditioned CPTP recoveries $\{\mathcal{R}_s\}$.

We quantify recovery by the entanglement fidelity $F_e$ for the maximally mixed logical input $I_L/2$, purified by the Bell state $|\Phi\rangle_{LR} = (|00\rangle + |11\rangle)/\sqrt{2}$~\cite{FletcherShorWin2007,Schumacher1996,Nielsen1996EntanglementFidelity}. 
For any candidate CPTP map $\mathcal{R}_s$ acting on the normalized conditional state $|\psi_s\rangle_{LR} = (K_s \otimes I_R)|\Phi\rangle_{LR}/\sqrt{D_s}$, the conditional entanglement fidelity is bounded for arbitrary Kraus operators by
\begin{equation}
F_e(s; \mathcal{R}_s) \le F_e^{\mathrm{opt}}(s) = \frac{(\mathrm{tr}|K_s|)^2}{4 D_s} = \frac{1 + \sqrt{1 - m_s^2}}{2},
\label{eq:main_recovery_fidelity_bound}
\end{equation}
where $|K_s| \equiv (K_s^\dagger K_s)^{1/2}$ (see Appendix~\ref{app:decoder_optimality}). The bound is saturated by the inverse unitary in the polar decomposition $K_s = W_s |K_s|$. 
Under even phase pinning, $W_s^\dagger$ reduces precisely (up to an overall phase) to the Pauli correction prescribed by MLD: $I_L$ when $|A_s^0| \ge |A_s^1|$ and $Z_L$ when $|A_s^1| > |A_s^0|$. 
Because the right-hand side of Eq.~\eqref{eq:main_recovery_fidelity_bound} matches $P_{\mathrm{succ}}(s)$ (Eq.~\eqref{eq:p_succ_ms}), MLD independently saturates the optimal fidelity for every supported syndrome: $F_e^{\mathrm{opt}}(s) = F_e^{\mathrm{MLD}}(s) = P_{\mathrm{succ}}(s)$.
Averaging over the Born distribution yields the exact ensemble equality
\begin{equation}
F_e^{\mathrm{opt}} \equiv \max_{\{\mathcal{R}_s\in\mathrm{CPTP}\}} F_e(\{\mathcal{R}_s\}) = F_e^{\mathrm{MLD}} = P_{\mathrm{succ}}.
\label{eq:even_optimal_recovery}
\end{equation}
This equality establishes that MLD and optimal quantum recovery share the same asymptotically correctable set, extending the common-threshold equality (Eq.~\eqref{eq:even_common_threshold}) to optimal deterministic CPTP operations. 

The unequal eigenvalues of $|K_s|$ correspond to non-degenerate Schmidt coefficients in the conditional state, reflecting irreversible logical $Z_L$ information acquired by the syndrome measurement. 
The minimum achievable entanglement infidelity is strictly $1 - P_{\mathrm{succ}}$. 
Consequently, the even-distance plateaus in Fig.~\ref{fig:measure_parity} represent the ultimate quantum-optimal performance in the finite size; no post-measurement quantum recovery can lift them.

\subsection{Fidelity at the maximum error angle}\label{subsec:swap_endpoint_recovery}
For the square rotated surface codes considered here, the logical operation at the maximum error angle \(\theta=\pi/4\) can be determined exactly. 
For every syndrome with \(D_s>0\), the Kraus operator takes the form, up to an overall phase,
\begin{equation}
\frac{K_s}{\sqrt{D_s}}
=\frac{1}{\sqrt2}
\begin{cases}
I_L\pm i Z_L,&d\ \mathrm{odd},\\
I_L\pm Z_L,&d\ \mathrm{even}.
\end{cases}
\label{eq:swap_endpoint_kraus}
\end{equation}
The sign \(\pm\) depends on the syndrome and the reference correction. 
These forms follow from the explicit stabilizer algebra at \(\theta=\pi/4\); the detailed operator proofs are given in Appendix~\ref{app:swap_endpoint_recovery}.

For both odd and even $d$, the two logical-sector amplitudes therefore have equal magnitudes,
\begin{equation}
    |A_0^s|^2=|A_1^s|^2=\frac{D_s}{2}.
\end{equation}
Consequently, \(q_0^s=q_1^s=1/2\) on every supported syndrome: neither Pauli correction is preferred, and both conditional Pauli measures equal \(1/2\). Born averaging gives the exact finite-distance result
\begin{equation}
    P_{\mathrm{succ}}=P_{\mathrm{Bayes}}=\frac12,
\end{equation}
accounting for the common values at $\pi/4$ in Fig.~\ref{fig:measure_parity}.

Despite these identical Pauli fidelities, quantum recoverability differs sharply between the two parities. 
At odd distance, each branch is proportional to the unitary \(e^{\pm i\pi Z_L/4}\), which can be reversed using the syndrome record, and \(I_c=\ln2\). 
At even distance, each branch is proportional to a rank-one logical \(Z_L\) projector: the syndrome measurement reveals the logical \(Z_L\) eigenvalue and \(I_c=0\). 
These values depend on the code geometry and boundary conditions. 
For example, in the cylindrical geometry studied by Ref.~\cite{YanBaoVijay2026}, both Pauli decoding measures equal either \(1/2\) or \(1\), depending on the circumference and length of the cylinder. 
In contrast, the square rotated patches considered here have \(P_{\mathrm{succ}}=P_{\mathrm{Bayes}}=1/2\) at every code distance. 
The corresponding geometry dependence reflects how syndrome projection and boundary conditions determine the logical operation at the maximum error angle $\pi/4$.

\section{From coherent amplitudes to a class-D Majorana network}\label{sec:rbim_maj_net}
The preceding logical-space analysis establishes exact relations between the recovery measures and constrains the relative phase of $A_0^s$ and $A_1^s$. 
Their magnitudes, and hence the remaining dependence of logical recovery on $\theta$ and $d$, require a microscopic evaluation of the coherent sums over physical error configurations.

Building on the statistical-mechanical formulation of surface-code decoding \cite{DennisEtAl2002, WANGPreskill2003}, we express the two logical-sector amplitudes as RBIM partition functions with complex couplings \cite{VennBehrendsBeri2023, YanBaoVijay2026}.
The two sectors are distinguished by a logical defect, while pairing the amplitudes with their complex conjugates gives the physical sector weights and logical interference.
We then construct the row-transfer matrices and apply a Jordan--Wigner transformation to obtain Majorana bilinear operators \cite{Chalker2002Mapping, GruzbergReadLudwig2001, ReadLudwig2000}.
This yields a free-fermion representation for evaluating both amplitudes and an associated Majorana network in symmetry class D \cite{VennBehrendsBeri2023, FavaEtAl2023, WangEtAl2025SelfDual}.
At $\theta=\pi/4$, the network exhibits deterministic signed-SWAP propagation.
We introduce a moving Majorana frame that makes the bulk propagation clean, providing a natural starting point for the bulk effective theory developed in Sec.~\ref{sec:nlsm}.

\subsection{Complex random-bond Ising representation}\label{subsec:complex_rbim}
For a syndrome $s$, let $C_s$ be the reference string and let $\Gamma$ represent the nontrivial logical string. A representative of logical sector $\mu\in\{0,1\}$ is $C^{s}_{\mu}=C_s\oplus\mu\Gamma$. Associate with it an Ising bond field $\eta^{s}_{\mu,\tilde{l}}=(-1)^{C^{s}_{\mu,l}}=\pm1$. 
Here \(l\) labels a physical-qubit edge and \(\tilde l=\langle ij\rangle\) is the corresponding dual edge; the Ising spins \(\sigma_i=\pm1\) reside on dual vertices.
We fix the sign convention through
\begin{equation}
    \begin{aligned}
    \mathcal{Z}[\eta^s_\mu,\beta_p]&=\sum_{\{\sigma\}}\exp(\beta_p\sum_{\langle ij\rangle}\eta^s_{\mu,ij}\sigma_i\sigma_j),
    \end{aligned}
\end{equation}
with Hamiltonian $H=-\sum_{\langle ij\rangle}\beta_p\eta^s_{\mu,ij}\sigma_i\sigma_j$ \cite{DennisEtAl2002}. We consider $\beta_p$ as the dimensionless coupling.
Corresponding to the stabilizer deformation of the error string in Eq.~\eqref{eq:error_string_deform}, the domain-wall expansion of $\mathcal{Z}[\eta^s_\mu,\beta_p]$ sums over all allowed local deformations of $C^s_\mu$
\begin{equation}
\mathcal{Z}[\eta^s_\mu,\beta_p]
=
e^{N\beta_p}
\sum_{B}
\left(e^{-2\beta_p}\right)^{|C^s_\mu\oplus B|},
\end{equation}
Thus, matching this expression to the error string expansion of amplitude $A_\mu^s=(\cos\theta)^N\sum_{E|[E\oplus C_s]=\mu}(i\tan\theta)^{|E|}$ in Eq.~\eqref{eq:A_mu_def}, the coherent error amplitudes are reproduced by choosing the complex coupling
\begin{equation}
    e^{-2\beta_p}=i\tan\theta .
\end{equation}
On the principal branch for $0<\theta\leq\pi/4$,
\begin{equation}
    \beta_p=\beta_R+i\beta_I,\quad\beta_R=-\frac12\ln\tan\theta,\quad\beta_I=-\frac{\pi}{4}.\label{eq:beta_p_r}
\end{equation}
Consequently, the Kraus amplitudes and the Ising partition functions are related by a common factor, $A_\mu^s=\mathcal N_\theta\mathcal{Z}[\eta^s_\mu,\beta_p],$ with $\mathcal N_\theta=(\cos\theta/e^{\beta_p})^N$ independent of $s$ and $\mu$ for a fixed lattice and boundary convention. 

Because $\beta_p$ is complex, a single partition function $\mathcal{Z}[\eta^s_\mu,\beta_p]$ represents a specific logical sector amplitude and cannot be interpreted as a probability. The syndrome probability instead contains forward and backward complex-conjugate amplitudes:
\begin{equation}
    \begin{aligned}
    p_s&=\sum_{\mu=0,1}|A_\mu^s|^2=
    |\mathcal N_\theta|^2
    \sum_{\mu=0,1}
    \mathcal{Z}[\eta^s_\mu,\beta_p]
    \mathcal{Z}[\eta^s_\mu,\beta_p^*]\\
    &=|\mathcal N_\theta|^2\sum_{\mu}\sum_{\{\sigma\},\{\tau\}}\exp\left[\sum_{\langle ij\rangle}\eta^s_{\mu,ij}(\beta_p\sigma_i\sigma_j+\beta^*_p\tau_i\tau_j)\right].
    \end{aligned}
\end{equation}
The corresponding Hamiltonian is
\begin{equation}
    H[\sigma,\tau]=-\sum_{\langle ij\rangle}\beta_p\eta_{ij}\sigma_i\sigma_j-\sum_{\langle ij\rangle}\beta^*_p\eta_{ij}\tau_i\tau_j.
\end{equation}
The Ising spins \(\sigma\) and \(\tau\) are summed independently, but the two layers share the same syndrome- and sector-dependent bond field $\eta$. The labels forward and backward refer to the amplitude and its complex conjugate. 

The double-layer structure is a direct consequence of the coherent error channel, which retains the interference between distinct error strings within a logical sector, like $A_\mu^s(A_\mu^s)^*$.
Interference between different logical sectors instead involves products $A_\mu^s(A_\nu^s)^*$ with $\mu\ne\nu$, for which the two layers share the syndrome but carry different logical defects. 
If the interferences are neglected (i.e., assume $\sigma_i\sigma_j=\tau_i\tau_j$), we get the Pauli twirling result: $\beta_p+\beta^*_p=2\beta_R=-\ln{\tan{\theta}}$ leads to the stochastic error rate $p=\sin^2\theta$ ($p_c\approx0.11\to\theta_c\approx0.10\pi$) \cite{BravyiEtAl2018, EmersonEtAl2007, SilvaEtAl2008}.

The two logical sectors correspond to bond configurations related by the logical string $\Gamma$, $\eta^s_{1,\tilde l}=\eta^s_{0,\tilde l}(-1)^{\Gamma_l}$, where $\Gamma_l\in\{0,1\}$ indicates whether the physical edge $l$ belongs to $\Gamma$. 
This change preserves the measured syndrome but introduces a logical defect. 
In the diagonal sector weight, the same defect is inserted into both conjugate layers. 
We define the domain-wall free energy (DWFE) of our double-layer complex RBIM by
\begin{equation}
    \Delta F_s
    =-\ln\frac{|A_1^s|^2}{|A_0^s|^2}
    =-\ln
    \frac{\mathcal Z[\eta^s_1,\beta_p]\mathcal Z[\eta^s_1,\beta_p^*]}
         {\mathcal Z[\eta^s_0,\beta_p]\mathcal Z[\eta^s_0,\beta_p^*]}.\label{eq:dwfe}
\end{equation}
The signed free energy $\Delta F_s$ favors the reference logical sector when positive and the opposite sector when negative. 
It is finite when both sector weights are nonzero; a vanishing sector weight leads to the corresponding infinite limit \cite{YanBaoVijay2026, DWRG1984, Chalker2002Mapping, BaoAnand2026, WanDaiZhu2026}.

Because $\Delta F_s = -\ln r_s^2$ with the magnitude ratio $r_s=|A^s_1/A^s_0|$, a large free-energy cost ($|\Delta F_s| \gg 1$) reflects sharp dominance of one logical sector ($\min(r_s,1/r_s) \to 0$), whereas $\Delta F_s \to 0$ indicates degenerate sector weights ($r_s \to 1$). 
Since both $P_{\mathrm{succ}}$ and $P_{\mathrm{Bayes}}$ relate to the magnitude ratio, the statistical-mechanical representation connects directly to the recovery measures.
We give the detailed analysis of DWFE and recovery measures in Sec.~\ref{sec:near_swap}.
The asymptotic recovery threshold thus coincides with the ordering transition where the domain-wall free energy diverges in the thermodynamic limit.

\subsection{Transfer matrix and Majorana network}\label{subsec:maj_network}
Following the established RBIM-to-fermion mapping~\cite{ChoFisher1997, Cho1997Thesis}, we express each logical-sector partition function as a product of local row-transfer operators. 
A Jordan--Wigner transformation converts these operators into Majorana bilinears, providing a free-fermion representation and an associated scattering-network description \cite{ChalkerCoddington1988, Chalker2002Mapping, GruzbergReadLudwig2001, YanBaoVijay2026}. 
We first construct the forward layer with coupling $\beta_p$; its complex-conjugate counterpart supplies the backward layer with coupling $\beta_p^*$ required by physical weights and interference. 

\begin{figure}[t]
    \centering
    \includegraphics[width=1.0\linewidth]{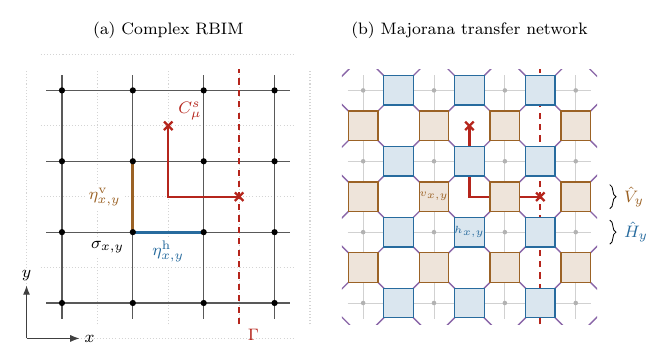}
\caption{\textbf{Majorana representation of the RBIM.} Complex RBIM and its Majorana network representation for a fixed syndrome \(s\) and logical sector \(\mu\). (a) Black vertices represent Ising spins \(\sigma_{x,y}\). The red reference error string \(C^s_\mu\) runs on the dual lattice, with crosses marking its syndrome endpoints; the Ising bonds it crosses carry \(\eta=-1\). The bond variables \(\eta^{\mathrm h}_{x,y}\) and \(\eta^{\mathrm v}_{x,y}\) label the rightward and upward links from \((x,y)\), respectively. The red dashed line \(\Gamma\) represents the logical defect associated with \(Z_\Gamma\): inserting it flips \(\eta^{\mathrm h}_{x,y}\to-\eta^{\mathrm h}_{x,y}\) on every horizontal bond crossed by \(\Gamma\). This is equivalent to replacing periodic with antiperiodic Ising boundary conditions in the \(x\) direction. (b) Chalker-Coddington network model of the fermion in symmetry class D~\cite{Chalker2002Mapping,GruzbergReadLudwig2001,FavaEtAl2023,WangEtAl2025SelfDual,JianEtAl2022,JianShapourianBauerLudwig2023}. The corresponding horizontal and vertical transfer factors are represented by blue \(h_{x,y}\) and brown \(v_{x,y}\) nodes at the respective bond midpoints. The corresponding red dashed line only passes beneath the blue nodes and implements the same sign reversal in their horizontal couplings, \(h_{x,y}(\eta^{\mathrm h}_{x,y})\to h_{x,y}(-\eta^{\mathrm h}_{x,y})\). Transfer proceeds upward along the $y$ direction, with each row unit \(\hat T_y=\hat V_y\hat H_y\) applying the horizontal layer first and then the vertical layer.
}
    \label{fig:rbim_majorana}
\end{figure}

For convenience, we construct the transfer matrix on an $L_x\times L_y$ cylinder, periodic in $x$ and open in $y$. This choice of boundary conditions is not expected to affect the bulk physics.
With fixed syndrome $s$ and logical sector $\mu$, we write $\eta=\eta^s_\mu$ and suppress these labels below for simplicity.
The partition function can be expressed as a product of row transfer matrices,
\begin{equation}
\mathcal Z[\eta,\beta_p]
=\mathcal A[\eta,\beta_p]
\langle\psi_f|\hat H_{L_y}
\hat T_{L_y-1}\cdots\hat T_1|\psi_i\rangle,\label{eq:partition_function_tm}
\end{equation}
with the transfer unit for a single row, $\hat T_y=\hat V_y\hat H_y$. 
$|\psi_i\rangle$ and $|\psi_f\rangle$ encode the chosen bottom and top boundary conditions. Here, we choose free boundary conditions $|\psi_i\rangle=|\psi_f\rangle=|+\rangle_{x}^{\otimes L_x}$.
The factor $\mathcal A[\eta,\beta_p]$ collects the scalar normalizations of the vertical-bond transfer weights. These factors do not affect conjugation of the Majorana operators, but the scalar factors must be retained in partition functions and logical-sector ratios. 
\(\eta^\mathrm{h}_{x,y}\) and \(\eta^\mathrm{v}_{x,y}\) denote the bond signs on the rightward and upward links from \((x,y)\), respectively (see Fig.~\ref{fig:rbim_majorana}(a)).
The horizontal and reduced vertical factors are
\begin{equation}
    \begin{aligned}
\hat H_y&=\exp\!\left[\sum_x\beta_p\eta^\mathrm{h}_{x,y}\sigma_{x}^z\sigma_{x+1}^z\right],\\
\hat V_y&=\exp\!\left[\sum_x\widetilde\beta_{x,y}\sigma_{x}^x\right].\label{eq:transfer_VH}
\end{aligned}
\end{equation}
Here $\sigma^{x,z}$ are Pauli operators in the auxiliary row-spin space with PBC in the $x$-direction. The vertical dual coupling is $\widetilde\beta_{x,y}=\operatorname{arctanh}(e^{-2\beta_p\eta^\mathrm{v}_{x,y}})$.

The Jordan--Wigner transformation gives
\begin{equation}
\sigma_{x}^x\longrightarrow i\gamma_{2x-1}\gamma_{2x},\quad
\sigma_{x}^z\sigma_{x+1}^z\longrightarrow i\gamma_{2x}\gamma_{2x+1}.
\end{equation}
$\gamma_{2x-1}$ and $\gamma_{2x}$ are Majorana operators in the $x$-th unit cell that act on the same auxiliary row space at every transfer step.
Then, we have the Majorana representation of the transfer matrices within a fixed fermion-parity sector \cite{VennBehrendsBeri2023, YanBaoVijay2026, SchultzMattisLieb1964, GruzbergReadLudwig2001, Chalker2002Mapping}
\begin{equation}
    \begin{aligned}
\hat{H}_y=&\prod_{x=1}^{L_x}h_{x,y}=\prod_{x=1}^{L_x}\exp{\left[\beta_p\eta^\mathrm{h}_{x,y}i\gamma_{2x}\gamma_{2x+1}\right]},\\
        \hat{V}_y=&\prod_{x=1}^{L_x}v_{x,y}=\prod_{x=1}^{L_x}\exp{\left[\widetilde{\beta}_{x,y}i\gamma_{2x-1}\gamma_{2x}\right]},\label{eq:transfer_node}
    \end{aligned}
\end{equation}
Here, the row transfer matrix $\hat{H}_y~(\hat{V}_y)$ can be factored into Majorana bilinear operations on horizontal (vertical) links, denoted as $h_{x,y}~(v_{x,y})$ (Fig.~\ref{fig:rbim_majorana}(b)).
Each local factor is Gaussian: its conjugation acts linearly on the Majorana operators. The fixed-syndrome transfer problem is therefore a free-fermion problem. 
The closing horizontal bond carries the fermion-parity factor associated with the periodic row, together with the logical-defect data \cite{Chalker2002Mapping, YanBaoVijay2026}.

The linear Majorana transfer relations can be recast as scattering relations between incoming and outgoing channels, defining a Majorana network with alternating horizontal and vertical nodes
~\cite{Chalker2002Mapping, VennBehrendsBeri2023}.
This reformulation does not require the transfer operators themselves to be unitary.
The associated network belongs to class D: it has particle--hole symmetry with $\mathcal C^2=+1$, but no independent time-reversal or chiral symmetry~\cite{Zirnbauer1996, AltlandZirnbauer1997, VennBehrendsBeri2023, YanBaoVijay2026}.

\subsection{SWAP point and clean-SWAP frame}\label{subsec:signed_swap}
\paragraph{Signed SWAP point}
At $\theta=\pi/4$, the elementary transfer factors exchange the participating Majoranas accompanied by local bond signs $\eta^\mathrm{h}_{x,y}$ and $\eta^\mathrm{v}_{x,y}$, yielding deterministic bulk propagation without local backscattering.
Thus, we call it the signed SWAP point.

For a fixed syndrome and logical sector, the horizontal and vertical transfer couplings in Eq.~\eqref{eq:transfer_VH} both separate exactly into their values at the SWAP point and remainders that vanish as \(\theta\to\pi/4\).
On the branch continuous from $\widetilde\beta_{x,y}=i\eta^\mathrm{v}_{x,y}\pi/4$ at $\theta=\pi/4$, they take the form
\begin{equation}
\begin{aligned}
    \beta_p\eta^\mathrm{h}_{x,y}
    &=-i\frac{\pi}{4}\eta^\mathrm{h}_{x,y}
      +\beta_R\eta^\mathrm{h}_{x,y},\\
    \widetilde\beta_{x,y}
    &=+i\frac{\pi}{4}\eta^\mathrm{v}_{x,y}
      +i(\theta-\frac{\pi}{4}),\label{eq:beta_parameter}
\end{aligned}
\end{equation}
where $\beta_R=-\tfrac12\ln\tan\theta$. The first terms in both equations specify the signed SWAP factors, while the remaining terms vanish at $\theta=\pi/4$. 

At the SWAP point $\theta=\pi/4$, the reduced horizontal and vertical transfer factors in Eq.~\eqref{eq:transfer_node} therefore become
\begin{equation}
    \begin{aligned}
        h_{x,y}^{(0)}=&\exp\!\left[+\frac{\pi}{4}\eta^\mathrm{h}_{x,y}\gamma_{2x}\gamma_{2x+1}\right],\\
        v_{x,y}^{(0)}=&\exp\!\left[-\frac{\pi}{4}\eta^\mathrm{v}_{x,y}\gamma_{2x-1}\gamma_{2x}\right].\label{eq:swap_node}
    \end{aligned}
\end{equation}
Each SWAP factor transfers either participating Majorana entirely to the other position, multiplied by a random sign $\eta^\mathrm{h}_{x,y}~(\eta^\mathrm{v}_{x,y})$.
For example, $U_\eta=\exp(\eta\pi\gamma_a\gamma_b/4)$ acts as $U_\eta\gamma_aU_\eta^{-1}=-\eta\gamma_b$ and $U_\eta\gamma_bU_\eta^{-1}=\eta\gamma_a$. 
Each SWAP factor exchanges the two participating Majoranas with orientation-dependent signs \(\pm\eta\) (Appendix~\ref{app:swap_intro}).
Composing the horizontal and vertical nodes produces two counterpropagating Majoranas with deterministic trajectories: each incoming Majorana follows a unique outgoing path without splitting or local backscattering.
The random bond signs accumulate along these trajectories, directly entering the boundary-to-boundary transfer amplitudes $\mathcal{Z}[\eta,\beta_p]$ (Eq.~\eqref{eq:partition_function_tm}). Thus, these signs must be retained when constructing physical weights and performing the syndrome average. 
We therefore introduce a clean-SWAP frame through a row-dependent redefinition of the Majorana signs. This frame makes the bulk SWAP propagation clean while retaining the boundary and logical defect information needed to reconstruct the logical amplitudes.

\paragraph{Clean-SWAP frame}
The deterministic trajectories provide a straightforward way to construct the clean-SWAP frame. 
To track propagation between rows, we label the Majoranas on each transfer cut by \(\gamma_{j,y}\). 
The row index \(y\) labels successive transfer cuts of the same auxiliary chain; it does not introduce additional quantum degrees of freedom. The clean-SWAP frame is specified separately on each cut by the accumulated signs \(u_{j,y}\).
Fix the Majorana signs on the initial row ($u_{j,0}=1$) and follow each Majorana through the SWAP network. Along its unique trajectory, record the product of the random signs relative to the corresponding clean-SWAP trajectory. Denoting this accumulated sign for mode $j$ on row $y$ by $u_{j,y}=\pm1$, we define the new Majorana basis
\begin{equation}
    \Gamma_{j,y}=u_{j,y}\gamma_{j,y}.\label{eq:clean_swap_frame}
\end{equation}
The accumulated signs are thereby absorbed into the definitions of the Majorana operators, so their bulk propagation follows the clean-SWAP rule. 
Tracking this transformation is essential when detuning away from $\theta=\pi/4$, as it determines how the syndrome-dependent signs enter the residual bilinears.
The explicit construction is given in Appendix~\ref{app:swap_basis}.

The logical-sector amplitudes depend on the complete boundary-to-boundary contraction, including the accumulated Majorana signs and logical defects. Unitary SWAP factors can therefore yield nonunitary logical Kraus operators after syndrome projection and boundary contraction, as demonstrated by the even-distance result at $\pi/4$ (Sec.~\ref{subsec:swap_endpoint_recovery}). 
The clean-SWAP frame retains these boundary and defect data while exposing the deterministic bulk propagation that provides the starting point for the detuning expansion below.

\section{Class-D field theory for diffusive regime}\label{sec:nlsm}
Building on the Majorana network established in the previous section, we develop a bulk effective field theory for uniform coherent errors by expanding about the signed-SWAP point \(\theta=\pi/4\). 
Our derivation treats spatially uniform rotations directly, without invoking Gaussian-distributed error angles~\cite{YanBaoVijay2026}. 

We first factor out the exact signed-SWAP evolution and identify the leading detuning vertices. 
Aligning the \(n\) forward and \(n\) complex-conjugate Majorana layers then exposes the \(O(2n)\) flavor symmetry. 
After averaging over the shared syndrome signs, the connected second cumulant generates an inter-flavor quartic interaction of order \(\beta_R^2\).
Next, the continuum construction retains a staggered channel compatible with clean propagation and describes its slowly varying collective orientations. 
Using antisymmetric Hubbard–Stratonovich fields and expanding about the saddle point, we obtain a class-D nonlinear sigma model (NL$\sigma$M) with target space \(SO(2n)/U(n)\) with bare inverse stiffness \(t_{\sigma,0}\propto\beta_R^2\). 
The field theory is valid beyond the ballistic-to-diffusive crossover scale $d \gg \lambda_{\mathrm{mfp}} \sim \beta_R^{-2}$, where $\lambda_{\mathrm{mfp}}$ represents the Majorana scattering mean free path. 

\subsection{Detuning expansion about signed SWAP}
\label{subsec:detuning_swap}
Near the signed-SWAP point, the uniform real detuning \(\beta_R=-\tfrac{1}{2}\ln\tan\theta\) controls both transfer layers. 
The horizontal coupling changes by \(\beta_R\eta^{\mathrm h}_{x,y}\), while the vertical residual is governed by \(\phi(\beta_R)\equiv\pi/4-\theta\) (Eq.~\eqref{eq:beta_parameter}).
The vertical detuning $\phi(\beta_R)$ has the following expansion:
\begin{equation}
    \phi(\beta_R)=\arctan(\tanh\beta_R)=\beta_R-\frac23\beta_R^3+O(\beta_R^5).
\end{equation}
Both horizontal and vertical generators begin at first order in $\beta_R$, which serves as the expansion parameter for the subsequent derivations of NL$\sigma$M.

The clean-SWAP frame established in Sec.~\ref{subsec:signed_swap} provides a ballistic background about which to expand the Majorana transfer network in $\beta_R$. 
We first factor out the SWAP evolution in the original Majorana basis and then express the remaining operators in the clean frame. 
Throughout this construction, the syndrome and logical sector are fixed. 
We work with the reduced bulk transfer operator, carrying the scalar normalization and logical-defect data separately.

\paragraph{Exact row factorization.}
With the horizontal layer acting first, the reduced row transfer operator is $\hat T_y=\hat V_y\hat H_y$. 
Substituting Eq.~\eqref{eq:beta_parameter} into Eq.~\eqref{eq:transfer_node}, the node decomposition gives
\begin{equation}
\begin{aligned}
    \hat H_y&=\hat H_y^{(0)}\delta\hat H_y,
    &\delta\hat H_y
    &=\exp\!\left[
        i\beta_R\sum_x\eta^{\mathrm h}_{x,y}
        \gamma_{2x}\gamma_{2x+1}
    \right],\\
    \hat V_y&=\hat V_y^{(0)}\delta\hat V_y,
    &\delta\hat V_y
    &=\exp\!\left[
        \phi(\beta_R)\sum_x
        \gamma_{2x-1}\gamma_{2x}
    \right],
\end{aligned}
\end{equation}
where the superscript $(0)$ denotes the SWAP point. $\delta\hat V_y$ has no explicit bond sign; \(\eta^{\mathrm v}_{x,y}\) remains in the signed-SWAP factor and subsequently enters the clean frame.
With $\hat T_y^{(0)}\equiv\hat V_y^{(0)}\hat H_y^{(0)}$, we obtain the exact factorization
\begin{equation}
    \hat T_y
    =\hat T_y^{(0)}
    \left[
        (\hat H_y^{(0)})^{-1}
        \delta\hat V_y
        \hat H_y^{(0)}
    \right]\delta\hat H_y.
    \label{eq:detuning_row_factorization}
\end{equation}
The vertical detuning should be transported through the entire horizontal SWAP layer. 
For a bulk bond, each of its two Majoranas passes through a neighboring horizontal node, giving
\begin{equation}
    (\hat H_y^{(0)})^{-1}
    \gamma_{2x-1}\gamma_{2x}
    \hat H_y^{(0)}=
    -\eta^{\mathrm h}_{x-1,y}\eta^{\mathrm h}_{x,y}
    \gamma_{2x-2}\gamma_{2x+1}.
\label{eq:transported_vertical_bilinear}
\end{equation}
This conjugation determines both the support and the sign of the transported vertical perturbation (see Appendix~\ref{app:swap_perturbation} for details).

\paragraph{Leading perturbation in the clean-SWAP frame.}
We now move to the clean frame $\Gamma_{j,y}=u_{j,y}\gamma_{j,y}$ defined in Eq.~\eqref{eq:clean_swap_frame}. Both detuning terms in Eq.~\eqref{eq:detuning_row_factorization} act on the input slice $y$ and therefore use the same frame signs.
In this frame, the detuning coefficients contain the signs 
\begin{equation}
\begin{aligned}
    \xi^{\mathrm h}_{x,y}
    &=\eta^{\mathrm h}_{x,y}u_{2x,y}u_{2x+1,y},\\
    \xi^{\mathrm v}_{x,y}
    &=\eta^{\mathrm h}_{x-1,y}\eta^{\mathrm h}_{x,y}
      u_{2x-2,y}u_{2x+1,y}.
\end{aligned}
\label{eq:detuning_frame_signs}
\end{equation}
Here $\xi^{\mathrm{h}(\mathrm{v})}_{x,y}$ labels the sign of the transported horizontal (vertical) part. These signs inherit the original bond configuration and the accumulated trajectory signs.
The clean frame removes the signs from the bulk SWAP propagation while retaining them in these detunings and in the transformed boundary data (see Appendix~\ref{app:swap_basis}).

In the clean frame, we define the Hermitian Majorana bilinears that appear in transported detuning terms
\begin{equation}
    B^{\mathrm h}_{x,y}
    =i\Gamma_{2x,y}\Gamma_{2x+1,y},\quad
    B^{\mathrm v}_{x,y}
    =i\Gamma_{2x-2,y}\Gamma_{2x+1,y}.
\label{eq:detuning_bilinears}
\end{equation}
Expressing the detuning part in Eq.~\eqref{eq:detuning_row_factorization} in the clean frame gives the exact ordered product
\begin{equation}
    e^{\mathcal P_y}=
    \exp\!\left[
        i\phi(\beta_R)\sum_x
        \xi^{\mathrm v}_{x,y}B^{\mathrm v}_{x,y}
    \right]
    \exp\!\left[
        \beta_R\sum_x
        \xi^{\mathrm h}_{x,y}B^{\mathrm h}_{x,y}
    \right].
\label{eq:exact_detuning_operator}
\end{equation}
Near \(\beta_R=0\), we choose \(\mathcal P_y\) continuously from \(\mathcal P_y=0\). 
Expanding around $\beta_R=0$ gives $\mathcal P_y=\mathcal P_y^{(1)}+O(\beta_R^2)$ with the leading-order perturbation
\begin{equation}
    \mathcal P_y^{(1)}
    =\beta_R\sum_x
      \xi^{\mathrm h}_{x,y}B^{\mathrm h}_{x,y}
      +i\beta_R\sum_x
      \xi^{\mathrm v}_{x,y}B^{\mathrm v}_{x,y}.
\label{eq:leading_detuning_operator}
\end{equation}
The $O(\beta_R^2)$ remainder begins with the Baker--Campbell--Hausdorff commutator between the two exponents. 
The second-order BCH term remains bilinear and must be accounted for separately when performing the sign average. Appendix~\ref{app:nlsm_derivation} shows that its mean vanishes under sign average.

The leading perturbation in Eq.~\eqref{eq:leading_detuning_operator} couples the counterpropagating SWAP trajectories through bilinear vertices whose coefficients are linear in \(\beta_R\) and retain the syndrome-dependent signs. The transfer problem remains Gaussian for each fixed syndrome and logical sector. 

\subsection{Rényi replicas and flavor symmetry}\label{subsec:replica_symmetry}
To connect the microscopic description of Majorana network to information measures, we evaluate the entropies of the physical output states with the classical syndrome record retained~\cite{GruzbergReadLudwig2001,JianYouVasseurLudwig2020,JianEtAl2022,JianShapourianBauerLudwig2023}. 
At positive integer $n$, each term in the density-matrix moments $\operatorname{tr}\rho^n$ contains $n$ forward and $n$ complex-conjugate backward layers sharing the same syndrome, giving $2n$ Majorana flavors.
Below, we derive these moments and their $n\to1$ limit, retaining the logical-sector weights and defect data needed for coherent information.
We then align the forward and conjugate Majorana bases to expose the $O(2n)$ flavor symmetry of the leading bulk vertices~\cite{BaoChoiAltman2021}.

\paragraph{Rényi $n$-replica construction.}
The replica construction is most transparent when it is derived directly from the coherent information. We follow the notation in Sec.~\ref{subsec:measures}.
Let $\rho_{SLR}$ denote the joint state of the measured logical output and its reference qubit, including the classical syndrome record, and denote its output by $\rho_{SL}\equiv\operatorname{tr}_R\rho_{SLR}$.
We define the Rényi entropy $I_c^{(n)}=S_n(\rho_{SL})-S_n(\rho_{SLR})=\tfrac{1}{1-n}\ln(\operatorname{tr}\rho_{SL}^n/\operatorname{tr}\rho_{SLR}^n)$.
The coherent information is obtained by taking the limit $n\to1$,
\begin{equation}
    I_c
=-\left.\partial_n\left[
\operatorname{tr}\rho_{SL}^n-
\operatorname{tr}\rho_{SLR}^n
\right]\right|_{n=1}.
\end{equation}
The density-matrix trace moments are
\begin{equation}
    \operatorname{tr}\rho_{SLR}^n=\sum_sD_s^n,\quad
    \operatorname{tr}\rho_{SL}^n=\frac{1}{2^n}\sum_s\sum_{\chi=\pm1}(T_\chi^s)^n,
\end{equation}
where we have $T^s_{\chi}=|\mathcal N_\theta|^2|\mathcal{Z}^s_{0}+\chi \mathcal{Z}^s_{1}|^2$ and $D_s=|\mathcal N_\theta|^2(|\mathcal{Z}^s_{0}|^2+|\mathcal{Z}^s_{1}|^2)$. 
$\mathcal N_\theta$ is the common factor relating the RBIM partition functions to the logical amplitudes (see Sec.~\ref{subsec:complex_rbim}). For simplicity, $\mathcal{Z}^s_\mu$ denotes $\mathcal{Z}[\eta^s_\mu,\beta_p]$.
Appendix~\ref{app:coherent_info} gives the replica derivation details.

For positive integer $n$, the moments admit a replicated transfer-network representation. 
Expanding their weights produces terms of the form
\begin{equation}
    \prod_{a=1}^{n}A_{\mu_a}^s(A_{\nu_a}^s)^*
    =
    |\mathcal N_\theta|^{2n}
    \prod_{a=1}^{n}
    \mathcal Z_{\mu_a}^s(\mathcal Z_{\nu_a}^s)^*.
    \label{eq:replica_sector_product}
\end{equation}
In $D_s^n$, the forward and backward sector labels are paired, $\nu_a=\mu_a$. 
In $(T_\chi^s)^n$, the two lists are independently summed with weights $\chi^{\sum_a(\mu_a+\nu_a)}$, retaining cross-sector interference. 
Every term thus contains exactly $n$ forward amplitudes at $\beta_p$ and $n$ conjugate amplitudes at $\beta_p^*$, giving $2n$ Majorana layer flavors. 
In the original transfer representation, sector assignments share the local operators away from the specified logical defects. 
A same-sector product $[\mathcal{Z}^s_\mu(\mathcal{Z}^{s}_\mu)^*]^n$ therefore identifies the common local bulk operator structure. The full moments retain the sector sums, character weights, logical defects, and boundary data, as detailed in Appendix~\ref{app:coherent_info}.
The syndrome weight is already contained in these moments; no additional Born-weight pair is introduced.

The replica construction applies to both even and odd cases, even though odd-distance coherent information is fixed under phase pinning (Eq.~\eqref{eq:odd_exact_coherent_information}), where $T_+^s=T_-^s=D_s$ and $\operatorname{tr}\rho_{SL}^n=2^{1-n}\operatorname{tr}\rho_{SLR}^n$ lead to $I_c=\ln2$.
This constant Rényi entropy does not make the underlying replicated theory trivial: the individual moments depend on the syndrome distribution through $\sum_sD_s^n$, which can vary with angle and system size, but cancels in their ratio for the odd case. 
Pauli-decoding measures probe different combinations of the sector amplitudes and can therefore remain nontrivial. 
For the even case, this cancellation is not enforced, and the exact bounds in Eq.~\eqref{eq:even_averaged_bounds} relate the coherent-information deficit to the Pauli-decoding deficits without requiring identical finite-size behavior. 
Thus, a common local replica structure is compatible with size-parity dependent observables. 
$P_{\mathrm{Bayes}}$ admits a related replica representation with marked sector-weight insertions in Ref.~\cite{YanBaoVijay2026}.

\paragraph{$SO(2n)$ Majorana flavor symmetry.}
Each term in the integer-replica moments (Eq.~\eqref{eq:replica_sector_product}) contains $n$ forward layers at $\beta_p=\beta_R-i\pi/4$ and $n$ conjugate layers at $\beta_p^*=\beta_R+i\pi/4$. 
We denote their clean-frame Majoranas by $\Gamma_{j,y}^{\lambda,a}$, where $j$ labels the site along the auxiliary chain, $y$ labels the input transfer cut, $\lambda\in\{F,B\}$ distinguishes the forward and backward layers, and $a=1,\ldots,n$ labels the replicas. 
We retain these labels until the forward and backward perturbations are expressed in a common basis. At fixed row $y$, we use the fermionic replica convention
\begin{equation}
    \{\Gamma_{j,y}^{\lambda,a},\Gamma_{k,y}^{\lambda',b}\}
    =2\delta_{jk}\delta_{\lambda\lambda'}\delta_{ab}.
\end{equation}
Thus distinct flavors anticommute, while their even operators commute.
Following the Hermitian bilinear in Eq.~\eqref{eq:detuning_bilinears}, define
\begin{equation}
    \begin{aligned}
    B_{x,y}^{\mathrm h;\lambda,a}
    &=i\Gamma_{2x,y}^{\lambda,a}\Gamma_{2x+1,y}^{\lambda,a},\\
    B_{x,y}^{\mathrm v;\lambda,a}
    &=i\Gamma_{2x-2,y}^{\lambda,a}\Gamma_{2x+1,y}^{\lambda,a}.
    \end{aligned}
\label{eq:replicated_bilinears}
\end{equation}
The labels $\mathrm h$ and $\mathrm v$ distinguish the horizontal and transported vertical perturbations, while \(x\) identifies the vertex along the chain, and \(y\) identifies its input cut. 
Both the horizontal and transported vertical bilinears are same-cut operators.
The semicolon separates the node type from the layer and replica labels. We suppress the fixed syndrome and external logical-sector labels. The local bulk coefficients $\xi^{\mathrm h}_{x,y}$ and $\xi^{\mathrm v}_{x,y}$ are shared by the replicas and therefore carry no replica index.

For each replica, the forward row generator is given by Eq.~\eqref{eq:leading_detuning_operator}. 
To fix the conjugation convention, choose the Jordan--Wigner representation in which odd-site Majoranas are real and even-site Majoranas are imaginary. 
The real clean-frame signs preserve this property, and both bilinears in Eq.~\eqref{eq:replicated_bilinears} have real matrix elements within a layer. 
Complex conjugation therefore preserves the horizontal coefficient and reverses the explicit $i$ in the vertical coefficient. 
Identifying the conjugate copy with the $B$-layer fields gives
\begin{equation}
\begin{aligned}
    \mathcal P_y^{(1);F,a}
    &=\beta_R\sum_x\xi^{\mathrm h}_{x,y}B_{x,y}^{\mathrm h;F,a}
      +i\beta_R\sum_x\xi^{\mathrm v}_{x,y}B_{x,y}^{\mathrm v;F,a},\\
    \mathcal P_y^{(1);B,a}
    &=\beta_R\sum_x\xi^{\mathrm h}_{x,y}B_{x,y}^{\mathrm h;B,a}
      -i\beta_R\sum_x\xi^{\mathrm v}_{x,y}B_{x,y}^{\mathrm v;B,a}.
\end{aligned}
\label{eq:forward_backward_perturbations}
\end{equation}
Summing over the replicas gives the horizontal and vertical contributions to the replicated row generator,
\begin{equation}
\begin{aligned}
    \mathcal P_{y,\mathrm h}^{(1)}
    &=\beta_R\sum_x\xi^{\mathrm h}_{x,y}
      \sum_{a=1}^{n}\left(
      B_{x,y}^{\mathrm h;F,a}+B_{x,y}^{\mathrm h;B,a}\right),\\
    \mathcal P_{y,\mathrm v}^{(1)}
    &=i\beta_R\sum_x\xi^{\mathrm v}_{x,y}
      \sum_{a=1}^{n}\left(
      B_{x,y}^{\mathrm v;F,a}-B_{x,y}^{\mathrm v;B,a}\right).
\end{aligned}
\label{eq:raw_replicated_vertices}
\end{equation}
Thus $\sum_{\lambda,a}\mathcal P_y^{(1);\lambda,a}
=\mathcal P_{y,\mathrm h}^{(1)}+\mathcal P_{y,\mathrm v}^{(1)}$. These vertices are invariant under independent global $O(n)$ rotations within forward and backward layers: $\Gamma_{j,y}^{\lambda,a}\longmapsto\sum_{b=1}^{n}(R_\lambda)_{ab}\Gamma_{j,y}^{\lambda,b}$, with $R_\lambda\in O(n)$.
Each sum $\sum_a B_{x,y}^{\mathrm h;\lambda,a}$ or $\sum_a B_{x,y}^{\mathrm v;\lambda,a}$ is invariant under the corresponding rotation. A common flavor sum over both layer types is not yet manifest because the vertical vertex carries their relative minus sign.

To expose the flavor symmetry in the replicated vertices, leave the forward fields unchanged and redefine the backward fields by
\begin{equation}
    \widetilde\Gamma_{j,y}^{B,a}
    =\Omega_{j,y}\Gamma_{j,y}^{B,a},
    \quad \Omega_{2x,y}=\Omega_{2x+1,y}=(-1)^{x+y}.
\label{eq:backward_flavor_relabeling}
\end{equation}
The spatial staggering preserves the horizontal bilinear and reverses the transported vertical bilinear.
The transformed bilinears are defined by substituting $\widetilde\Gamma$ into Eq.~\eqref{eq:replicated_bilinears}. Since $\Omega_{2x,y}\Omega_{2x+1,y}=1$ and $\Omega_{2x-2,y}\Omega_{2x+1,y}=-1$, they obey
\begin{equation}
    \widetilde B_{x,y}^{\mathrm h;B,a}
    =B_{x,y}^{\mathrm h;B,a},
    \qquad
    \widetilde B_{x,y}^{\mathrm v;B,a}
    =-B_{x,y}^{\mathrm v;B,a}.
\end{equation}
The vertical difference in Eq.~\eqref{eq:raw_replicated_vertices} therefore becomes a sum in the aligned basis. We now use a combined flavor index $A=1,\ldots,2n$, with
\begin{equation}
    \Gamma_{j,y}^{A}
    =\begin{cases}
        \Gamma_{j,y}^{F,a},&A=a,\\
        \widetilde\Gamma_{j,y}^{B,a},&A=n+a,
    \end{cases}
    \qquad a=1,\ldots,n.
\label{eq:combined_majorana_flavors}
\end{equation}
The bilinears $B_{x,y}^{\mathrm h;A}$ and $B_{x,y}^{\mathrm v;A}$ are defined by the same site pairs as in Eq.~\eqref{eq:replicated_bilinears}, using these aligned fields. 
Now, the replicated row vertices take the compact form
\begin{equation}
\begin{aligned}
    \mathcal P_{y,\mathrm h}^{(1)}
    &=\beta_R\sum_x\xi^{\mathrm h}_{x,y}
      \sum_{A=1}^{2n}B_{x,y}^{\mathrm h;A},\\
    \mathcal P_{y,\mathrm v}^{(1)}
    &=i\beta_R\sum_x\xi^{\mathrm v}_{x,y}
      \sum_{A=1}^{2n}B_{x,y}^{\mathrm v;A}.
\end{aligned}
\label{eq:aligned_replicated_vertices}
\end{equation}
The flavor contractions are invariant under the same global rotation at every site, 
 $\Gamma_{j,y}^{A}\mapsto\sum_{C=1}^{2n}R_{AC}\Gamma_{j,y}^{C}$ with $R\in O(2n)$, because $\sum_A R_{AC}R_{AD}=\delta_{CD}$. 
This basis makes the $O(2n)$ flavor symmetry of the quartic interaction manifest. 
Its connected subgroup $SO(2n)$ is used in the subsequent saddle-manifold construction. 

The deterministic transformation in Eq.~\eqref{eq:backward_flavor_relabeling} supplements the syndrome-dependent clean-SWAP frame. 
Its spatial staggering aligns the vertices, while the row factor $(-1)^y$ cancels within each same-row bilinear and ensures $\Omega_{j,y}\Omega_{j\pm2,y+1}=1$ along either bulk trajectory. 
The aligned fields therefore share the same clean bulk routing and leading vertex structure. 
Details are in Appendix~\ref{app:backward_clean_frame}.

\subsection{Class-D NL$\sigma$M}\label{subsec:nlsm_rgflow}
Starting from the aligned vertices in Eq.~\eqref{eq:aligned_replicated_vertices}, we first average the shared syndrome signs to obtain the leading inter-flavor interaction. 
We then identify the surviving collective channel and derive its long-wavelength class-D NL\(\sigma\)M, which supplies the starting point for the RG and logical-observable analysis in the next section.

\paragraph{Syndrome averaging and the leading quartic interaction.}
The \(2n\) replicated layers share the same syndrome record, so summing over the syndrome couples their Majorana flavors. 
Accounting for gauge redundancy rewrites this sum as uniform counting over allowed bond-sign configurations, with logical defects, scalar factors, and transformed boundaries retained. 
The replicated amplitudes already contain the physical weights; no additional Born factor is inserted. 
For each retained moment, a uniform local average \(\langle\cdot\rangle_{\mathrm{loc}}\) applies when every sign assignment on its full bond support, including trajectory histories, occurs equally often in the globally allowed configurations (Appendix~\ref{app:random_average}). The physical angle \(\theta\) remains uniform.

Let $\ell=(x,y,\alpha)$ label a spacetime node, with $\alpha=\mathrm h,\mathrm v$. 
Using the aligned bilinear $B_\ell^A=B_{x,y}^{\alpha;A}$, the leading replicated row perturbation in Eq.~\eqref{eq:aligned_replicated_vertices} is
\begin{equation}
\begin{gathered}
    \mathcal P_y^{(1)}=\sum_{\ell\in y}c_\ell\xi_\ell\mathcal B_\ell,
    \qquad \mathcal B_\ell=\sum_{A=1}^{2n}B_\ell^A,
\end{gathered}
\label{eq:replicated_sign_vertices}
\end{equation}
with $c_{\mathrm h}=\beta_R$, $c_{\mathrm v}=i\beta_R$, and $\ell\in y$ restricting the sum to row $y$. 
Each $\xi_\ell$ is a product of fundamental bond signs. 
For the retained bulk vertices, the support rule and cross-row conditioning of Appendix~\ref{app:random_average} give
\begin{equation}
\begin{gathered}
    \langle\xi_\ell\rangle_{\mathrm{loc}}=0,\qquad
K_{\ell\ell'}\equiv\langle\xi_\ell\xi_{\ell'}\rangle_{\mathrm{loc}}=\delta_{\ell\ell'}.
\end{gathered}
\label{eq:syndrome_sign_covariance}
\end{equation}

Let $\mathcal U_I$ denote the ordered residual row product in the interaction picture of $\hat T_0$, the clean SWAP row over all $2n$ aligned flavors. 
For $\ell\in y$, write $\mathcal B_{\ell,I}=\hat T_0^{-y}\mathcal B_\ell\hat T_0^y$. 
The first moment and the mean second-order BCH term vanish under the same counting measure, leaving
\begin{equation}
\begin{gathered}
    \ln\langle\mathcal U_I\rangle_{\mathrm{loc}}
    =\Delta\mathcal L_I^{(2)}+O(\beta_R^3),\\
    \Delta\mathcal L_I^{(2)}
    =\frac12\sum_{\ell,\ell'}c_\ell c_{\ell'}
    K_{\ell\ell'}\,
    \mathcal T\!\left[\mathcal B_{\ell,I}\mathcal B_{\ell',I}\right].
\end{gathered}
\label{eq:syndrome_second_cumulant}
\end{equation}
The sums run over retained bulk nodes; $\mathcal T$ puts later rows to the left and symmetrizes equal-row products. 
Appendix~\ref{app:sign_second_cumulant} gives the ordered expansion without factorizing row averages or assuming vanishing higher sign cumulants.

The diagonal covariance leaves only same-node squares. 
Expanding the flavor sum gives $\frac12c_\ell^2\mathcal B_\ell^2=n c_\ell^2 I+c_\ell^2\sum_{A<B}B_\ell^AB_\ell^B$. 
For complete bulk rows, the constants cancel between equally many horizontal and vertical nodes because $c_{\mathrm h}^2+c_{\mathrm v}^2=0$. 
The remaining row insertion is
\begin{equation}
    \mathcal Q_y^{(2)}=\beta_R^2\sum_x\sum_{1\leq A<B\leq 2n}
    \left[B_{x,y}^{\mathrm h;A}B_{x,y}^{\mathrm h;B}
    -B_{x,y}^{\mathrm v;A}B_{x,y}^{\mathrm v;B}\right].
\label{eq:local_hv_quartic}
\end{equation}
It contributes $\Delta\mathcal L_I^{(2)}=\sum_y\hat T_0^{-y}\mathcal Q_y^{(2)}\hat T_0^y$. 
The relative minus sign comes from $c_{\mathrm v}^2=-\beta_R^2$. 
Each $\mathcal B_\ell$ is invariant under a common global $O(2n)$ rotation, so the quartic has this bulk flavor symmetry. 
Fixed-syndrome transfers remain Gaussian; the interaction arises from averaging their shared signs at uniform $\theta$. 
Ref.~\cite{YanBaoVijay2026} obtains a related quartic from spatial angle fluctuations. 

\paragraph{Staggered channel and class-D NL$\sigma$M.}
To construct a collective-field description, we express the replicated transfer weight as a fermionic path integral. 
Physical Majorana insertions are represented by Grassmann fields $\chi_R^A,\chi_L^A$, whose same-chirality bilinears $M_\rho^{AB}=\chi_\rho^A\chi_\rho^B$ define antisymmetric flavor currents, with $\rho=R,L$ and $1\leq A<B\leq2n$ (Appendix~\ref{app:eft_derivations}). 
The horizontal and vertical quartics in Eq.~\eqref{eq:local_hv_quartic} couple a common left current to right currents separated by one cell, with opposite signs. 
With the Grassmann normalization (Appendix~\ref{app:eft_derivations}), the retained second cumulant has the coupling $g=4\beta_R^2$ and enters the weight as $e^{-S_0+\Delta_g}$, where $S_0$ is the free lattice action for the SWAP point.
The displaced quartic Majorana interaction, coupling right- and left-moving flavor currents, is illustrated below:
\begin{center}
\includegraphics[width=\linewidth]{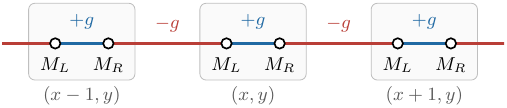}
\end{center}
\begin{equation}
\begin{aligned}
\Delta_g={}&g\sum_{x,y}\sum_{A<B}\bigl[M_R^{AB}(x,y)-M_R^{AB}(x-1,y)\bigr]M_L^{AB}(x,y)\\
={}&g\sum_Q\sum_{A<B}f(K)M_R^{AB}(Q)M_L^{AB}(-Q).
\end{aligned}
\label{eq:bulk_current_form_factor}
\end{equation}
Here $Q=(K,\Omega)$ is the current momentum, the sum of its constituent fermion momenta, with $\Omega$ conjugate to the discrete transfer row. 
The displacement produces $f(K)=1-e^{-iK}$: the two contributions cancel for spatially uniform currents, $f(0)=0$, but reinforce for alternating currents, $f(\pi)=2$. 
The same cancellation affects a literal coincident-current reading of the continuum Hubbard--Stratonovich (HS) action and the subsequent saddle-point construction in Ref.~\cite{YanBaoVijay2026}; its displaced lattice identities remain valid (Appendix~\ref{app:eft_derivations}).

Clean propagation supplies the temporal condition. Currents formed from on-shell right and left movers obey $\Omega+K=0$ and $\Omega-K=0$, respectively, modulo $2\pi$. 
The simultaneous clean-shell candidates are therefore $Q=(0,0)$ and $\Pi=(\pi,\pi)$; the form factor removes the former. 
The spacetime staggering factor $s_{x,y}=(-1)^{x+y}$ stays constant along either clean step $(x,y)\mapsto(x\pm1,y+1)$, because the row sign compensates the spatial sign change. 
A source at $\Pi$ thus preserves trajectory momentum and couples near-shell fermions to near-shell fermions. 
This motivates a continuum theory for a slowly varying envelope about the staggered channel.

We introduce separate antisymmetric horizontal and vertical HS matrices $H(x,y),V(x,y)$, retaining their displaced couplings to the fermions. 
The Gaussian cost penalizes their amplitudes, while fermion integration lowers the effective action in the staggered channel. 
With the full lattice dispersion retained, their balance in the regulated long-trajectory limit gives a stationary saddle with radius $q_\star\propto g$ and positive radial curvature (Appendix~\ref{app:diffusive_saddle}). 
The radius fixes the collective self-energy amplitude, while its flavor orientation $J$ remains degenerate. 
Global $SO(2n)$ rotations relate equivalent orientations, and $U(n)$ preserves a reference orientation, giving the connected class-D manifold~\cite{BocquetSerbanZirnbauer2000,SenthilFisher2000,EversMirlin2008, FavaEtAl2023,GruzbergReadLudwig2001,YanBaoVijay2026}. The uniform saddle family is
\begin{equation}
\begin{gathered}
H(x,y)=-V(x,y)=\frac{q_\star}{2}s_{x,y}J,\\
J^T=-J,\quad J^2=-I_{2n},\quad J\in SO(2n)/U(n).
\end{gathered}
\label{eq:classD_target}
\end{equation}
The staggering belongs to these collective fields; the physical error angle remains uniform.

Allowing the orientation to vary slowly defines $J(\mathbf r)$. Uniform rotations cost no potential, while orientation gradients induce a response from the massive radial and complementary bond modes. Integrating out these modes yields a positive gradient cost, while keeping the spacetime staggering factor $s_{x,y}$ exact. After rescaling coordinates to absorb the leading anisotropy, the resulting NL$\sigma$M action and bare matching are
\begin{equation}
\begin{gathered}
S_{\mathrm{bulk}}[J]
=\frac{1}{2t_{\sigma,0}}\int d^2\mathbf r\,\sum_{\mu=1,2}
\operatorname{tr}\!\left[(\partial_\mu J)^T(\partial_\mu J)\right]\\
t_{\sigma,0}
=8 g+O(g^2)
=32\beta_R^2+o(\beta^2_R)=O(\cos^2 2\theta).
\end{gathered}
\label{eq:bulk_nlsm}
\end{equation}
Here $t_{\sigma,0}$ is the bare inverse stiffness and $\operatorname{tr}$ is the ordinary flavor trace. 
The quadratic detuning dependence follows from the lattice fluctuation calculation in Appendix~\ref{app:diffusive_saddle}, including the response of the massive modes.

The resulting NL$\sigma$M describes the long-distance collective correlations of the replicated Majorana amplitudes generated by syndrome averaging. 
Its field $J(\mathbf r)$ specifies the local flavor orientation of the HS self-energy, whose magnitude is fixed by the saddle. 
The spacetime staggering supplies the underlying lattice pattern, while $J$ describes its slowly varying orientation. 
Uniform orientations are degenerate; spatial variations cost gradient action, with $1/t_{\sigma,0}$ measuring their stiffness. 
Thus, within the assumed diffusive regime, the microscopic propagation and interaction are reduced to a theory of soft collective orientations. 
This description applies beyond the scattering scale; at the exact SWAP point, the saddle radius vanishes, and propagation remains ballistic. 
We connect this bulk effective field theory to logical recovery measures in the next section with an RG-flow-motivated finite-size scaling variable.

\begin{figure*}[t]
    \centering
    \includegraphics[width=1.0\linewidth]{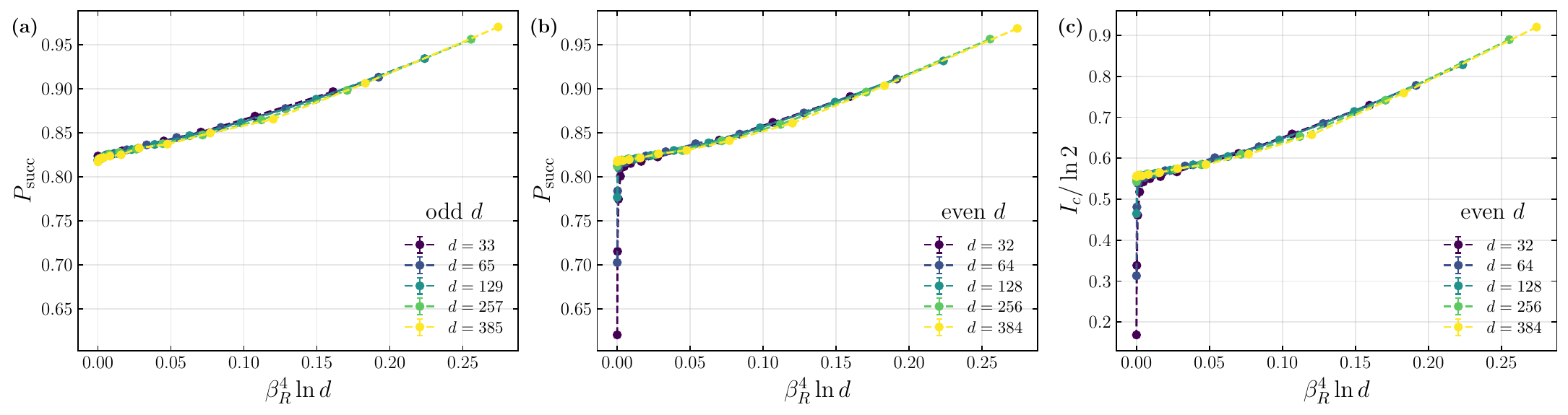}
    \caption{\textbf{Finite-size collapse of $P_{\mathrm{succ}}$ and $I_c$ for the RG crossover between the high-fidelity regime and the finite-size diffusive regime.} Panels (a) and (b) show $P_{\mathrm{succ}}$ for odd and even distances, respectively; panel (c) shows $I_c/\ln2$ for even distances. Data from Fig.~\ref{fig:measure_parity} with $0.12\leq\theta/\pi\leq0.23$ are replotted against the proposed scaling variable $X_4=\beta_R^4\ln d$. Error bars indicate one standard error; dashed lines connect points at fixed $d$ and are guides to the eye.
    }
    \label{fig:nlsm_fitting}
\end{figure*}
\section{Logical observables and finite-size scaling}\label{sec:numerical_nlsm}
In the \(n\to1\) replica limit, the leading weak-coupling RG flow of the NL$\sigma$M combined with the quadratic bare matching gives a relative-running variable proportional to \(\beta_R^4\ln(L/L_0)\), motivating the finite-size scaling variable \(X_4=\beta_R^4\ln d\). 
We test $X_4$ against the Born-averaged recovery measures in Fig.~\ref{fig:nlsm_fitting} and examine its implications for finite-size threshold drift toward $\theta_c = \pi/4$.

\subsection{RG flow and crossover scales.}
Long-wavelength fluctuations of $J(\mathbf{r})$ renormalize the stiffness as the observation scale increases. 
In the replica limit $n\to1$, the one-loop contribution vanishes and the leading flow is cubic~\cite{Hikami1981,GruzbergReadLudwig2001,YanBaoVijay2026}:
\begin{equation}
\frac{dt_\sigma}{d\ell}
=4t_\sigma^3+O(t_\sigma^4),\quad \ell=\ln(L/L_0),
\label{eq:classD_rg}
\end{equation}
where $L_0$ is a matching length within the assumed diffusive regime and $t_{\sigma,0}=t_\sigma(L_0)\ll1$. 
The positive flow increases the inverse stiffness, strengthening orientation fluctuations at larger scales.
Integrating the leading cubic term gives
\begin{equation}
t_\sigma(L)=\frac{t_{\sigma,0}}{\sqrt{1-w(L)}},\quad
w(L)\equiv8t_{\sigma,0}^2\ln\frac{L}{L_0}.
\label{eq:classD_running}
\end{equation}
Here, $w$ measures the accumulated relative change of the coupling: for $w\ll1$, $[t_\sigma(L)-t_{\sigma,0}]/t_{\sigma,0}=w/2+O(w^2)$. The running solution applies for $L\geq L_0$ while $t_\sigma(L)\ll1$.

The quadratic matching $t_{\sigma,0}=32\beta_R^2[1+o(1)]$ for the retained lattice action in Eq.~\eqref{eq:bulk_nlsm} gives $w(L)\propto\beta_R^4\ln(L/L_0)$ near SWAP. 
The fourth power arises because the cubic flow makes relative running proportional to $t_{\sigma,0}^2$, while the bare coupling is quadratic in $\beta_R$. 
A fixed fractional change of the weak coupling defines the much longer RG scale $d_{\mathrm{RG}}$ introduced earlier, with
\begin{equation}
\ln\frac{d_{\mathrm{RG}}}{L_0}
=\frac{C}{\beta_R^4}[1+o(1)].
\label{eq:rg_crossover_scale}
\end{equation}
Here $C>0$ depends on the chosen fractional change and bare matching. 
$d_{\mathrm{RG}}$ measures its subsequent slow RG evolution. 

For fixed-aspect-ratio codes, taking $L\propto d$ and absorbing the length conversion into $L_0$ motivates the finite-size scaling variable
\begin{equation}
X_4=\beta_R^4\ln d=\left(-\frac{1}{2}\ln\tan\theta\right)^4\ln d.
\label{eq:logarithmic_collapse_coordinate}
\end{equation}
The constant prefactor is absorbed into the scaling functions. 
The omitted correction $-\beta_R^4\ln L_0(\theta)$ is relatively small when $|\ln L_0|/\ln d\ll1$~\cite{SlevinOhtsuki1999}.  
Even at fixed $w$, the running coupling retains its bare-coupling dependence, so collapse of logical observables onto functions of $X_4$ alone is an additional scaling hypothesis, tested below.

\subsection{Finite-size scaling}
We test $P_{\mathrm{succ}}(d,\theta)\simeq F_{\mathrm{succ}}(X_4)$ for square codes in $0.12\leq\theta/\pi\leq0.23$, allowing a separate scaling function for each code-distance parity without imposing its form.
Figs.~\ref{fig:nlsm_fitting}(a) and (b) show approximate alignment along the rising branches, with substantial residual size dependence near SWAP, especially for smaller even sizes.
Panel (c) shows the corresponding even-distance $I_c/\ln2$ collapse; Fig.~\ref{fig:nlsm_fitting_ic_bayes} in Appendix~\ref{app:collapse_power} gives the $P_{\mathrm{Bayes}}$ comparison for both parities.

The alignment supports the theory-motivated finite-size scaling variable over the sampled range, with residual drift limiting quantitative validation.
Appendix~\ref{app:collapse_power} compares alternative powers: $a=4$ improves alignment over $a=2,3$ along the resolved rising branches.
The data therefore provide a consistency test of the proposed scaling without uniquely determining an asymptotic power.

Along a rising scaling branch, increasing $d$ at fixed $\theta$ moves toward larger $X_4$ and higher $P_\mathrm{succ}$ and $I_c$.
If the collapse persists to larger sizes, a fixed fidelity corresponds to a fixed $X_\star>0$, giving
\begin{equation*}
\frac{\pi}{4}-\theta(d)\simeq\beta_R(d)
\simeq\left(\frac{X_\star}{\ln d}\right)^{1/4}.
\end{equation*}
The predicted $(\ln d)^{-1/4}$ drift of fixed-performance contours remains to be tested at larger sizes.
For every fixed $0<\theta<\pi/4$, $X_4\to\infty$ as $d\to\infty$.
If the infrared evolution drives the Born-averaged logical-sector ambiguity to zero throughout this interval for both even and odd sizes, both Pauli fidelities approach unity; the even-parity bounds of Sec.~\ref{sec:measure_parity} then also give $I_c\to\ln2$.
Under this additional assumption, the Pauli-decoding and even-parity coherent-information thresholds lie at SWAP, $\theta_c=\pi/4$, as proposed in Refs.~\cite{YanBaoVijay2026, YangLudwigJian2026}; odd coherent information remains maximal independently.
This threshold scenario requires logical-defect and infrared matching beyond the weak-coupling flow, and comparisons with earlier extrapolations must match the decoder, observable, and geometry~\cite{BravyiEtAl2018, PatoStaplesBrown2025, VennBehrendsBeri2023, EcksteinEtAl2025}.

The slow logarithmic evolution poses a severe numerical challenge near SWAP. 
For example, at \(\theta/\pi=0.22\), \(\beta_R^4\simeq8.08\times10^{-5}\), so a tenfold increase in code distance changes \(X_4\) by only \(1.86\times10^{-4}\). 
Increasing \(X_4\) by \(10^{-3}\) would require a size ratio of approximately \(2.4\times10^5\). Thus, accessible sizes can probe only a narrow interval of the proposed scaling curve at fixed angle, limiting direct tests of its long-distance evolution.
Approaching SWAP at fixed $d$ instead eventually leaves the diffusive regime; in the next section, we examine this parity-dependent response in the nearly-ballistic regime.

\section{Nearly-ballistic scaling and plateau }\label{sec:near_swap}
The bulk NL$\sigma$M applies in the diffusive regime ($d \gg \lambda_{\text{mfp}} \sim \beta_R^{-2}$). 
Approaching SWAP ($\theta \to \pi/4$) at fixed code distance instead probes a nearly ballistic regime where the scattering length diverges, and the diffusive saddle vanishes, requiring an analysis beyond bulk RG. 
For square surface codes, this nearly-ballistic response depends strongly on code-distance parity: even-distance decoding curves approach SWAP smoothly, whereas odd distances display a pronounced pre-SWAP bump and a cusp at SWAP.

Below, we first identify the common scattering scale and derive a complement relation constraining the logical amplitudes.
We then demonstrate that the nearly-ballistic response collapses as $\beta_R^2 d$ for even distances and as $\beta_R d$ for odd distances, where the Born-weighted logical-angle distribution accounts for the pre-SWAP bump.
Finally, we establish uniform-angle plateau benchmarks for the decoding measures.

\subsection{Nearly-ballistic scales and complement relation}
\label{subsec:swap_variables_parity}
Both Pauli measures depend on the relative magnitudes of the amplitudes $A_s^0$ and $A_s^1$. 
While phase pinning distinguishes an odd logical rotation from an even logical weak measurement, it leaves the detuning dependence of this magnitude ratio undetermined. 
Below, we first identify the common bulk scattering scale, and then derive a complement relation that constrains the logical response.

\paragraph{Bulk scattering scale and logical-response scaling variables.}
For the square rotated planar codes considered here, we introduce the parity-dependent logical-response variables
\begin{equation}
    u\equiv\beta_R^2d,\qquad v\equiv\beta_Rd.
\end{equation}
The approximate collapses presented below motivate the near-ballistic response scales \(d_{\mathrm{ba}}\sim\beta_R^{-\alpha}\), with \(\alpha=2\) for even distance and \(\alpha=1\) for odd distance (see also the crossover scales in the phase diagram of Fig.~\ref{fig:setup_phase_diagram}). 
These scales organize the departure of the logical observables from their SWAP values.

A separate bulk diagnostic is provided by the scalar-free, single-layer transfer calculation on long strips with periodic transverse boundaries in Appendix~\ref{app:bulk_lyapunov}. 
Its width-extrapolated Lyapunov rate, \(\Gamma\simeq2\beta_R^2\), defines the transfer-growth length \(\ell_\Gamma\equiv\Gamma^{-1}\). 
We assume that the bulk mean free path satisfies \(\lambda_{\mathrm{mfp}}\sim\ell_\Gamma\propto\beta_R^{-2}\), common to both code-distance parities. This matching supports the scattering scale beyond which the diffusive NL\(\sigma\)M is applicable.
Its transport interpretation, proportionality coefficient, and matching to the planar Born-weighted problem remain to be established.

For even $d$, \(d_{\mathrm{ba}}\sim \beta^{-2}_R\) and the proposed bulk scattering length have the same parametric dependence. 
For odd $d$, \(d_{\mathrm{ba}}\sim \beta^{-1}_R\) can occur well before bulk scattering, leaving the window $    \beta_R^{-1}\ll d\ll\beta^{-2}_R\to u\ll1\ll v.$
This separation may account for the plateau extending closer to SWAP than in the even case. 
Establishing the plateau’s formation requires the Born-weighted logical-angle statistics discussed below.



\paragraph{Complement relation of the logical amplitudes.}
Consider the square surface code with \(N=d^2\) physical qubits and one logical qubit, as in Fig.~\ref{fig:setup_phase_diagram}. 
Let \(\Omega\) contain all physical qubits and define \(Z_\Omega=\prod_{j\in\Omega}Z_j\). Multiplication by \(Z_\Omega\) complements an error support: \(Z_\Omega Z_E=Z_{E^c}\), where \(E^c=\Omega\setminus E\). 
The operator \(Z_\Omega\) commutes with the coherent rotation \(U(\theta)=\Pi_{j\in\Omega}\exp(i\theta Z_j)\) and it also commutes with each syndrome projector \(\Pi_s\) whenever every \(X\) stabilizer has even weight. 
Thus, \(E\) and \(E^c\) have the same syndrome. With reference corrections and logical-sector labels fixed, complementation therefore preserves the logical sector for even \(d\) and exchanges the two sectors for odd \(d\):
\begin{equation}
\mu_s(E^c)\equiv\mu_s(E)+d\pmod2.
\label{eq:swap_complement_sector}
\end{equation}
Syndrome preservation follows from the even weight of the \(X\) stabilizers, while the sector shift follows from the weight-\(d\) logical-\(X\). 

To relate the corresponding coherent amplitudes under complementation, extend \(\beta_R\) to either sign through \(\theta(\beta_R)=\arctan e^{-2\beta_R}\), so that \(\theta(-\beta_R)=\pi/2-\theta(\beta_R)\). 
The single-qubit identity \(e^{i(\pi/2-\theta)Z}=iZ e^{-i\theta Z}\) gives the exact relation 
\begin{equation}
    U(\frac{\pi}{2}-\theta)=i^N Z_\Omega U(\theta)^*,
\end{equation}
where complex conjugation is taken in the computational $Z$ basis. 
This identity pairs the coefficients of complementary error supports. 
Explicitly, an error support \(E\) of weight \(k\) has coefficient $a_E(\theta)=i^k(\cos\theta)^{N-k}(\sin\theta)^k$, while its complement satisfies
\begin{equation}
    a_{E^c}(\frac{\pi}{2}-\theta)
=i^{N-k}(\cos\theta)^{N-k}(\sin\theta)^k
=i^N[a_E(\theta)]^*.
\label{eq:swap_complement_coefficient}
\end{equation}
Summing over the paired configurations at fixed syndrome therefore yields
\begin{equation}
    A_{(\mu+d)\bmod2}^s(-\beta_R)
=i^N[A_\mu^s(\beta_R)]^*,
\quad \mu=0,1.
\label{eq:swap_complement_amplitudes}
\end{equation}
This exact relation under detuning reversal—which we term the \emph{complement relation}—relates syndrome-resolved amplitudes at opposite detunings about the SWAP fixed point ($\beta_R = 0$).

\paragraph{Constraints of the logical response.}
We now translate the complement relation into constraints on the logical response, including DWFE and decoding fidelities.

With the reference correction and sector labels held fixed as the angle changes, we find that for the DWFE $\Delta F_s=-\ln{|A_1^s|^2/|A_0^s|^2}$ (Eq.~\eqref{eq:dwfe}), Eq.~\eqref{eq:swap_complement_amplitudes} gives 
\begin{equation}
\Delta F_s(-\beta_R)=
\begin{cases}
\Delta F_s(\beta_R),&d\ \text{even},\\
-\Delta F_s(\beta_R),&d\ \text{odd},
\end{cases}
\label{eq:swap_logratio_symmetry}
\end{equation}
wherever both amplitudes are nonzero.
At fixed $d$, the amplitudes are analytic functions of $\beta_R$. If $A_0^s(0)A_1^s(0)\neq0$, then $\Delta F_s$ is analytic in a neighborhood of the SWAP point, and it has the following expansion
\begin{equation}
\Delta F_s(\beta_R)=
\begin{cases}
\frac12\Delta F_s''(0)\beta_R^2+O(\beta_R^4),
&d\ \text{even},\\
\Delta F_s'(0)\beta_R+O(\beta_R^3),
&d\ \text{odd},
\end{cases}
\label{eq:swap_logratio_response}
\end{equation}
where primes denote derivatives with respect to $\beta_R$, and $\Delta F_s(0)=0$ exactly, because the amplitudes have equal magnitudes at $\beta_R=0$ (see Sec.~\ref{subsec:swap_endpoint_recovery}).
Thus, even $d$ forbids the odd-order terms and odd $d$ permits a linear response. 

Furthermore, the complement relation for the amplitudes implies $D_s(-\beta_R)=D_s(\beta_R)$, ensuring that the Born weights are even in detuning for both even and odd $d$. 
The conditional MLD fidelity depends only on the magnitude of the DWFE,
\begin{equation}
    P_{\mathrm{succ}}(s)-\frac{1}{2}=\frac{1}{2}\tanh{\frac{|\Delta F_s|}{2}}=\frac{|\Delta F_s|}{4}+O(|\Delta F_s|^3).\label{eq:Psucc_dwfe}
\end{equation}
Consequently, both $P_\mathrm{succ}$  and its Born average are even in detuning for even and odd $d$. 

\paragraph{Compatibility with the bulk scaling.}
The bulk scaling variable \(X_4=\beta_R^4\ln d\) is motivated by the quadratic bare matching and cubic RG flow discussed in Sec.~\ref{sec:numerical_nlsm}; it does not require a quartic scaling of the logical fidelity near SWAP. 
The nearly-ballistic expansion and proposed bulk collapse concern different limits: at fixed \(d\), taking \(\beta_R\to0\) gives \(u=\beta_R^2d\to0\), whereas holding \(X_4>0\) fixed as \(d\to\infty\) gives \(u=d\sqrt{X_4/\ln d}\to\infty\).

A common bulk scaling can therefore coexist with different parity-dependent nearly-ballistic responses. 
The complement relation fixes the allowed detuning powers, while the distance dependence of their coefficients and the resulting response scaling require separate analysis. 
We examine these even- and odd-distance responses below.

\begin{figure}[t]
    \centering
    \includegraphics[width=1.0\linewidth]{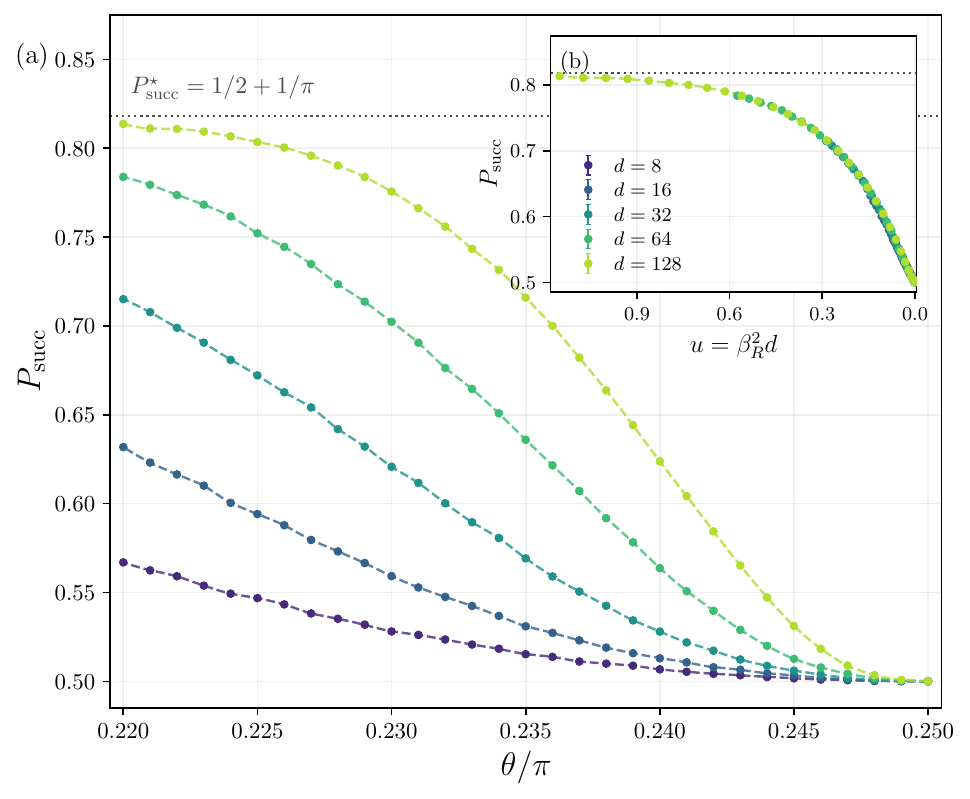}
    \caption{\textbf{Even-distance $P_{\mathrm{succ}}$ near SWAP.}
    Born-averaged $P_{\mathrm{succ}}$ for square rotated planar codes with $d=8,16,32,64,128$, plotted versus (a) $\theta/\pi$ and (b) $u=\beta_R^2d$, where $\beta_R=-\tfrac12\ln\tan\theta$.
    Each distance uses 31 angles, $\theta/\pi=0.220,0.221,\ldots,0.250$, with $50{,}000$ syndrome samples per point. Error bars indicate one standard error. 
    The rescaled curves approximately collapse, with residual finite-size deviations. 
    The horizontal axis in (b) is reversed to match the direction of approach to SWAP in (a).
    Dashed lines are guides to the eye.
    }
    \label{fig:swap_scaling_even}
\end{figure}
\subsection{Even distance: smooth response and $\beta_R^2 d$ scaling}
\label{subsec:swap_even_scaling}
The even-distance curves in Fig.~\ref{fig:measure_parity} approach SWAP smoothly: both Pauli fidelities tend to $1/2$, while $I_c$ tends to 0.
Fig.~\ref{fig:swap_scaling_even} resolves this nearly-ballistic response.
Replotting $P_{\mathrm{succ}}$ against $u=\beta_R^2d$ approximately aligns the curves, with a nearly linear response at small $u$ and bending toward the plateau when $u>1$.

At SWAP, each supported even-distance outcome gives $K_s\propto I_L\pm Z_L$ (Sec.~\ref{subsec:swap_endpoint_recovery}), a projective logical measurement with balanced sector weights. 
Detuning can make these weights unequal and soften the projection, increasing the conditional Pauli fidelity.
For such an outcome, the complement relation permits a quadratic DWFE, $\Delta F_s=\tfrac12\Delta F_s''(0)\beta_R^2+O(\beta_R^4)$. 
Then, at fixed $d$, Eq.~\eqref{eq:Psucc_dwfe} gives
\begin{equation}
P_{\mathrm{succ}}(s)-\tfrac12=\frac{|\Delta F_s''(0)|}{8}\beta_R^2+O(\beta_R^4).
\end{equation}
The Born average also includes outcomes absent at SWAP. 
These can contribute quadratically when detuning gives them $O(\beta_R^2)$ probability and unequal leading sector weights, leaving a finite conditional advantage over $1/2$.
Appendix~\ref{app:endpoint_coefficients} derives the full expansion,
\begin{equation}
\begin{gathered}
 P_{\mathrm{succ}}-\tfrac12
 =C_{\mathrm{even}}(d)\beta_R^2+O(\beta_R^4),\\
 C_{\mathrm{even}}(d)
 =\tfrac18\langle|\Delta F_s''(0)|\rangle_{p_0}+C_{\mathrm{new}}(d),
\end{gathered}
\end{equation}
where $\langle\cdot\rangle_{p_0}$ averages over the SWAP Born distribution on supported outcomes and $C_{\mathrm{new}}\geq0$ accounts for newly populated outcomes.
Such outcomes are especially important for the other measures: on supported branches, the $P_{\mathrm{Bayes}}-1/2$ and $I_c$ responses begin beyond quadratic order of $\beta_R$.

The approximate collapse and nearly linear small-$u$ response in Fig.~\ref{fig:swap_scaling_even}(b) are consistent with growth of the total coefficient $C_{\mathrm{even}}(d)\propto d$ over the sampled distances, reflecting the ballistic accumulation of local quadratic detuning over propagation length $d$. 
While precise determination of the coefficients requires controlled small-detuning extrapolation, the observed collapse supports $u$ as the natural nearly-ballistic scaling variable for even distances.

\subsection{Odd distance: bump, cusp and $\beta_R d$ scaling}
\label{subsec:swap_scaling_odd}
\begin{figure}[t]
    \centering
    \includegraphics[width=1.0\linewidth]{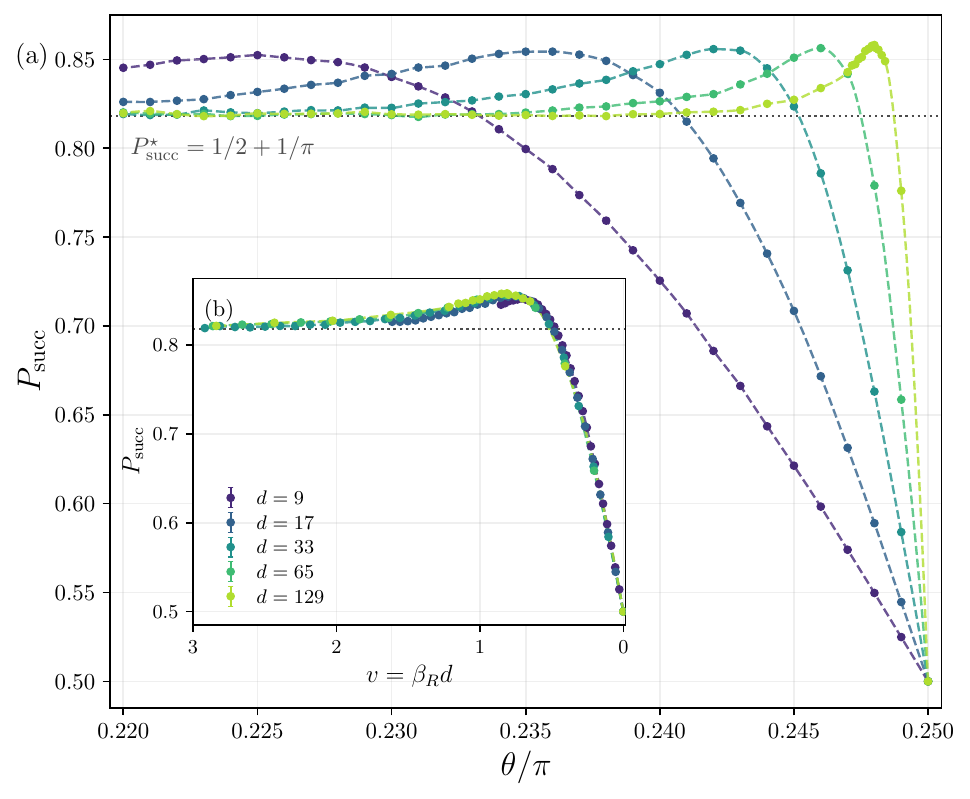}
    \caption{\textbf{Odd distance $P_{\mathrm{succ}}$ and pre-SWAP bump.}
    Born-averaged $P_{\mathrm{succ}}$ for the rotated planar codes with
    $d=9,17,33,65,129$ versus (a) $\theta/\pi$ and (b) $v=\beta_Rd\leq 3.0$ where $\beta_R=-\tfrac12\ln\tan\theta$.
    Both panels include eleven additional $d=129$ points at
    $v=0.65,0.70,\ldots,1.15$. Each point uses $50{,}000$ syndrome samples;
    error bars show one standard error of the mean. 
    The horizontal axis in (b) is reversed to match the direction of approach to SWAP in (a). Dashed lines are guides to the eye. 
    }
    \label{fig:swap_scaling_odd}
\end{figure}

Fig.~\ref{fig:measure_parity} shows a non-monotonic odd-distance response: as the physical angle approaches SWAP, the Pauli-decoding measures rise above their intermediate-angle plateau, reach a maximum, and then fall to \(1/2\) at the SWAP point. 
Plotting $P_{\mathrm{succ}}$ against \(v=\beta_Rd\) with finer resolution approximately aligns the nearly-ballistic curves across the sampled distances and places the peak positions in a common order-one range, as shown in Fig.~\ref{fig:swap_scaling_odd}. We first examine the origin of the scaling variable and then interpret the bump through the distribution of logical rotation angles.

\paragraph{Interpretation of the $\beta_Rd$ scaling and cusp.}
At SWAP, equal sector magnitudes and phase pinning give $K_s\propto I_L\pm iZ_L$ (Sec.~\ref{subsec:swap_endpoint_recovery}), yielding $P_{\mathrm{succ}}=1/2$. 
Detuning changes the conditional rotation angle and can make the logical weights unequal, improving Pauli recovery. 

The complement relation in Eq.~\eqref{eq:swap_logratio_response} gives the odd-$d$ expansion $\Delta F_s(\beta_R)=\Delta F_s'(0)\beta_R+O(\beta_R^3)$. Then, we have
\begin{equation}
    P_{\mathrm{succ}}(s)-\frac{1}{2}=\frac{|\Delta F_s'(0)|}{4}\,|\beta_R|+ O(|\beta_R|^3).
\end{equation}
Therefore, the conditional fidelity can rise linearly in $|\beta_R|$. 
The Born weights are even in detuning, so their variation first modifies the supported linear response at cubic order. Outcomes absent at SWAP also contribute only at cubic or higher order, as shown in Appendix~\ref{app:endpoint_coefficients}.
Averaging gives
\begin{equation}
P_{\mathrm{succ}}-\tfrac12
=C_{\mathrm{odd}}(d)|\beta_R|+O(|\beta_R|^3),
\label{eq:odd_endpoint_ensemble_onset}
\end{equation}
where $C_{\mathrm{odd}}(d)
=\langle|\Delta F_s'(0)|\rangle_{p_0}/4$.
For the sampled odd cases, \(C_{\mathrm{odd}}(d)>0\), so the signed-detuning extension of $P_{\mathrm{succ}}$ has a cusp at $\theta=\pi/4$.
The cusp is specific to the maximization in \(P_{\mathrm{succ}}\). For a linear odd-distance DWFE, the posterior purity instead satisfies \(P_{\mathrm{Bayes}}(s)-1/2=[\Delta F_s'(0)]^2\beta_R^2/8+O(\beta_R^4)\), giving a smooth quadratic response at fixed distance. 
Both measures nevertheless exhibit a near-SWAP bump and approximate \(\beta_Rd\) scaling. These shared features reflect the evolving distribution of posterior imbalance and should be distinguished from the nonanalytic endpoint onset of \(P_{\mathrm{succ}}\).

This scaling behavior is demonstrated in Fig.~\ref{fig:swap_scaling_odd}: while $P_{\text{succ}}$ in panel (a) peaks at angles that drift toward SWAP with increasing $d$, rescaling the horizontal axis by $v = \beta_R d$ in panel (b) produces an approximate collapse across all sampled odd distances, aligning both the initial linear slope and the pre-SWAP peaks within a common $v \sim O(1)$ window. 
This collapse is corroborated by endpoint-susceptibility measurements, which find that the Born-averaged absolute susceptibility grows approximately linearly with system size, $C_{\text{odd}}(d) \propto d$, across the sampled sizes. 
Together with the linear detuning response, this linear coefficient growth motivates $v = \beta_R d$ as the natural odd-distance scaling variable, consistent with ballistic accumulation along the network. 

Because $u = v^2/d \to 0$ as $d \to \infty$ at fixed $v$, an order-one odd logical response develops well before substantial bulk scattering sets in ($u \sim 1$). 
A microscopic derivation of $C_{\text{odd}}(d) \propto d$ remains an open problem; symmetry dictates only the permitted powers of $\beta_R$, while the full non-monotonic curve at finite $v$ reflects higher-order conditional nonlinearity and shifting Born weights.

The last nontrivial crossings of \(P_{\mathrm{succ}}\) between adjacent sampled odd distances in Fig.~\ref{fig:swap_scaling_odd}(a) also drift toward SWAP: their positions increase from \(\theta_\times/\pi\simeq0.230\) for \((d_1,d_2)=(9,17)\) to \(0.247\) for \((65,129)\). This approximately inverse-size drift is compatible with the nearly-ballistic crossover organized by \(\beta_Rd\). A nonmonotonic crossover can itself generate such moving crossings, so they do not independently locate the thermodynamic threshold. Quantitative extrapolation remains sensitive to angular resolution and finite-size corrections.

\paragraph{Angular interpretation of the bump.}
For a conditional logical rotation, $|A_1^s/A_0^s|=|\tan\theta_s|$, so $\Delta F_s=-2\ln|\tan\theta_s|$ wherever both amplitudes are nonzero. 
Then $P_\mathrm{succ}$ can be written as 
\begin{equation}
P_{\mathrm{succ}}=\tfrac12+\tfrac12\langle|\cos(2\theta_s)|\rangle_s.
\label{eq:odd_bump_harmonics}
\end{equation}
Logical rotations near $\theta_s=0,\pm\pi/2$ favor Pauli recovery, whereas balanced rotations at $\theta_s=\pm\pi/4$ give conditional fidelity $1/2$. 
A uniform Born-weighted angle distribution gives the plateau benchmark $P_{\mathrm{succ}}=1/2+1/\pi\simeq0.8183$; the bump is the rise above this value.

The second row of Fig.~\ref{fig:angle_d128_129} shows the Born-weighted logical angle distributions for $d=129$. 
As the physical angle approaches SWAP, the distribution first reaches approximate uniformity at $\theta/\pi=0.22$ [panel (f)]. 
Then at physical angle \(\theta/\pi=0.2480\) [panel (g)], the distribution has enhanced weight near the logical Pauli rotations \(\theta_s=0,\pm\pi/2\). 
Closer to SWAP, at \(\theta/\pi=0.2495\) [panel (h)], it concentrates near the balanced rotations \(\theta_s=\pm\pi/4\).
Through Eq.~\eqref{eq:odd_bump_harmonics}, this redistribution provides a distributional account of the plateau--bump--SWAP sequence. 
A finer scan for $d=129$ finds a sampled maximum $P_{\mathrm{succ}}=0.8577(6)$ at $v=\beta_Rd=0.85$ ($\theta/\pi\simeq0.24790$), near panel (g) of Fig.~\ref{fig:angle_d128_129}. Thus, the bump happens at $v=\beta_Rd=O(1)$. 

Extending the linear response $\Delta F_s = \Delta F'_s(0)\beta_R$ to finite $v$ under fixed Born weights yields an average fidelity that is strictly nondecreasing in $v$; the pre-SWAP bump therefore requires higher-order conditional nonlinearity, drifting Born weights, or both. 
Disentangling their microscopic contributions remains an open problem.

These findings reveal a general mechanism for enhanced Pauli recovery near the complement-symmetric point $\theta = \pi/4$. 
In CSS codes with even-weight $X$-stabilizers where the complement action $Z_\Omega$ swaps the logical sectors, the complement relation strictly enforces equal sector magnitudes and $P_{\text{succ}} = 1/2$ at SWAP. 
Detuning lifts this balance, and if the logical susceptibility scales with system size, substantial Pauli recovery gains can emerge within an angular window that shrinks with $d$. 
While the complement relation rigorously fixes the permitted perturbation orders across both even and odd code distances, the microscopic origin of the divergent susceptibility and the full non-monotonic scaling functions motivate further exploration in broader CSS code families.

\subsection{Logical-angle statistics of plateau}\label{subsec:plateau}
\begin{figure*}[t]
    \noindent
    \centering
    \includegraphics[width=1.0\linewidth]{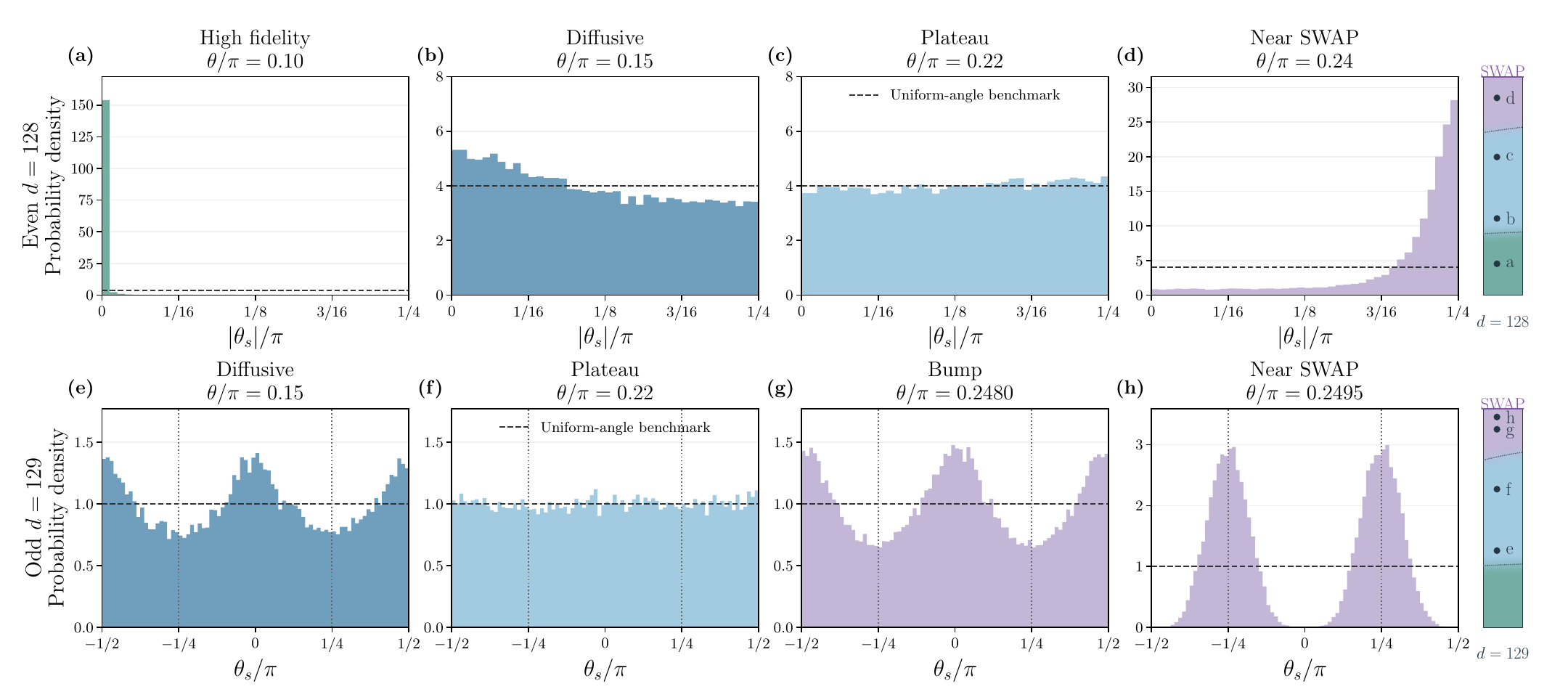}
    \caption{\textbf{Logical-angle distributions across finite-size crossover regimes.} 
Top row (a–d): Born-sampled distributions of the logical tilt-angle magnitude \(|\theta_s|/\pi\in[0,1/4]\) (Eq.~\eqref{eq:even_logical_tilt}) for even distance \(d=128\); the tilt sign is not reconstructed. Bottom row (e–h): distributions of the logical rotation angle \(\theta_s/\pi\in[-1/2,1/2)\) (Eq.~\eqref{eq:odd_logical_rotation}) for odd distance \(d=129\). Physical error angles are indicated above each panel. Each histogram uses \(50{,}000\) syndrome samples. Horizontal dashed lines mark uniform-angle benchmarks of density \(4\) and \(1\), respectively.    
 Panels (c,f) show approximately uniform angular plateaus, while (g) samples the odd-distance bump. Near SWAP, panels (d,h) show concentration toward the equal-sector-magnitude angles \(|\theta_s|=\pi/4\) and \(\theta_s=\pm\pi/4\), respectively; vertical dotted lines mark the latter. The right-hand strips schematically locate the eight examples in the crossover diagram of Fig.~\ref{fig:setup_phase_diagram}(b), with lettered markers corresponding to the distribution panels. Shading follows the regime colors of Fig.~\ref{fig:setup_phase_diagram}; dotted curves indicate the RG and ballistic crossover guides, and the upper solid boundary marks the signed-SWAP point \(\theta=\pi/4\). 
 }
    \label{fig:angle_d128_129}
\end{figure*}
When the respective scaling variable exceeds order unity---$u=\beta^2_Rd\gg1$ for even $d$ and $v=\beta_Rd\gg 1$ for odd $d$---the logical response approaches an apparent plateau (see Figs.~\ref{fig:swap_scaling_even}, \ref{fig:swap_scaling_odd}, and~\ref{fig:measure_parity} for finite-size plateaus in both odd and even sizes).
Each Pauli measure approaches the same value, while coherent information distinguishes the two cases: odd $I_c$ remains at $\ln2$, whereas even $I_c$ develops a plateau below this ceiling. 
The dashed lines mark the uniform-angle benchmarks
\begin{equation}
P_{\mathrm{succ}}^*=\frac12+\frac1\pi,\quad
P_{\mathrm{Bayes}}^*=\frac34,\quad
I_c^{*,\mathrm{even}}=2\ln2-1.
\label{eq:uniform_plateau_values}
\end{equation}
The benchmark $P_{\mathrm{succ}}^*=1/2+1/\pi$ in the odd case is equivalent to the uniform-angle infidelity estimate derived in Ref.~\cite{VennBeri2020}.
These values are exact under the angular-distribution assumption below; the numerical curves approach them approximately.

For even code distances, the logical channel acts as a weak measurement of strength $\alpha_s$, with the logical tilt angle $\theta_s$ parametrized by $m_s = \tanh(2\alpha_s) = \sin(2\theta_s)$ (Sec.~\ref{subsubsec:even_parity}). 
The first row of Fig.~\ref{fig:angle_d128_129} displays the distribution of the tilt magnitude $|\theta_s| \in [0, \pi/4]$ in different regimes. 
This folded angle removes the sign (north/south polarity) of the tilt.
Because $|\theta_s|$ is determined solely by the ratio of sector magnitudes without altering homology labels, all three even-distance decoding measures depend exclusively on this unsigned distribution through $|m_s|$.

At fixed error angle $\theta$ and code distance $d$, define the distribution of Born-weighted logical-tilt angle by
\begin{equation}
\rho(t)
=\sum_{s:D_s>0}D_s\,\delta(t-|\theta_s|).
\label{eq:even_born_weighted_angle_distribution}
\end{equation}
As a benchmark, take the tilt-angle magnitude to be uniformly distributed on $[0,\pi/4]$, with $\rho(t)=4/\pi$. 
Under $m=\sin(2t)$, this induces the density of the interference magnitude on $[0,1]$,
\begin{equation}
\rho(t)=\frac{4}{\pi}\xrightarrow{m=\sin(2t)}
\sigma(m)=\frac{2}{\pi\sqrt{1-m^2}}.
\label{eq:uniform_angle_interference_density}
\end{equation}
Averaging $I_c(m)=h_2(\tfrac{1+m}{2})$ (Eq.~\eqref{eq:Ic_ms_def}) with this density yields
\begin{equation}
I_c^{*,\mathrm{even}}
=\int_0^1\!dm\,\sigma(m)
 h_2\!\left(\frac{1+m}{2}\right)
=2\ln2-1.
\label{eq:even_uniform_plateau_integrals}
\end{equation}
Similarly, uniform-angle averaging of the conditional Pauli measures $P_\mathrm{succ}(m)=\tfrac{1+\sqrt{1-m^2}}{2}$ and $P_\mathrm{Bayes}(m)=1-\tfrac{m^2}{2}$ (Eq.~\eqref{eq:p_succ_ms}) gives the two Pauli benchmarks in Eq.~\eqref{eq:uniform_plateau_values}. Details of the integral and the odd distance averages are given in Appendix~\ref{app:uniform_angle_plateau}. The approximately flat distribution of $|\theta_s|$ in Fig.~\ref{fig:angle_d128_129}(c) supports these benchmarks for the sign-insensitive observables, without determining the signed tilt distribution.

While both code-distance parities share identical Pauli plateau benchmarks, their physical origins differ fundamentally. 
For odd distance, each conditional branch is an exact logical unitary rotation; hence $I_c = \ln 2$ holds across the entire plateau, and the Pauli deficits arise solely from restricting Pauli recoveries.
For even distance, the channel acts as a logical weak measurement along $Z_L$, so the reduced coherent information $I_c^{*,\mathrm{even}} = 2\ln 2 - 1$ represents irreversible measurement backaction.

Furthermore, although their asymptotic Pauli benchmarks coincide, the plateaus emerge over disparate angular windows for even and odd $d$ (see Figs.~\ref{fig:swap_scaling_even}(a) and~\ref{fig:swap_scaling_odd}(a)). 
For $d = 129$, the plateau spans a broader interval and extends much closer to SWAP than for $d = 128$, separated from $\theta=\pi/4$ only by the pre-SWAP bump. 
This hierarchy directly reflects the two response scales established in Sec.~\ref{subsec:swap_variables_parity}: the odd-distance response develops on the scale $\beta_R \sim d^{-1}$, whereas the even-distance response is governed by the bulk scattering scale at $\beta_R \sim d^{-1/2}$. 
In the intermediate window $d^{-1} \ll \beta_R \ll d^{-1/2}$, the odd case has already relaxed into its angular plateau while the network remains in the nearly ballistic regime, extending the plateau toward SWAP without requiring the bulk diffusion.

Ref.~\cite{VennBehrendsBeri2023} observed behavior consistent with a power-law approach to a nonzero extrapolated MLD infidelity below $1 - P_{\mathrm{succ}}^*$. 
However, such angular benchmarks reflect finite-size plateaus rather than thermodynamic limits. 
The benchmark values suppose a uniform Born-weighted distribution of both logical rotation and tilt angles, whose realization is sensitive to boundary conditions and aspect ratio; for instance, on a long cylinder, Ref.~\cite{YanBaoVijay2026} observed a plateau for $P_{\mathrm{Bayes}}$ exceeding $3/4$. 
We note that related numerical work found the syndrome-resolved projected ensemble of logical unitaries converging toward a unitary design \cite{ChengEtAl2025}. 
A possible physical origin of the approximate angular uniformity is the emergence of a finite-size metallic regime in the associated Majorana network. 
Studies of logical snapshots connect metallic behavior with broadly distributed logical ensembles \cite{EcksteinEtAl2025}, while the quantum-dot description provides a route from sufficiently mixing class-D scattering to a symmetry-constrained uniform logical-angle distribution \cite{BejanEtAl2026}. 
Such angular randomization could account for the plateau benchmarks even when metallic behavior persists only over a finite range of sizes. 
Applying this picture to the present geometry requires matching the boundary conditions and Born-weighted ensemble; its extension to the odd-distance plateau within the nearly ballistic window remains open.

\section{Discussion and outlook}
In this paper, we organize finite-size decoding under uniform coherent \(Z\) rotations through algebraic nearly-ballistic scales $d_{\rm ba}\sim 1/\beta^\alpha_R$ and a parametrically longer RG scale, \(d_{\mathrm{RG}}\sim\exp{(C/\beta_R^{4})}\). 
This hierarchy provides a picture for the approximate plateaus, finite-size drift, and parity-dependent near-SWAP behavior summarized in Fig.~\ref{fig:setup_phase_diagram} and Table~\ref{tab:bulk_logical_regimes}.
Microscopically, averaging shared syndrome signs generates an inter-flavor quartic interaction of order \(\beta_R^2\). 
This quartic interaction leads to a class-D NL\(\sigma\)M for the intermediate diffusive regime. 
Its quadratic bare matching and cubic RG flow motivate \(X_4=\beta_R^4\ln d\), which approximately organizes the numerical decoding data, consistent with $\theta_c=\pi/4$ \cite{YanBaoVijay2026, YangLudwigJian2026}.
The exact logical-channel results establish the recovery limits underlying this picture. 
For the even-distance square codes considered here, MLD maximizes entanglement fidelity for maximally mixed logical input over all CPTP quantum recoveries. 
Exact bounds also show that \(P_{\mathrm{succ}}\), \(P_{\mathrm{Bayes}}\), and \(I_c\) define the same asymptotically correctable set. 
At odd distance, the conditional logical operation is unitary, permitting perfect quantum recovery despite imperfect Pauli decoding. 
Thus, the parity-dependent decoding curves distinguish limitations of Pauli recovery from information lost through syndrome measurement.

Whether optimal Pauli decoding succeeds asymptotically for every fixed \(\theta<\pi/4\) remains open. 
Establishing this from the bulk effective field theory requires matching the microscopic logical defect to continuum observables and determining their infrared behavior beyond weak coupling; perturbative RG alone does not locate the decoding threshold. 
A complementary test is the even-distance Rényi entropy \(I_c^{(2)}\), defined by the syndrome-flagged trace moments in Appendix~\ref{app:coherent_info}. 
Within the proposed continuum construction, its connected target is \(SO(4)/U(2)\simeq S^2\). 
The nonvanishing one-loop flow, together with quadratic bare-coupling matching, motivates \(X_2=\beta_R^2\ln d\), contrasting with \(X_4\) at \(n\to1\). 
Agreement would support the proposed replica-dependent scaling.

Near SWAP, the complement relation constrains the SWAP expansion but does not determine the distance dependence of its coefficients or the observed scaling curves. 
For even distance, the unresolved issue is the relative contribution of syndromes supported at SWAP and those populated by detuning. 
For odd distance, it is how nonlinear conditional response and changing Born weights produce the bump and determine its position and fidelity value. 
A microscopic theory of the Born-weighted logical-angle distributions could address these questions and their dependence on geometry. 
Scattering descriptions of projected logical ensembles offer a possible route~\cite{BejanEtAl2026}, but require matching the Born ensemble and boundary conditions; their random-matrix predictions also need separate justification in the nearly ballistic regime.

Our exact parity results can be extended to known, spatially varying real \(Z\)-rotation angles. 
With even-weight \(Z\) stabilizers and an even-weight logical \(Z\), Pauli MLD remains optimal for entanglement fidelity, and the common-correctability bounds are also applicable. 
This includes Gaussian-distributed angles and their disorder averages~\cite{YanBaoVijay2026}. 
Unobserved angle fluctuations, mixed coherent–incoherent noise, and imperfect syndrome measurements require reassessing recovery using the resulting conditional channels and available measurement records~\cite{behrends2025surface, MartonAsboth2023, LeBlondEtAl2025}. 
Changes to stabilizer parity or lattice geometry likewise require reexamining phase pinning and the effective theory. 
The central question is when general CPTP recovery outperforms optimal Pauli recovery with the same available record, and whether this advantage produces distinct asymptotically correctable regions.
We leave these open problems for future study.

\section*{Acknowledgment}
We thank Yimu Bao for helpful discussions.
We thank Yimu Bao and Fei Song for helpful comments on an earlier version of the manuscript.
XDD acknowledges the support from the New Cornerstone Science Foundation.
XD is supported by the CRF award and a fellowship from the Research Grants Council of the Hong Kong Special Administrative Region, China (Project No. HKUST SRFS2324-6S01), and the New Cornerstone Science Foundation.
GYZ would like to thank A. Ludwig and S. Trebst for collaboration on related topic, and thank C. Jian, S. Yan, Y. Bao and S. Vijay for helpful discussions, and acknowledges the support of the National Natural Science Foundation of China---Young Scientists Fund (Grant No.~12504181), the Start-up Fund of HKUST(GZ) (Grant No.~G0101000221), the Guangdong provincial project (Grant No.~2024QN11X201), and the Guangdong Basic and Applied Basic Research Foundation (Grant No.~2026A1515010965). 
The Flatiron Institute is a division of the Simons Foundation.
The numerical simulation is based on the msim simulator in Ref.~\cite{PatoStaplesMsim} and performed on the HPC4 platform of the Hong Kong University of Science and Technology. 
We acknowledge the use of the AI tool ChatGPT for assistance with implementing and executing numerical simulations, figure preparation, checking derivations, and manuscript revision. 
The authors take full responsibility for the results and the final manuscript.

\newpage

\appendix
\makeatletter
\addtocontents{toc}{%
  \let\protect\l@subsection\protect\@gobbletwo
  \let\protect\l@subsubsection\protect\@gobbletwo
}
\makeatother

\section{Logical error channel}\label{app:logical_channel}

\subsection{Completeness of the syndrome-resolved Kraus operators}\label{app:kraus_completeness}

Let $V:\mathcal H_L\rightarrow\mathcal H_{\mathrm{phys}}$ be the encoding isometry and
\begin{equation}
    P_{\mathrm{code}}=VV^\dagger,\qquad V^\dagger V=I_L,
\end{equation}
where $P_{\mathrm{code}}$ projects onto $\mathcal H_{\mathrm c}$ as defined in Sec.~\ref{subsec:channel_setup}. With the syndrome projectors $\Pi_s$ and reference corrections $R_s=Z(C_s)$ defined there, define $F_s=R_s\Pi_sU(\theta)V:\mathcal H_L\to\mathcal H_{\mathrm{phys}}$. The unnormalized corrected physical state is $F_s\rho_LF_s^\dagger$.
The condition $\partial C_s=s$ gives $A_vR_s=s_vR_sA_v$, while $A_v\Pi_s=s_v\Pi_s$. Hence
\begin{equation}
    A_vF_s=s_vR_sA_v\Pi_sU(\theta)V=F_s.
\end{equation}
Moreover, $U(\theta)$, $\Pi_s$, and $R_s$ commute with every $Z$-type stabilizer and therefore preserve its $+1$ eigenvalue on encoded inputs. Thus the image of $F_s$ lies in the simultaneous $+1$ eigenspace of all stabilizers, so $P_{\mathrm{code}}F_s=F_s$. The logical Kraus operator is precisely that of Eq.~\eqref{eq:conditional_state}:
\begin{equation}
K_s=V^\dagger F_s
=V^\dagger R_s\Pi_sU(\theta)V.
\end{equation}
Using this code-space property, $R_s^\dagger R_s=I$, and $\Pi_s^2=\Pi_s$, we obtain
\begin{equation}
K_s^\dagger K_s
=F_s^\dagger P_{\mathrm{code}}F_s
=F_s^\dagger F_s
=V^\dagger U^\dagger(\theta)\Pi_sU(\theta)V.
\end{equation}
Using $\sum_s\Pi_s=I$ from Sec.~\ref{subsec:channel_setup} and the unitarity of $U(\theta)$, summing over $s$ gives
\begin{equation}
\begin{aligned}
\sum_sK_s^\dagger K_s
=&V^\dagger U^\dagger(\theta)
\left(\sum_s\Pi_s\right)
U(\theta)V\\
=&V^\dagger U^\dagger(\theta)U(\theta)V
=I_L.
\end{aligned}
\end{equation}
Consequently, for any normalized logical input $\rho_L$,
\begin{equation}
    p_s(\rho_L)
=\mathrm{Tr}\!\left(K_s\rho_LK_s^\dagger\right),
\qquad
\sum_sp_s(\rho_L)=1.
\end{equation}
Each branch map $\rho_L\mapsto K_s\rho_LK_s^\dagger$ is completely positive and trace-nonincreasing. Completeness makes the syndrome-record channel $\mathcal M_\theta$ in Eq.~\eqref{eq:instrument} trace-preserving.

\subsection{Consequences in the two-sector representation}\label{app:logical_input}
Using the two-sector representation of Sec.~\ref{subsec:logical_kraus},
\begin{equation}
    K_s=A_0^s I_L+A_1^s Z_L,
    \qquad Z_L^2=I_L,\quad \mathrm{Tr}\,Z_L=0,
\end{equation}
where the physical amplitudes $A_0^s,A_1^s$ include the channel normalization. With $D_s=|A_0^s|^2+|A_1^s|^2$ and $m_s$ defined in Eq.~\eqref{eq:m_s_def}, direct multiplication gives, for $D_s>0$,
\begin{equation}
K_s^\dagger K_s=D_s\left(I_L+m_sZ_L\right).
\end{equation}
If $D_s=0$, then $K_s=0$ and the outcome has zero probability for every input; $m_s$ is left undefined. The completeness relation proved in Appendix~\ref{app:kraus_completeness} therefore implies
\begin{equation}
    \sum_sD_s=1,
    \qquad
    \sum_{s:D_s>0}D_sm_s=0.
\end{equation}
For any normalized logical input $\rho_L$ and any outcome with $D_s>0$,
\begin{equation}
p_s(\rho_L)
=\mathrm{Tr}\!\left(\rho_LK_s^\dagger K_s\right)
=D_s\left[1+m_s\langle Z_L\rangle_{\rho_L}\right].
\end{equation}
Thus the syndrome probabilities depend on the input only through its logical $Z_L$ polarization. 
The sum rules ensure $\sum_sp_s(\rho_L)=1$: the input-dependent interference contribution cancels in the total probability, without eliminating conditional measurement backaction.

For $Z_L|0_L\rangle=|0_L\rangle$ and $Z_L|1_L\rangle=-|1_L\rangle$, the corresponding probabilities are
\begin{equation}
\begin{aligned}
p_s\!\left(|0_L\rangle\langle0_L|\right)=&\left|A_0^s+A_1^s\right|^2=D_s(1+m_s),\\
p_s\!\left(|1_L\rangle\langle1_L|\right)=&\left|A_0^s-A_1^s\right|^2=D_s(1-m_s).
\end{aligned}
\end{equation}
The two logical-sector amplitudes remain coherent within a fixed syndrome, and the logical $Z_L$ eigenvalue selects their relative sign. 
Hence $m_s$ is the normalized probability contrast,
\begin{equation}
m_s=\frac{p_s(|0_L\rangle\langle0_L|)-p_s(|1_L\rangle\langle1_L|)}
{p_s(|0_L\rangle\langle0_L|)+p_s(|1_L\rangle\langle1_L|)}.
\end{equation}
Its sign indicates which eigenstate favors outcome $s$, while $|m_s|$ measures their relative probability imbalance. 

All inputs with $\langle Z_L\rangle_{\rho_L}=0$ share this distribution, including the logical $X_L$ eigenstates $\ket{\pm}$, but their conditional outputs can differ and undergo nonunitary weak measurements. 
The full syndrome distribution is independent of the logical input if and only if $m_s=0$ for every $s$ with $D_s>0$. 
Then $K_s/\sqrt{D_s}$ is unitary, allowing perfect recovery with the syndrome record retained.

\subsection{Bounds relating the three measures}\label{app:measure_inequalities}
\paragraph{Common-threshold equality}
We prove Eq.~\eqref{eq:even_averaged_bounds} for the phase-pinned even-distance family of Sec.~\ref{subsec:common_threshold}.
The Born weights are $p_s=D_s$, and we write $\langle f_s\rangle_s=\sum_{s:D_s>0}D_sf_s$.
All conditional expressions below assume $D_s>0$. Phase pinning gives $m_s^2=4q^s_{0}q^s_{1}$; together with $q^s_{0}+q^s_{1}=1$, this implies $|q^s_{0}-q^s_{1}|=\sqrt{1-m_s^2}$. Hence
\begin{equation}
\begin{gathered}
1-P_{\mathrm{succ}}(s)=\frac{1-\sqrt{1-m_s^2}}{2},\\
1-P_{\mathrm{Bayes}}(s)=\frac{m_s^2}{2},\\
\ln2-I_c(s)=\ln2-h_2\!\left(\frac{1+m_s}{2}\right).
\label{eq:even_exact_kernels}
\end{gathered}
\end{equation}
Rationalizing the first expression gives
\begin{equation}
\begin{aligned}
\frac14m_s^2
\leq1-P_{\mathrm{succ}}(s)
=\frac{m_s^2}{2\left(1+\sqrt{1-m_s^2}\right)}
\leq\frac12m_s^2.
\end{aligned}
\end{equation}
For the deficit of $I_c$, set $m=|m_s|\in[0,1]$ and define
\begin{equation}
\begin{aligned}
f(m)=&\ln2-h_2\!\left(\frac{1+m}{2}\right)\\
=&\frac12\left[(1+m)\ln(1+m)+(1-m)\ln(1-m)\right],
\end{aligned}
\end{equation}
with $0\ln0=0$. 
Since $f(0)=f'(0)=0$, $f'(m)=\tfrac{1}{2}\ln{\tfrac{1+m}{1-m}}$ and $f''(m)=\tfrac{1}{1-m^2}$, integrating twice and setting $t=mu$ gives
\begin{equation}
g(m)\equiv\frac{f(m)}{m^2}
=\frac1{m^2}\int_0^m\frac{m-t}{1-t^2}\,dt
=\int_0^1\frac{1-u}{1-m^2u^2}\,du.
\end{equation}
For $0\leq u<1$, the integrand obeys
\begin{equation}
1-u\leq\frac{1-u}{1-m^2u^2}\leq\frac1{1+u}.
\end{equation}
Integration and continuous extension to $m=0,1$ yield $1/2\leq g(m)\leq\ln2$, and hence
\begin{equation}
\frac12m_s^2\leq\ln2-I_c(s)\leq(\ln2)m_s^2.
\end{equation}
Both constants are sharp: $g(0)=1/2$ and $g(1)=\ln2$. This entropy bound does not require phase pinning; phase pinning is needed to express the Pauli deficits through the same $m_s^2$.

Born averaging gives the bounds in Eq.~\eqref{eq:even_averaged_bounds}. 
Since the comparison constants are positive and independent of code size, at fixed physical angle $\theta$ as $d\to\infty$ through even distances,
\begin{equation}
\begin{gathered}
P_{\mathrm{succ}}\to1
\quad\Longleftrightarrow\quad
P_{\mathrm{Bayes}}\to1
\quad\Longleftrightarrow\quad\\
I_c\to\ln2
\quad\Longleftrightarrow\quad
\langle m_s^2\rangle_s\to0.
\end{gathered}
\label{eq:even_common_zero_equality}
\end{equation}
Thus the three measures define the same asymptotically correctable set and the same threshold, whenever that set has a well-defined threshold boundary. These bounds do not locate the threshold.


\paragraph{Sharp bounds between the Pauli measures.}
For arbitrary relative phase, define $e_s=\min(q^s_{0},q^s_{1})\in[0,1/2]$. Given that two weights sum to one,
\begin{equation}
 1-P_{\mathrm{succ}}(s)=e_s,\quad
 1-P_{\mathrm{Bayes}}(s)=2e_s(1-e_s).
 \label{eq:app_pauli_deficit_identity}
\end{equation}
With $1\leq2(1-e_s)\leq2$, we have
\begin{equation}
 e_s\leq2e_s(1-e_s)\leq2e_s.
\end{equation}
Multiplying by the Born weights $D_s$ and summing gives
\begin{equation}
 \frac12(1-P_{\mathrm{Bayes}})
 \leq1-P_{\mathrm{succ}}
 \leq1-P_{\mathrm{Bayes}}.
 \label{eq:app_sharp_pauli_bounds}
\end{equation}
Both constants are sharp: the ratio $(1-P_{\mathrm{succ}}(s))/(1-P_{\mathrm{Bayes}}(s))$ approaches $1/2$ as $e_s\to0^+$ and equals $1$ at $e_s=1/2$. 
Setting $e_s = e$ identically across all supported syndromes establishes that both the upper bound ($e = 1/2$) and lower bound ($e \to 0^+$) are sharp for arbitrary binary posterior ensembles.
This proves the bound in Eq.~\eqref{eq:even_pauli_bounds} without phase pinning or a code distance restriction. 
For one logical qubit, the lower coefficient $1/2$ strengthens the bound $1/(4\ln2)$ in Appendix A of Ref.~\cite{YanBaoVijay2026}. 
The two Pauli measures therefore share the same decoding threshold, even when coherent information defines a different one.

\subsection{Odd-weight stabilizers and the limits of the common-threshold equality}\label{app:odd_weight_limits}
For CSS codes encoding one logical qubit under coherent $Z$ rotations, the Kraus operators retain the two-sector form $K_s = A_s^0 I_L + A_s^1 Z_L$. However, if an odd-weight $Z$-type stabilizer generator $B$ exists ($|B| \equiv 1 \pmod 2$), strings within the same sector can differ in weight parity,
\begin{equation}
|E \oplus B| = |E| + |B| - 2|E \cap B| \equiv |E| + 1 \pmod 2.
\end{equation}
Configurations within the same homology sector then accumulate relative factors of $\pm i$, invalidating the global phase-pinning argument of Sec.~\ref{subsec:logical_parity} without precluding accidental alignment. (Odd-weight $X$-checks do not affect phase pinning for coherent $Z$ noise.)

To characterize arbitrary relative phases, we define the normalized interference parameter on the unit disk:
\begin{equation}
u_s \equiv \frac{2(A_s^0)^* A_s^1}{D_s} = x_s + i y_s, \quad x_s = m_s,
\label{eq:complex_u_def}
\end{equation}
where $|u_s|^2 = 4 q_s^0 q_s^1 \le 1$. 
The phase-pinned even and odd square families correspond to the real ($y_s = 0$) and imaginary ($x_s = 0$) diameters, respectively. 
Since $|q_s^0 - q_s^1| = \sqrt{1 - |u_s|^2}$, the conditional deficits are
\begin{gather}
1 - P_{\text{succ}}(s) = \frac{1 - \sqrt{1 - |u_s|^2}}{2}, \quad 1 - P_{\text{Bayes}}(s) = \frac{|u_s|^2}{2}, \nonumber \\
\ln 2 - I_c(s) = \ln 2 - h_2\left(\frac{1 + x_s}{2}\right).
\label{eq:conditional_deficits_unpinned}
\end{gather}
While the Pauli measures depend on the total magnitude $|u_s|$, coherent information depends exclusively on the real component $x_s$, which governs nonunitary weak measurement: $K_s^\dagger K_s = D_s (I_L + x_s Z_L)$ and $\rho_{L|s} = \frac{1}{2}(I_L + x_s Z_L)$. 
In particular, when $x_s = 0$, $K_s/\sqrt{D_s}$ remains unitary ($I_c(s) = \ln 2$) even if $|u_s| > 0$ and Pauli recovery is imperfect.

Averaging over the Born distribution $p_s = D_s$ and defining the ensemble second moments
\begin{equation}
X^2 \equiv \langle x_s^2 \rangle_s, \quad Y^2 \equiv \langle y_s^2 \rangle_s, \quad R^2 \equiv \langle |u_s|^2 \rangle_s = X^2 + Y^2,
\end{equation}
the single-variable bounds of Appendix~\ref{app:measure_inequalities} directly imply
\begin{equation}
\begin{aligned}
\frac{1}{4} R^2 &\le 1 - P_{\text{succ}} \le \frac{1}{2} R^2 = 1 - P_{\text{Bayes}}, \\ \frac{1}{2} X^2 &\le \ln 2 - I_c \le (\ln 2) X^2.
\label{eq:unpinned_moment_bounds}
\end{aligned}
\end{equation}
As $d \to \infty$ at fixed $\theta$, successful Pauli recovery requires $R^2 \to 0$ (forcing both $X^2 \to 0$ and $Y^2 \to 0$), whereas $I_c \to \ln 2$ requires only $X^2 \to 0$. 
Because $X^2 \le R^2$, any Pauli-correctable regime is coherent-information correctable, yielding the threshold hierarchy
\begin{equation}
\theta_c^{P_{\text{succ}}} = \theta_c^{P_{\text{Bayes}}} \le \theta_c^{I_c}.
\end{equation}

The thresholds coincide if and only if $X^2 \to 0 \implies Y^2 \to 0$ across the correctable interval (satisfied, e.g., if $Y^2 \le C X^2$ with size-independent $C$). 
Conversely, a split threshold requires $X^2 \to 0$ while $Y^2$ remains nonvanishing in the thermodynamic limit. 
A finite limiting ratio $Y^2/R^2$ alone is insufficient; for example, $X^2 = Y^2 \propto d^{-2}$ gives $Y^2/R^2 = 1/2$ even though all deficits vanish asymptotically. 
Whether boundary geometries with odd-weight checks exhibit split thresholds thus depends on the absolute asymptotic scaling of $Y^2$, rather than the mere absence of phase pinning.

\subsection{Maximum-likelihood decoding optimality}
\label{app:decoder_optimality}
We consider the one-logical-qubit instrument $K_s = A_s^0 I_L + A_s^1 Z_L$ and optimize over all deterministic, syndrome-conditioned CPTP recovery maps $\mathcal{R}_s$ acting on the logical state after syndrome measurement. 
Entanglement fidelity is evaluated for the maximally mixed state $I_L/2$, purified by the Bell state $|\Phi\rangle_{LR} = (|00\rangle + |11\rangle)/\sqrt{2}$ with an isolated reference $R$. 
All conditional quantities are evaluated on supported syndromes ($D_s > 0$) with Born weights $p_s = D_s$.

\paragraph{Recovery objective and conditional fidelity}
For each outcome $s$, a general CPTP recovery map takes the Kraus form
\begin{equation}
\mathcal{R}_s(\rho) = \sum_a L_{s,a}\rho L_{s,a}^\dagger, \quad \sum_a L_{s,a}^\dagger L_{s,a} = I_L.
\label{eq:app_recovery_kraus}
\end{equation}
Acting on the normalized conditional state $|\psi_s\rangle_{LR} = (K_s \otimes I_R)|\Phi\rangle_{LR}/\sqrt{D_s}$, the conditional entanglement fidelity is
\begin{equation}
\begin{aligned}
F_e(s; \mathcal{R}_s) =& \langle \Phi | (\mathcal{R}_s \otimes \mathrm{id}_R)(|\psi_s\rangle\langle\psi_s|) |\Phi\rangle \\
= &\frac{1}{4 D_s} \sum_a |\!\operatorname{tr}(L_{s,a} K_s)|^2,
\label{eq:app_fe_overlap}
\end{aligned}
\end{equation}
where we used $\langle \Phi | (M \otimes I_R) |\Phi\rangle = \frac{1}{2}\operatorname{tr} M$. 
Selecting $\mathcal{R}_s$ according to the syndrome record yields the trace-preserving recovered channel $\mathcal{E}_{\mathrm{rec}}(\rho_L) = \sum_s \mathcal{R}_s(K_s \rho_L K_s^\dagger)$ and the ensemble fidelity $F_e(\{\mathcal{R}_s\}) = \sum_{s: D_s > 0} D_s F_e(s; \mathcal{R}_s)$. 

Evaluating Eq.~\eqref{eq:app_fe_overlap} for unitary Pauli recoveries ($L_{s,1} = U$) gives $F_e(s; I_L) = q_s^0$ and $F_e(s; Z_L) = q_s^1$, whereas $X_L$ and $Y_L$ yield zero. Thus, standard maximum-likelihood Pauli decoding achieves $F_e^{\mathrm{MLD}}(s) = \max(q_s^0, q_s^1) = P_{\mathrm{succ}}(s)$ for arbitrary relative phase.

\paragraph{Universal quantum upper bound}
To maximize $F_e(s; \mathcal{R}_s)$ over all CPTP maps, we employ the polar decomposition $K_s = W_s |K_s|$, where $|K_s| = (K_s^\dagger K_s)^{1/2} \ge 0$ and $W_s$ is unitary (extending the polar isometry to a unitary on the kernel if $K_s$ is rank-deficient). 
Defining $T_{s,a} \equiv L_{s,a} W_s$, trace preservation implies $\sum_a T_{s,a}^\dagger T_{s,a} = I_L$. 
Applying the Cauchy--Schwarz inequality for the Hilbert--Schmidt inner product to $|K_s|^{1/2}$ and $T_{s,a} |K_s|^{1/2}$ yields
\begin{equation}
|\!\operatorname{tr}(T_{s,a} |K_s|)|^2 \le \operatorname{tr}|K_s| \cdot \operatorname{tr}(|K_s| T_{s,a}^\dagger T_{s,a}).
\end{equation}
Summing over the recovery Kraus index $a$, we obtain the universal operator upper bound
\begin{equation}
\begin{aligned}
\sum_a |\!\operatorname{tr}(L_{s,a} K_s)|^2 &\le \operatorname{tr}|K_s| \cdot \operatorname{tr}\!\left(|K_s| \sum_a T_{s,a}^\dagger T_{s,a}\right)\\ 
&= (\operatorname{tr}|K_s|)^2.
\label{eq:app_cs_bound}
\end{aligned}
\end{equation}
The bound is saturated by the unitary recovery $\mathcal{R}_s(\rho) = W_s^\dagger \rho W_s$ ($L_{s,1} = W_s^\dagger$), since $\operatorname{tr}(W_s^\dagger K_s) = \operatorname{tr}|K_s|$. 
This establishes that the optimal recovery is always unitary, holding for full-rank and rank-one Kraus operators alike without requiring matrix inversion.

\paragraph{Singular-value reduction and closed-form bound}
The bound is evaluated by diagonalizing $|K_s|^2 = K_s^\dagger K_s = D_s(I_L + m_s Z_L)$ in the logical $Z_L$ eigenbasis. 
The singular values of $K_s$ are
\begin{equation}
\sigma_{s,\pm} = |A_s^0 \pm A_s^1| = \sqrt{D_s(1 \pm m_s)},
\label{eq:app_singular_values}
\end{equation}
where $\pm$ denotes the $Z_L$ eigenvalue. Substituting $\operatorname{tr}|K_s| = \sigma_{s,+} + \sigma_{s,-}$ into Eq.~\eqref{eq:app_cs_bound} yields the closed-form conditional optimum
\begin{equation}
\begin{aligned}
F_e^{\mathrm{opt}}(s) &\equiv \max_{\mathcal{R}_s \in \mathrm{CPTP}} F_e(s; \mathcal{R}_s) = \frac{(\operatorname{tr}|K_s|)^2}{4 D_s} \\
&= \frac{(\sqrt{1+m_s} + \sqrt{1-m_s})^2}{4} = \frac{1 + \sqrt{1 - m_s^2}}{2}.
\label{eq:app_fe_opt_closed}
\end{aligned}
\end{equation}
Because the recovery map can be optimized independently for each syndrome, the optimal ensemble entanglement fidelity is $F_e^{\mathrm{opt}} = \sum_{s: D_s > 0} D_s F_e^{\mathrm{opt}}(s)$.

\paragraph{Even phase pinning and Pauli MLD optimality}
For the even-distance square surface codes, factoring out a common global phase makes both sector amplitudes $A_s^0$ and $A_s^1$ real. 
Then $|A_s^0 + A_s^1| + |A_s^0 - A_s^1| = 2\max(|A_s^0|, |A_s^1|)$, and Eq.~\eqref{eq:app_fe_opt_closed} reduces to
\begin{equation}
F_e^{\mathrm{opt}}(s) = \frac{\max(|A_s^0|^2, |A_s^1|^2)}{D_s} = P_{\mathrm{succ}}(s).
\label{eq:app_even_mld_optimal_fidelity}
\end{equation}
The optimal polar recovery unitary $W_s^\dagger$ reduces precisely (up to an overall sign) to $I_L$ when $|A_s^0| \ge |A_s^1|$ and to $Z_L$ when $|A_s^1| > |A_s^0|$; at equal magnitudes, either choice achieves the conditional optimum $1/2$. Because MLD independently saturates the optimal fidelity for every supported outcome, averaging over the Born distribution yields the exact finite-distance equality
\begin{equation}
F_e^{\mathrm{opt}} \equiv \max_{\{\mathcal{R}_s\}} F_e(\{\mathcal{R}_s\}) = P_{\mathrm{succ}}.
\label{eq:app_even_ensemble_opt}
\end{equation}
This identity establishes the optimality of Pauli MLD at finite code distance independently of the thermodynamic limit. It demonstrates that the even-distance fidelity plateaus in Fig.~\ref{fig:measure_parity} represent the ultimate performance attainable by any quantum recovery.

\subsection{Fidelities at the maximum error angle}
\label{app:swap_endpoint_recovery}
We prove the endpoint Kraus forms of Sec.~\ref{subsec:swap_endpoint_recovery} for the square rotated surface code (Fig.~\ref{fig:setup_phase_diagram}) with $N = d^2$ physical qubits and one encoded logical qubit. 
Throughout this subsection, operators are evaluated at the maximum error angle $\theta = \pi/4$, where $U_* = \Pi_j e^{i\pi Z_j/4}$. 
Retaining the encoding isometry $V$, syndrome projector $\Pi_s$, and reference correction $R_s = Z(C_s)$ of Sec.~\ref{subsec:channel_setup}, the Kraus operator is $K_s = V^\dagger R_s \Pi_s U_* V$. 
On each physical qubit, $U_*$ satisfies the Clifford exchange
\begin{equation}
U_*^\dagger X_j U_* = Y_j = i X_j Z_j \implies X_j U_* = U_* Y_j.
\label{eq:app_endpoint_clifford}
\end{equation}

\paragraph{Odd distance}
Let $\Omega$ denote the set of all $N = d^2$ physical qubits, with global Pauli strings $X_\Omega = \Pi_{j\in\Omega} X_j$, $Z_\Omega = \Pi_{j\in\Omega} Z_j$, and $Y_\Omega = \Pi_{j\in\Omega} Y_j$. 
Because all stabilizer generators have even weight, $X_\Omega$ and $Z_\Omega$ preserve the code space. 
For odd $d$, $X_\Omega$ anticommutes with the weight-$d$ logical representative $Z_L$, and $Z_\Omega$ anticommutes with $X_L$. 
For the standard rotated patch of Fig.~\ref{fig:setup_phase_diagram} with $N \equiv 1 \pmod 4$, this gives
\begin{equation}
\begin{gathered}
X_\Omega V = V X_L, \quad Z_\Omega V = V Z_L,\\
Y_\Omega V = i^N X_\Omega Z_\Omega V = V Y_L,
\label{eq:app_odd_global_pauli}
\end{gathered}
\end{equation}
where $Y_L = i X_L Z_L$. 
Equation~\eqref{eq:app_endpoint_clifford} then implies $X_\Omega U_* V = U_* Y_\Omega V = U_* V Y_L$. 
Furthermore, $X_\Omega$ commutes with all stabilizer generators (since each check has even weight) and hence $[X_\Omega, \Pi_s] = 0$. 
Since $X_\Omega R_s = c_s R_s X_\Omega$ with sign $c_s = (-1)^{|C_s|}$, we obtain
\begin{equation}
\begin{aligned}
X_L K_s =& V^\dagger X_\Omega R_s \Pi_s U_* V \\
=& c_s V^\dagger R_s \Pi_s X_\Omega U_* V = c_s K_s Y_L.
\label{eq:app_odd_intertwiner}
\end{aligned}
\end{equation}
Substituting $K_s = A_s^0 I_L + A_s^1 Z_L$ and $Y_L = i X_L Z_L$ into Eq.~\eqref{eq:app_odd_intertwiner} and matching coefficients of $X_L$ and $X_L Z_L$ yields
\begin{equation}
A_s^1 = i c_s A_s^0 \to \frac{K_s}{\sqrt{2}A^0_s} =  \frac{I_L + i c_s Z_L}{\sqrt{2}} = e^{i c_s \pi Z_L / 4}.
\label{eq:app_odd_kraus_endpoint}
\end{equation}
Thus, for every supported syndrome, $K_s/\sqrt{D_s}$ is proportional to the deterministic logical rotation $e^{i c_s \pi Z_L / 4}$.

\paragraph{Even distance}
Let $\Gamma$ be the top-boundary support of the logical operator $Z_L$, containing $d$ physical qubits numbered $1, \dots, d$. 
For even $d$, the $d/2$ weight-two boundary checks $A_{v_j} = X_{2j-1} X_{2j}$ ($j = 1, \dots, d/2$) partition $\Gamma$ without overlap. 
Their product and its measured syndrome eigenvalue are
\begin{equation}
B_\Gamma = \prod_{j=1}^{d/2} A_{v_j} = X_\Gamma, \quad b_s = \prod_{j=1}^{d/2} s_{v_j} \in \{\pm 1\},
\end{equation}
so that $B_\Gamma V = V$ and $\Pi_s B_\Gamma = b_s \Pi_s$. 
Applying Eq.~\eqref{eq:app_endpoint_clifford} across $\Gamma$ gives
\begin{equation}
\begin{gathered}
    U_*^\dagger B_\Gamma U_* = \prod_{k\in\Gamma} Y_k = i^d X_\Gamma Z_\Gamma = (-1)^{d/2} B_\Gamma Z_\Gamma\\
    \implies B_\Gamma U_* V = (-1)^{d/2} U_* V Z_L.
\end{gathered}
\end{equation}
Multiplying on the left by $V^\dagger R_s \Pi_s$ and using $\Pi_s B_\Gamma = b_s \Pi_s$ yields
\begin{equation}
b_s K_s = (-1)^{d/2} K_s Z_L \implies K_s Z_L = \epsilon_s K_s,
\label{eq:app_even_eigenvalue}
\end{equation}
where $\epsilon_s \equiv (-1)^{d/2} b_s$.
Comparing the coefficients of $I_L$ and $Z_L$ enforces $A_s^1 = \epsilon_s A_s^0$, which gives
\begin{equation}
K_s = A_s^0 (I_L + \epsilon_s Z_L).
\label{eq:app_even_kraus_endpoint}
\end{equation}
Thus, each supported branch collapses to a rank-one projector $(I_L \pm Z_L)/2$; the boundary parity $b_s$ directly measures the logical $Z_L$ eigenvalue.

\paragraph{Normalization and recovery measures}
For both cases, Eqs.~\eqref{eq:app_odd_kraus_endpoint} and~\eqref{eq:app_even_kraus_endpoint} imply that
\begin{equation}
|A_s^0|^2 = |A_s^1|^2 = \frac{D_s}{2} \implies q_s^0 = q_s^1 = \frac{1}{2}.
\end{equation}
Consequently, both conditional Pauli measures equal $1/2$ on every supported syndrome, yielding the exact ensemble averages
\begin{equation}
P_{\text{succ}} = P_{\text{Bayes}} = \frac{1}{2}, \quad \theta = \frac{\pi}{4}.
\end{equation}
The interference parameter $m_s$ vanishes identically for odd $d$ and equals $\pm 1$ for even $d$. Evaluating the coherent information therefore gives
\begin{equation}
I_c = \begin{cases}
\ln 2, & d\text{ odd}, \\
0, & d\text{ even},
\end{cases}
\end{equation}
confirming the exact values at $\theta=\pi/4$ in Fig.~\ref{fig:measure_parity}.

\subsection{SWAP coefficients and newly populated syndromes}
\label{app:endpoint_coefficients}
We derive the near-SWAP expansion of $P_{\mathrm{succ}}$ at finite distance $d$, holding the reference correction and logical-sector labels fixed. 
The normalized physical amplitudes give $D_s=|A_0^s|^2+|A_1^s|^2$ and $\sum_sD_s=1$, so
\begin{equation}
 P_{\mathrm{succ}}-\tfrac12
 =\sum_{s}D_s\bigl[P_{\mathrm{succ}}(s)-\tfrac12\bigr]
 =\tfrac12\sum_s\left||A_1^s|^2-|A_0^s|^2\right|.
 \label{eq:app_endpoint_success_identity}
\end{equation}
The last expression remains defined when both amplitudes vanish. 
Set $p_{0,s}=D_s(0)$ and separate the outcomes supported at SWAP, $\mathcal S_+=\{s:p_{0,s}>0\}$, from those absent there, $\mathcal S_0=\{s:p_{0,s}=0\}$. 
In $\mathcal S_0$, both SWAP-point amplitudes vanish; this set also includes outcomes that remain absent under detuning. Write $\langle f_s\rangle_{p_0}=\sum_{s\in\mathcal S_+}p_{0,s}f_s$.
All derivatives are with respect to $\beta_R$, using the signed-detuning extension of Sec.~\ref{subsec:swap_variables_parity}; the physical regime has $\beta_R\geq0$.

\paragraph{Even distance.}
For the even square surface codes, Appendix~\ref{app:swap_endpoint_recovery} establishes $|A_0^s(0)|=|A_1^s(0)|$ on $\mathcal S_+$. 
The complement relation makes each sector weight even. 
Hence the DWFE $\Delta F_s=-\ln(|A_1^s|^2/|A_0^s|^2)$ is analytic and even near zero. Using $P_{\mathrm{succ}}(s)-\tfrac12=\tfrac12\tanh(|\Delta F_s|/2)$ gives
\begin{equation}
\begin{gathered}
 \Delta F_s(\beta_R)=\tfrac12\Delta F_s''(0)\beta_R^2+O(\beta_R^4),\\
 D_s(\beta_R)=p_{0,s}+O(\beta_R^2),\\
 D_s\bigl[P_{\mathrm{succ}}(s)-\tfrac12\bigr]
 =\frac{p_{0,s}|\Delta F_s''(0)|}{8}\beta_R^2+O(\beta_R^4).
 \label{eq:app_supported_even_coefficient}
\end{gathered}
\end{equation}
Changes in the Born weight multiply a quadratic conditional excess and first contribute at fourth order.

For $s\in\mathcal S_0$, since the endpoint ratio is undefined, we need to define $a_{\mu,s}=\left.\partial_{\beta_R}A_\mu^s\right|_0$, with $\mu=0,1$, and expand the amplitudes directly. 
Evenness of each sector weight removes odd powers from the resulting probabilities:
\begin{equation}
\begin{gathered}
 A_\mu^s(\beta_R)=a_{\mu,s}\beta_R+O(\beta_R^2),\\
D_s(\beta_R)=\bigl(|a_{0,s}|^2+|a_{1,s}|^2\bigr)\beta_R^2+O(\beta_R^4),\\
 \tfrac12\left||A_1^s|^2-|A_0^s|^2\right|
 =\tfrac12\left||a_{1,s}|^2-|a_{0,s}|^2\right|\beta_R^2+O(\beta_R^4).
\label{eq:app_absent_amplitude_expansion}
\end{gathered}
\end{equation}
Thus, a newly populated outcome can contribute quadratically despite its zero weight at SWAP. 
Summing both sets in Eq.~\eqref{eq:app_endpoint_success_identity} gives the expansion used in Sec.~\ref{subsec:swap_even_scaling}:
\begin{equation}
\begin{gathered}
 P_{\mathrm{succ}}-\tfrac12=C_{\mathrm{even}}(d)\beta_R^2+O(\beta_R^4),\\
 C_{\mathrm{even}}(d)=\tfrac18\langle|\Delta F_s''(0)|\rangle_{p_0}+C_{\mathrm{new}}(d),\\
 C_{\mathrm{new}}(d)=\tfrac12\sum_{s\in\mathcal S_0}
 \left||a_{1,s}|^2-|a_{0,s}|^2\right|.
\end{gathered}
 \label{eq:app_even_endpoint_coefficient}
\end{equation}
The quadratic probability involves $|a_{0,s}|^2+|a_{1,s}|^2$, while the success excess involves their absolute difference. 
Only unequal leading weights contribute to $C_{\mathrm{new}}$; balanced weights or vanishing first derivatives give an $O(\beta_R^4)$ or smaller excess. 
Since $p_{0,s}=0$ on $\mathcal S_0$, the nonnegative $C_{\mathrm{new}}$ is missed by an endpoint average alone.

\paragraph{Odd distance.}
The complement relation exchanges the sector weights under $\beta_R\mapsto-\beta_R$. 
On $\mathcal S_+$, it enforces endpoint balance and an odd DWFE, while $D_s=p_{0,s}+O(\beta_R^2)$ is even. 
The same fidelity kernel as above gives
\begin{equation}
\begin{aligned}
 \Delta F_s(\beta_R)&=\Delta F_s'(0)\beta_R+O(\beta_R^3),\\
 D_s\bigl[P_{\mathrm{succ}}(s)-\tfrac12\bigr]
 &=\frac{p_{0,s}|\Delta F_s'(0)|}{4}|\beta_R|+O(|\beta_R|^3).
 \end{aligned}
 \label{eq:app_supported_odd_expansion}
\end{equation}
For $s\in\mathcal S_0$, Eq.~\eqref{eq:app_absent_amplitude_expansion} and sector exchange give $|a_{0,s}|^2=|a_{1,s}|^2$. 
Any quadratic Born weight is therefore balanced. 
The sector-weight difference is odd and starts at cubic order or higher, so
\begin{equation}
 \tfrac12\left||A_1^s|^2-|A_0^s|^2\right|=O(|\beta_R|^3).
 \label{eq:app_absent_odd_order}
\end{equation}
Thus, outcomes absent at SWAP do not enter the linear coefficient. 
Summing Eq.~\eqref{eq:app_endpoint_success_identity} gives
\begin{equation}
\begin{gathered}
 P_{\mathrm{succ}}-\tfrac12=C_{\mathrm{odd}}(d)|\beta_R|+O(|\beta_R|^3),\\
 C_{\mathrm{odd}}(d)=\tfrac14\langle|\Delta F_s'(0)|\rangle_{p_0}.
 \end{gathered}
 \label{eq:app_odd_endpoint_coefficient}
\end{equation}
For $C_{\mathrm{odd}}(d)>0$, this gives a cusp in the signed-detuning extension and a one-sided linear response in the physical regime. 

These expansions determine the leading response of $P_{\mathrm{succ}}$ at fixed $d$. 
The even finite-size collapse is consistent with $C_{\mathrm{even}}(d)\propto d$ over the sampled sizes, while the odd collapse and SWAP-susceptibility evidence support an approximately extensive $C_{\mathrm{odd}}(d)$. 
These trends motivate $\beta_R^2d$ and $\beta_Rd$, respectively.
For even distance, the trend concerns the total coefficient, not its two contributions separately. 

\subsection{Uniform-angle plateau benchmarks}\label{app:uniform_angle_plateau}
We derive the plateau benchmarks of Sec.~\ref{subsec:plateau} and express them in the folded angle used near SWAP. 
Throughout, angular distributions are Born-weighted with $p_s=D_s$ and zero-weight branches are omitted. 
Angular uniformity is a benchmark assumption; it does not follow from phase pinning.

\paragraph{Even-distance coherent information.}
Under even phase pinning, the logical tilt angle $\theta_s \in (-\pi/4, \pi/4)$ defined in Eq.~\eqref{eq:even_logical_tilt} yields the unordered sector weights $\cos^2 t_s$ and $\sin^2 t_s$ with magnitude $t_s = |\theta_s|$, and $|m_s| = \sin(2t_s)$. 
Assuming a uniform Born distribution for $t \in [0, \pi/4]$, the average coherent information in Eq.~\eqref{eq:even_uniform_plateau_integrals} is
\begin{equation}
I_c^{*,\mathrm{even}}
=\frac4\pi\int_0^{\pi/4}
h_2\!\left(\frac{1+\sin(2t)}2\right)dt
=2\ln2-1.
\label{eq:even_uniform_entropy_integral}
\end{equation}
Using the identity $\frac{1 + \sin(2t)}{2} = \cos^2(t - \pi/4)$ and the substitution $u = \pi/4 - t$, the integrand becomes $h_2(\cos^2 u) = -\cos^2 u \ln(\cos^2 u) - \sin^2 u \ln(\sin^2 u)$. 
By reflection across $u = \pi/4$, the integral over $[0, \pi/4]$ combines into the quarter-circle integral
\begin{equation}
\int_0^{\tfrac{\pi}{4}} h_2(\cos^2 u)\, du = -\int_0^{\tfrac{\pi}{2}} \cos^2 x \ln(\cos^2 x)\, dx = -J'(1),
\end{equation}
where $J(a) \equiv \int_0^{\pi/2} \cos^{2a}\!x\, dx = \tfrac{\sqrt{\pi}\,\Gamma(a + 1/2)}{2\,\Gamma(a + 1)}$. 
Differentiating at $a = 1$ yields $J'(1) = \frac{\pi}{4}(1 - 2\ln 2)$, which directly gives
\begin{equation}
I_c^{*,\mathrm{even}} = -\frac{4}{\pi} J'(1) = 2\ln 2 - 1.
\end{equation}

\paragraph{Odd-distance averages.}
For odd distance, the logical rotation angle $\theta_s$ of Eq.~\eqref{eq:odd_logical_rotation} is defined modulo $\pi$ on $[-\pi/2, \pi/2)$. 
At fixed physical angle and code distance, its Born-weighted distribution is
\begin{equation}
\rho_d^{\mathrm{odd}}(\vartheta)
=\sum_{s:D_s>0}p_s\,\delta(\vartheta-\theta_s),
\qquad p_s=D_s.
\label{eq:born_weighted_angle_distribution}
\end{equation}
As a benchmark, replace this distribution by the uniform density $1/\pi$. 
The conditional Pauli measures in Eq.~\eqref{eq:pauli_meas_angle} then give
\begin{equation}
\begin{aligned}
&P_{\mathrm{succ}}^*
=\frac1\pi\int_{-\pi/2}^{\pi/2}
\frac{1+|\cos(2\vartheta)|}{2}\,d\vartheta
=\frac12+\frac1\pi,\\
&P_{\mathrm{Bayes}}^*
=\frac1\pi\int_{-\pi/2}^{\pi/2}
\left(\cos^4\vartheta+\sin^4\vartheta\right)d\vartheta
=\frac34.
\end{aligned}
\label{eq:odd_uniform_plateau_integrals}
\end{equation}
The two Pauli averages coincide with the even-distance values because the sector weights have the same angular dependence. 
In contrast, odd phase pinning gives $m_s=0$ and $I_c(s)=\ln2$ for every branch, independently of angular uniformity.

\section{Coherent information and R\'enyi replicas}\label{app:coherent_info}
This appendix derives the two trace moments used in Sec.~\ref{subsec:replica_symmetry} from the syndrome-resolved logical channel. 
We consider one logical qubit with maximally mixed input $I_L/2$ and retain the classical syndrome record. The logical Kraus operators are
\begin{equation}
 K_s=A_0^sI_L+A_1^sZ_L,
 \qquad A_\mu^s=\mathcal N_\theta\mathcal Z_\mu^s,
 \quad \mu=0,1,
 \label{eq:app_renyi_kraus}
\end{equation}
where $\mathcal N_\theta$ is the common normalization factor defined in Sec.~\ref{subsec:complex_rbim}. 
The physical amplitudes satisfy $\sum_sK_s^\dagger K_s=I_L$. 

\subsection{R\'enyi entropy  and coherent information}
\label{app:renyi_coherent_info}
Purify the logical input $I_L/2$ with an untouched reference qubit $R$:
\begin{equation}
 |\Phi\rangle_{LR}=\frac{|0_L0_R\rangle+|1_L1_R\rangle}{\sqrt2},
 \quad \operatorname{tr}_R|\Phi\rangle\langle\Phi|=\frac{I_L}{2}.
 \label{eq:app_renyi_purification}
\end{equation}
For outcome $s$, the unnormalized logical-reference vector $|\Psi_s\rangle=(K_s\otimes I_R)|\Phi\rangle$ is
\begin{equation}
 |\Psi_s\rangle=\frac{1}{\sqrt2}\big[(A_0^s+A_1^s)|0_L0_R\rangle+(A_0^s-A_1^s)|1_L1_R\rangle\big].
 \label{eq:app_renyi_conditional_vector}
\end{equation}
Its squared norm gives the Born probability:
\begin{equation}
 p_s=\langle\Psi_s|\Psi_s\rangle
 =\frac12\operatorname{tr}(K_s^\dagger K_s)
 =|A_0^s|^2+|A_1^s|^2
 \equiv D_s,
 \label{eq:app_renyi_born}
\end{equation}
with $\sum_sD_s=1$.
Thus $D_s$ includes the factor $|\mathcal N_\theta|^2$ multiplying the unnormalized RBIM weights. 
Conditional states are defined only for $D_s>0$, with normalized joint vector $|\Psi_s\rangle/\sqrt{D_s}$. 
Zero-weight outcomes contribute nothing to unconditional sums.

With orthogonal syndrome flags, the joint output and its reduction over the reference qubit are
\begin{equation}
\begin{gathered}
 \rho_{SLR}=\sum_s|s\rangle\langle s|\otimes|\Psi_s\rangle\langle\Psi_s|,\\
 \rho_{SL}=\operatorname{tr}_R\rho_{SLR}
 =\sum_s|s\rangle\langle s|\otimes\frac12K_sK_s^\dagger.
 \end{gathered}
 \label{eq:app_renyi_joint_state}
\end{equation}
In the logical-$Z$ eigenbasis, $Z_L|\chi\rangle_L=\chi|\chi\rangle_L$ with $\chi=\pm1$, the eigenvalue of $K_s$ is $A_0^s+\chi A_1^s$. 
The unnormalized output block therefore has eigenvalues $T_\chi^s/2$, where
\begin{equation}
 T_\chi^s\equiv|A_0^s+\chi A_1^s|^2
 =|\mathcal N_\theta|^2|\mathcal Z_0^s+\chi\mathcal Z_1^s|^2.
 \label{eq:app_renyi_character_weights}
\end{equation}
Using the interference parameter $m_s$ from Eq.~\eqref{eq:m_s_def}, $T_\chi^s=D_s(1+\chi m_s)$ for $D_s>0$; both character weights vanish when $D_s=0$. 
The character $\chi$ labels the logical eigenstate.

Both states are block diagonal in the syndrome record, even though the conditional vectors $|\Psi_s\rangle$ need not be mutually orthogonal. Each nonzero joint-state block has a single eigenvalue $D_s$, while the output block has eigenvalues $T_\chi^s/2$. Their trace moments are therefore
\begin{equation}
 \operatorname{tr}\rho_{SLR}^n=\sum_sD_s^n,
 \quad \operatorname{tr}\rho_{SL}^n=\frac{1}{2^n}\sum_s\sum_{\chi=\pm1}(T_\chi^s)^n.
 \label{eq:app_renyi_output_moment}
\end{equation}
The factor $2^{-n}$ comes from the maximally mixed input, equivalently the Bell-pair normalization. Since $T_+^s+T_-^s=2D_s$, both moments are normalized at $n=1$:
\begin{equation}
 \operatorname{tr}\rho_{SLR}=\sum_sD_s=1,
 \quad \operatorname{tr}\rho_{SL}=\frac12\sum_{s,\chi}T_\chi^s=1.
 \label{eq:app_renyi_trace_normalization}
\end{equation}
These are exact spectral identities for real $n>0$. For positive integer $n$, orthogonality of the syndrome projectors ensures that all copies in a surviving replica term carry the same $s$, so the syndrome sum lies outside the $n$th power.

For $n>0$, $n\ne1$, let $S_n(\rho)=(1-n)^{-1}\ln\operatorname{tr}\rho^n$. We define the finite-$n$ replica family as the difference of the output and joint R\'enyi entropies,
\begin{equation}
 I_c^{(n)}\equiv S_n(\rho_{SL})-S_n(\rho_{SLR})
 =\frac{1}{1-n}
 \ln\frac{\tfrac{1}{2^n}\sum_{s,\chi}(T_\chi^s)^n}{\sum_sD_s^n}.
 \label{eq:app_renyi_entropy_difference}
\end{equation}
For $D_s>0$, the conditional output $\rho_{SL|s}=K_sK_s^\dagger/(2D_s)$ has eigenvalues $\lambda_{\chi|s}=T_\chi^s/(2D_s)=(1+\chi m_s)/2$, so equivalently
\begin{equation}
 I_c^{(n)}
 =\frac{1}{1-n}
 \ln\frac{\sum_{s}D_s^n\sum_\chi\lambda_{\chi|s}^n}
 {\sum_{s}D_s^n}.
 \label{eq:app_renyi_conditional_moments}
\end{equation}
The normalized syndrome weights are proportional to $D_s^n$, and the logarithm is taken after summing the conditional trace moments.

The von Neumann limit recovers the Born average. We use $0\ln0=0$ throughout. 
Thus
\begin{equation}
\begin{aligned}
 I_c=&\lim_{n\to1}I_c^{(n)}
 =-\left.\partial_n\left[\operatorname{tr}\rho_{SL}^n-\operatorname{tr}\rho_{SLR}^n\right]\right|_{n=1}\\
 =&\ln2-\frac12\sum_{s,\chi}T_\chi^s\ln T_\chi^s
 +\sum_sD_s\ln D_s.
 \end{aligned}
 \label{eq:app_renyi_von_neumann_limit}
\end{equation}
Substituting $T_\chi^s=D_s(1+\chi m_s)$ on the nonzero branches cancels the classical syndrome entropy, leaving
\begin{equation}
 I_c=\sum_{s}D_s\,h_2\!\left(\frac{1+m_s}{2}\right),
 \label{eq:app_renyi_conditional_entropy}
\end{equation}
where $h_2(x)=-x\ln x-(1-x)\ln(1-x)$.
The coherent information is therefore the Born-weighted entanglement entropy of the pure conditional logical-reference states.

\subsection{Syndrome record}\label{app:syndrome_record}
The cancellation of classical syndrome entropy in Eq.~\eqref{eq:app_renyi_conditional_entropy} relies on retaining the syndrome record. 
If the classical syndrome register is discarded after applying the reference correction $R_s$, the logical channel reduces to the unconditioned trace-preserving mixture
\begin{equation}
\begin{aligned}
    \overline{\mathcal{M}}_\theta(\rho_L) &\equiv \operatorname{tr}_S \mathcal{M}_\theta(\rho_L) = \sum_s K_s \rho_L K_s^\dagger \\
    &= (1 - p)\rho_L + p Z_L \rho_L Z_L - i\kappa [\rho_L, Z_L],
\end{aligned}
\label{eq:app_unconditioned_channel}
\end{equation}
where $p \equiv \sum_s |A_s^1|^2$ is the total flipped weight, and $\kappa \equiv \operatorname{Im}\sum_s (A_s^0)^* A_s^1$ governs coherent phase precession ($\operatorname{Re}\sum_s (A_s^0)^* A_s^1 = 0$ by completeness $\sum_s K_s^\dagger K_s = I_L$). 
Thus, $\overline{\mathcal{M}}_\theta$ preserves diagonal populations while suppressing off-diagonal coherence:
\begin{equation}
\rho_{00} \to \rho_{00}, \quad \rho_{11} \to \rho_{11}, \quad \rho_{01} \to \gamma_\theta \rho_{01}, 
\end{equation}
where $\gamma_\theta \equiv 1 - 2p + 2i\kappa$.
For even $d$, phase pinning ensures that $(A_s^0)^* A_s^1$ is real on every supported branch, so $\kappa = 0$; discarding the syndrome turns $\overline{\mathcal{M}}_\theta$ into an exact Pauli dephasing channel. 
For odd $d$, where each branch is a unitary rotation $e^{i\theta_s Z_L}$, averaging over the unknown rotation angles likewise induces dephasing ($|\gamma_\theta| < 1$).

For the maximally mixed logical input purified by $|\Phi\rangle_{LR}$, discarding the syndrome flags leaves the joint state $\bar{\rho}_{LR} = \sum_s |\Psi_s\rangle\langle\Psi_s|$, which incoherently superposes the conditional states without orthogonal syndrome tags. 
While the logical output remains fully mixed ($\bar{\rho}_L = I_L/2$), $\bar{\rho}_{LR}$ has non-zero eigenvalues $(1 \pm |\gamma_\theta|)/2$. 
The resulting coherent information is
\begin{equation}
\bar{I}_c = \ln 2 - h_2\!\left(\frac{1 + |\gamma_\theta|}{2}\right) \le I_c,
\label{eq:app_unflagged_ic}
\end{equation}
where the inequality follows from quantum data processing~\cite{SchumacherNielsen1996}. 

This degradation can be severe: $\bar{I}_c < I_c$ can occur even when every conditional branch is individually unitary ($I_c = \ln 2$). 
An explicit classical register $S$ is strictly required whenever the reference correction $R_s$ has already mapped all conditional states back into the single code space $\mathcal{H}_c$. 
Alternatively, if $R_s$ is deferred, the mutually orthogonal stabilizer eigenspaces $\Pi_s \mathcal{H}_{\mathrm{phys}}$ themselves act as physical syndrome flags, retaining full coherent information before correction.

\subsection{Physical double layers, replicas, and logical interference}

For positive integer $n$, the trace moments specify the amplitudes to be replicated. The joint-state weight expands as
\begin{equation}
 D_s^n=|\mathcal N_\theta|^{2n}
 \sum_{\boldsymbol{\mu}\in\{0,1\}}
 \prod_{a=1}^n\mathcal Z_{\mu_a}^s(\mathcal Z_{\mu_a}^s)^*,
 \label{eq:app_renyi_joint_expansion}
\end{equation}
where $\boldsymbol{\mu}=\{\mu_1,\ldots,\mu_n\}$ and each $\mu_a$ takes $0$ or $1$.
The output weight instead contains independent forward and backward sector sums:
\begin{equation}
\begin{aligned}
 (T_\chi^s)^n
 =&|\mathcal N_\theta|^{2n}
 (\mathcal Z_0^s+\chi\mathcal Z_1^s)^n
 \bigl[(\mathcal Z_0^s)^*+\chi(\mathcal Z_1^s)^*\bigr]^n\\
 =&|\mathcal N_\theta|^{2n}
 \sum_{\boldsymbol{\mu},\boldsymbol{\nu}}
 \chi^{\sum_a(\mu_a+\nu_a)}
 \prod_{a=1}^n\mathcal Z_{\mu_a}^s(\mathcal Z_{\nu_a}^s)^*.
 \label{eq:app_renyi_output_expansion}
 \end{aligned}
\end{equation}
A physical double layer consists of one forward amplitude at $\beta_p$ and one conjugate amplitude at $\beta_p^*$. Their sector labels match in $D_s^n$ but need not match in $(T_\chi^s)^n$. Every term contains $n$ such pairs sharing the same syndrome $s$, hence $2n$ Majorana layer flavors.

The syndrome weight is already included in this count. 
For $D_s>0$, the unnormalized output block is $D_s\rho_{SL|s}$, so its $n$th power carries $D_s^n$:
\begin{equation}
 D_s^n\operatorname{tr}\rho_{SL|s}^n
 =D_s^n\sum_\chi\left(\frac{T_\chi^s}{2D_s}\right)^n
 =\frac{1}{2^n}\sum_\chi(T_\chi^s)^n.
 \label{eq:app_renyi_no_extra_born}
\end{equation}
Thus the trace moment contains $n$ physical double layers; an additional factor $D_s$ would change the observable.

A single character $\chi$ is shared by all $n$ pairs. Its sum in Eq.~\eqref{eq:app_renyi_output_moment} imposes a parity constraint on the summed sector labels:
\begin{equation}
 \sum_{\chi=\pm1}\chi^{\sum_a(\mu_a+\nu_a)}
 =\begin{cases}
 2,&\sum_a(\mu_a+\nu_a)\ \text{even},\\
 0,&\sum_a(\mu_a+\nu_a)\ \text{odd}.
 \end{cases}
 \label{eq:app_renyi_character_constraint}
\end{equation}
At $n=1$, this enforces $\mu_1=\nu_1$. For $n\ge2$, it permits unmatched pairs and hence cross-sector interference; for example, $\boldsymbol\mu=(0,0)$ and $\boldsymbol\nu=(1,1)$ survive at $n=2$. Correspondingly, $T_\pm^s=D_s(1\pm m_s)$ gives
\begin{equation}
 (T_{+1}^s)^2+(T_{-1}^s)^2=2D_s^2(1+m_s^2).
 \label{eq:app_renyi_interference_check}
\end{equation}
At fixed sector assignments, these expansions require a shared syndrome sum. Expressing it as a local sign average requires the syndrome-to-bond-sign mapping, including its gauge multiplicities, constraints, and boundary conventions, as treated in Appendix~\ref{app:random_average}. The external logical defects remain fixed during that average.

In the original transfer representation, sector assignments differ only on the specified defect supports. A representative fixed-sector product such as $[\mathcal Z_0^s(\mathcal Z_0^s)^*]^n$ therefore identifies the common local bulk operator structure. In the clean frame of Appendix~\ref{app:swap_basis}, the accumulated frame signs also carry sector dependence, which can extend along trajectories beyond the original defect supports. The full moments retain this transformed data together with the sector sums, character weights, and boundary contractions in Eqs.~\eqref{eq:app_renyi_joint_expansion} and \eqref{eq:app_renyi_output_expansion}.

\section{Clean Majorana SWAP propagation}\label{app:swap_intro}

At $\theta=\pi/4$, the reduced Majorana transfer factors of Sec.~\ref{subsec:maj_network} become signed exchanges. For one forward amplitude layer, we derive the clean bulk routing and its long-wavelength transport kernel, providing the starting point for Appendices~\ref{app:swap_basis} and~\ref{app:eft_derivations}.

\subsection{Elementary Majorana rotation and signed exchange}

For Hermitian Majoranas satisfying $\{\gamma_i,\gamma_j\}=2\delta_{ij}$, define
\begin{equation}
 R_{ij}(\alpha)=e^{\alpha\gamma_i\gamma_j},
 \qquad i\ne j,\quad \alpha\in\mathbb R,
 \label{eq:app_swap_rotation_definition}
\end{equation}
with active conjugation $\mathcal O\mapsto R_{ij}\mathcal O R_{ij}^{-1}$.
Since $(\gamma_i\gamma_j)^\dagger=-\gamma_i\gamma_j$ and $(\gamma_i\gamma_j)^2=-1$, this operator is unitary and equals $\cos\alpha+\gamma_i\gamma_j\sin\alpha$. Using $[\gamma_i\gamma_j,\gamma_i]=-2\gamma_j$ and $[\gamma_i\gamma_j,\gamma_j]=2\gamma_i$ gives
\begin{equation}
\begin{aligned}
        R_{ij}(\alpha)\gamma_iR_{ij}^{-1}(\alpha)
        =&\gamma_i\cos2\alpha-\gamma_j\sin2\alpha,\\
    R_{ij}(\alpha)\gamma_jR_{ij}^{-1}(\alpha)
        =&\gamma_i\sin2\alpha+\gamma_j\cos2\alpha.
\label{eq:app_swap_rotation}
\end{aligned}
\end{equation}
The rotation angle is $2\alpha$, so $\alpha=\varepsilon\pi/4$, with $\varepsilon=\pm1$, gives a signed quarter-turn:
\begin{equation}
    \gamma_i\longmapsto-\varepsilon\gamma_j,
    \qquad
    \gamma_j\longmapsto\varepsilon\gamma_i.
\label{eq:app_signed_swap}
\end{equation}
The opposite signs make the exchange a proper orthogonal rotation. Here \emph{SWAP} denotes this exchange of auxiliary Majorana operators, rather than a two-qubit gate applied to the code.

The reduced endpoint nodes are therefore unitary. Scalar transfer factors cancel in conjugation but remain part of the partition-function amplitudes.

\subsection{Clean row and ballistic propagation}

Suppress the row label and denote the two Majorana operators in cell $x$ by
\begin{equation}
 \gamma_{L,x}=\gamma_{2x-1},\qquad \gamma_{R,x}=\gamma_{2x}.
 \label{eq:app_swap_cell_modes}
\end{equation}
Setting the local bond signs to $+1$ in Eq.~\eqref{eq:swap_node} defines the clean forward-layer reference at $\beta_R=0$:
\begin{equation}
    \begin{gathered}
            h_x^{(0)}=\exp\!\left[\frac{\pi}{4}\gamma_{R,x}\gamma_{L,x+1}\right],\\
        v_x^{(0)}=\exp\!\left[-\frac{\pi}{4}\gamma_{L,x}\gamma_{R,x}\right],\\
        \hat T_0=\hat V_0\hat H_0
=\left(\prod_xv_x^{(0)}\right)\left(\prod_xh_x^{(0)}\right).
    \label{eq:app_clean_swap_row}
    \end{gathered}
\end{equation}
Closing-bond parity and twist factors are carried separately. Appendix~\ref{app:swap_basis} constructs the frame relating syndrome-dependent signed routing to this clean reference. With active conjugation $\mathcal O\mapsto\hat T_0\mathcal O\hat T_0^{-1}$, the horizontal layer acts first. Applying Eq.~\eqref{eq:app_signed_swap} twice gives
\begin{equation}
 \begin{gathered}
    \gamma_{L,x}\xrightarrow{\ \hat H_0\ }\gamma_{R,x-1}
         \xrightarrow{\ \hat V_0\ }-\gamma_{L,x-1},\\
    \gamma_{R,x}\xrightarrow{\ \hat H_0\ }-\gamma_{L,x+1}
         \xrightarrow{\ \hat V_0\ }-\gamma_{R,x+1}.
 \end{gathered}
\label{eq:app_swap_routes}
\end{equation}
Each complete row routes an $L$ mode one cell to the left and an $R$ mode one cell to the right, with a common minus sign. Each input follows one outgoing path without splitting or backscattering in the bulk: this is the ballistic SWAP endpoint.

\subsection{Continuum transport and clean kinetic action}
To track propagation, expand an operator linear in the Majoranas in the fixed row basis,
\begin{equation}
 \begin{aligned}
 \mathcal O_y=\sum_x\left[f_{L,x}(y)\gamma_{L,x}+f_{R,x}(y)\gamma_{R,x}\right],
 \end{aligned}
 \label{eq:app_swap_operator_profile}
\end{equation}
and $\mathcal O_{y+1}=\hat T_0\mathcal O_y\hat T_0^{-1}$.
Relabeling the sums in Eq.~\eqref{eq:app_swap_routes} gives the exact profile recurrences
\begin{equation}
\begin{aligned}
     f_{L,x}(y+1)=&-f_{L,x+1}(y),\\
 f_{R,x}(y+1)=&-f_{R,x-1}(y).
 \label{eq:app_swap_profile}
\end{aligned}
\end{equation}
The shifted arguments identify the input feeding position $x$, so these profiles move in the directions shown above. Write $f_{L/R,x}(y)=(-1)^y\widetilde f_{L/R,x}(y)$ to remove the common row sign. This deterministic staggering is distinct from the syndrome-dependent frame of Appendix~\ref{app:swap_basis}, which retains the clean minus sign. The staggering factors at the endpoints remain in the boundary data.

Let $a_x$ and $a_y$ denote the cell and row spacings, with $X=a_xx$ and $Y=a_yy$. For smooth envelopes $\widetilde f_{L/R}(X,Y)$, the leading gradient expansion gives
\begin{equation}
 (\partial_Y-v\partial_X)\widetilde f_L=0,\qquad
 (\partial_Y+v\partial_X)\widetilde f_R=0,
 \label{eq:app_swap_continuum}
\end{equation}
where $v=a_x/a_y$.
Here $Y$ measures distance along the transfer direction, not the physical time of the coherent rotation.
These differential operators define a clean kinetic functional for Majorana Grassmann fields $\chi=(\chi_R,\chi_L)^T$, in the ordering used in Appendix~\ref{app:eft_derivations}:
\begin{equation}
\begin{aligned}
     S_{\mathrm{step},0}
 =&\frac12\int dX\,dY\,\chi^T\mathcal D_{\mathrm{step},0}\chi,\\
 \mathcal D_{\mathrm{step},0}
 =&\operatorname{diag}(\partial_Y+v\partial_X,\partial_Y-v\partial_X).
 \label{eq:app_swap_step_action}
\end{aligned}
\end{equation}
The derivatives are antisymmetric under integration by parts, with boundary terms treated separately. The factor $1/2$ fixes our quadratic Grassmann convention. This functional encodes the clean transport equations.

For the Euclidean convention used in Appendix~\ref{app:eft_derivations}, continue $Y=-i\tau_E$ and normalize the temporal derivative:
\begin{equation}
\begin{aligned}
    S_{0,E}=&\frac12\int dX\,d\tau_E\,\chi_E^T\mathcal D_{0,E}\chi_E,\\
 \mathcal D_{0,E}=&\operatorname{diag}(\partial_{\tau_E}-iv\partial_X,\partial_{\tau_E}+iv\partial_X),
 \label{eq:app_swap_euclidean_action}
\end{aligned}
\end{equation}
where $\mathcal D_{0,E}=-i\mathcal D_{\mathrm{step},0}\big|_{\partial_Y=i\partial_{\tau_E}}$.
This continuation specifies a continuum kinetic convention; it does not identify the unitary endpoint node with a Euclidean step $e^{-a_EH}$ for a Hermitian Hamiltonian. For the $2n$ aligned layer flavors of Sec.~\ref{subsec:replica_symmetry}, the clean action is $\sum_{A=1}^{2n}S_{0,E}[\chi_E^A]$. Appendix~\ref{app:eft_derivations} states the continuum matching assumptions and develops the interacting theory beyond this ballistic starting point.

Boundary contractions and logical twists remain necessary for decoding observables; clean bulk routing alone does not determine a threshold.

\section{Clean-SWAP frame and detuning operators}\label{app:swap_basis}
We construct the clean frame for a fixed syndrome and logical sector, using the active-conjugation convention of Appendix~\ref{app:swap_intro}. 
This construction fixes the frame in which we express the detuning operators and the backward-layer alignment. 
Scalar transfer normalizations and boundary or logical defect data are retained separately.

\subsection{Signed routing and the moving frame}
\label{app:signed_routing_frame}
For the cylinder geometry in Sec.~\ref{subsec:maj_network}, spatial indices are understood modulo $L_x$, with fixed closing-bond factors carried separately in both the signed and clean reference rows. 
Label the successive cuts by $y=0,\ldots,N_y$, where $N_y=L_y-1$ counts the complete row steps in the transfer matrix representation of the partition function $\mathcal Z[\eta,\beta_p]
=\mathcal A[\eta,\beta_p]
\langle\psi_f|\hat H_{L_y}
\hat T_{L_y-1}\cdots\hat T_1|\psi_i\rangle$.
Row and bond labels below use this shifted origin.  
The cut label $y$ tracks successive representations of the same auxiliary Majorana chain.

Write $\gamma_{L,x,y}=\gamma_{2x-1,y}$ and $\gamma_{R,x,y}=\gamma_{2x,y}$. 
At $\beta_R=0$, the reduced row at $\theta=\pi/4$ is $\hat T_y^{(0)}=\hat V_y^{(0)}\hat H_y^{(0)}$, with the signed nodes of Eq.~\eqref{eq:swap_node}. Suppressing the row label on the Majoranas during conjugation, the bulk routes are
\begin{equation}
\begin{aligned}
     &\gamma_{L,x}\xrightarrow{\ \hat H_y^{(0)}\ }
 \eta^{\mathrm h}_{x-1,y}\gamma_{R,x-1}
 \xrightarrow{\ \hat V_y^{(0)}\ }
 -\eta^{\mathrm h}_{x-1,y}\eta^{\mathrm v}_{x-1,y}\gamma_{L,x-1},\\
 &\gamma_{R,x}\xrightarrow{\ \hat H_y^{(0)}\ }
 -\eta^{\mathrm h}_{x,y}\gamma_{L,x+1}
 \xrightarrow{\ \hat V_y^{(0)}\ }
 -\eta^{\mathrm h}_{x,y}\eta^{\mathrm v}_{x+1,y}\gamma_{R,x+1}.
\end{aligned}
\label{eq:app_signed_row_routes}
\end{equation}
The bond signs affect the accumulated sign but leave the one-cell shift from cut $y$ to cut $y+1$ unchanged.

Define the clean frame with identity initial signs by
\begin{equation}
 \Gamma_{2x-1,y}=u^L_{x,y}\gamma_{L,x,y},
 \quad \Gamma_{2x,y}=u^R_{x,y}\gamma_{R,x,y},
 \label{eq:app_moving_frame_definition}
\end{equation}
where $u^{L/R}_{x,y}\in\{\pm1\}$ with the initial condition $u^{L/R}_{x,0}=1$.
This is the basis  transformation in Eq.~\eqref{eq:clean_swap_frame} with $u_{2x-1,y}=u^L_{x,y}$ and $u_{2x,y}=u^R_{x,y}$. 
Requiring the transformed route to retain only the clean minus sign fixes the recurrences
\begin{equation}
\begin{aligned}
 u^L_{x-1,y+1}&=\eta^{\mathrm h}_{x-1,y}\eta^{\mathrm v}_{x-1,y}u^L_{x,y},\\
 u^R_{x+1,y+1}&=\eta^{\mathrm h}_{x,y}\eta^{\mathrm v}_{x+1,y}u^R_{x,y}.
 \label{eq:app_moving_frame_recursion}
 \end{aligned}
\end{equation}
For example, the left route has coefficient
$-\eta^{\mathrm h}_{x-1,y}\eta^{\mathrm v}_{x-1,y}u^L_{x,y}u^L_{x-1,y+1}=-1$.
Thus each $u$ accumulates the bond signs along its incoming trajectory relative to the clean route, giving
\begin{equation}
 \Gamma_{2x-1,y}\longmapsto-\Gamma_{2x-3,y+1},
 \quad \Gamma_{2x,y}\longmapsto-\Gamma_{2x+2,y+1}.
 \label{eq:app_moving_frame_clean_routes}
\end{equation}
The clean minus signs of Appendix~\ref{app:swap_intro} are retained. 
With $u^{L/R}_{x,0}=1$ fixed, the recurrences uniquely determine every later frame from the trajectory signs, including the final frame at $y=N_y$. 

Multiplying the recurrences over the periodic row gives
\begin{equation}
 \frac{\prod_xu^L_{x,y+1}u^R_{x,y+1}}{\prod_xu^L_{x,y}u^R_{x,y}}
 =\left(\prod_x\eta^{\mathrm h}_{x,y}\eta^{\mathrm v}_{x,y}\right)
  \left(\prod_x\eta^{\mathrm h}_{x,y}\eta^{\mathrm v}_{x+1,y}\right)=1.
\label{eq:app_moving_frame_determinant}
\end{equation}
Each bond sign occurs twice, so $\prod_j u_{j,y}=+1$ on every cut.
Thus the frame flips an even number of Majorana signs and can be implemented on auxiliary row states by a parity-preserving Gaussian unitary $G_y$, generated by Majorana bilinears. 
In the fixed row basis, it satisfies
\begin{equation}
 G_y\gamma_jG_y^{-1}=u_{j,y}\gamma_j,
 \qquad G_0=I.
 \label{eq:app_moving_frame_lift}
\end{equation}
This implementation is called a \emph{Spin lift}: it lifts the determinant-one Majorana basis transformation to a unitary acting on the auxiliary row Hilbert space. 
The two lifts $G_y$ and $-G_y$ act identically on all Majoranas.
Using $G_y$ to express the row states in the clean frame also transforms the transfer operators for $\ket{\psi_{y+1}}=\hat{T}_y\ket{\psi_y}$:
\begin{equation}
|\psi_y^{\mathrm{cl}}\rangle=G_y|\psi_y\rangle,
 \qquad \hat T_y^{\mathrm{cl}}=G_{y+1}\hat T_yG_y^{-1},
\label{eq:app_moving_frame_state_coordinates}
\end{equation}
where the superscript ``$\mathrm{cl}$" denotes the clean frame.
The factors $G_{y+1}$ and $G_y^{-1}$, acting on the row modes, account for the frames on the output and input cuts, respectively. 
At the SWAP point, the recurrences imply
\begin{equation}
 \hat T_y^{(0)}\doteq G_{y+1}^{-1}\hat T_0G_y.
 \label{eq:app_moving_frame_row}
\end{equation}
Here $\hat T_0$ always denotes the clean reference row of Eq.~\eqref{eq:app_clean_swap_row}, with the same fixed closing-bond convention; $\hat T_y^{(0)}$ denotes the signed SWAP row, including the first row $\hat T_0^{(0)}$. 
The symbol $\doteq$ denotes equality up to an overall Spin-lift sign. 

Over $N_y$ steps, the intermediate frames cancel:
\begin{equation}
 \hat T_{N_y-1}^{(0)}\cdots\hat T_0^{(0)}
 \doteq G_{N_y}^{-1}(\hat T_0)^{N_y}G_0.
 \label{eq:app_moving_frame_product}
\end{equation}
Including the terminal horizontal layer of $
\langle\psi_f|\hat H_{L_y}
\hat T_{L_y-1}\cdots\hat T_1|\psi_i\rangle$, denoted by $\hat H_{\mathrm{end}}^{(0)}$ ($\hat H_{L_y}^{(0)}$ in the original row labels), gives the clean-frame boundary states
\begin{equation}
 |\psi_i^{\mathrm{cl}}\rangle
 =G_0|\psi_i\rangle=|\psi_i\rangle,\quad
 \langle\psi_f^{\mathrm{cl}}|
 =\langle\psi_f|\hat H_{\mathrm{end}}^{(0)}G_{N_y}^{-1}.
 \label{eq:app_moving_frame_boundaries}
\end{equation}
The contraction $\langle\psi_f^{\mathrm{cl}}|(\hat T_0)^{N_y}|\psi_i^{\mathrm{cl}}\rangle$, multiplied by the original scalar factors and accumulated lift sign, reconstructs the amplitude at $\theta=\pi/4$. 

The clean frame rewrites the original partition function using clean bulk propagation, while its syndrome and logical-sector dependence remains in the transformed boundary data and scalar factors.

\subsection{Exact detuning factorization and leading vertices}
\label{app:swap_perturbation}
Keep the syndrome and logical sector fixed and use the forward-layer branch of Eq.~\eqref{eq:beta_parameter}. The frame signs $u_{j,y}$ and their lifts $G_y$ are constructed from the signed-SWAP reference and held fixed as $\beta_R$ varies.

The reduced layers separate exactly into their signed-SWAP factors and residual operators:
\begin{equation}
\begin{aligned}
 \hat H_y&=\hat H_y^{(0)}\delta\hat H_y,
 &\delta\hat H_y&=\exp\!\left[i\beta_R\sum_x\eta^{\mathrm h}_{x,y}\gamma_{2x,y}\gamma_{2x+1,y}\right],\\
 \hat V_y&=\hat V_y^{(0)}\delta\hat V_y,
 &\delta\hat V_y&=\exp\!\left[\phi(\beta_R)\sum_x\gamma_{2x-1,y}\gamma_{2x,y}\right],
\end{aligned}
\label{eq:app_detuning_layer_split}
\end{equation}
where $\phi(\beta_R)=\frac\pi4-\theta=\arctan(\tanh\beta_R) =\beta_R-\frac23\beta_R^3+O(\beta_R^5)$.
Both residual operators equal the identity at $\beta_R=0$. 
The vertical residual has no explicit $\eta^{\mathrm v}_{x,y}$ factor; the vertical bond sign remains in $\hat V_y^{(0)}$ and in the trajectory frame of the preceding subsection.

With $\hat T_y=\hat V_y\hat H_y$ and $\hat T_y^{(0)}=\hat V_y^{(0)}\hat H_y^{(0)}$, moving the vertical residual to the right of the endpoint row gives
\begin{equation}
 \hat T_y=\hat T_y^{(0)}
 \left[(\hat H_y^{(0)})^{-1}\delta\hat V_y\hat H_y^{(0)}\right]\delta\hat H_y
 =\hat T_y^{(0)}e^{A_{\mathrm v,y}}e^{A_{\mathrm h,y}}.
 \label{eq:app_detuning_row_split}
\end{equation}
The conjugation involves the entire horizontal layer. In the bulk, its inverse sends the two Majoranas of a vertical node through different neighboring SWAP nodes:
\begin{equation}
\begin{aligned}
 (\hat H_y^{(0)})^{-1}\gamma_{2x-1,y}\hat H_y^{(0)}
 &=-\eta^{\mathrm h}_{x-1,y}\gamma_{2x-2,y},\\
 \quad
 (\hat H_y^{(0)})^{-1}\gamma_{2x,y}\hat H_y^{(0)}
 &=\eta^{\mathrm h}_{x,y}\gamma_{2x+1,y}.
\label{eq:app_detuning_inverse_horizontal_routes}
 \end{aligned}
\end{equation}
Multiplying these expressions fixes the support and minus sign of the transported vertical term. Thus the exact residual exponents are
\begin{equation}
\begin{aligned}
 A_{\mathrm h,y}&=i\beta_R\sum_x\eta^{\mathrm h}_{x,y}\gamma_{2x,y}\gamma_{2x+1,y},
 \\
 A_{\mathrm v,y}&=-\phi(\beta_R)\sum_x
 \eta^{\mathrm h}_{x-1,y}\eta^{\mathrm h}_{x,y}\gamma_{2x-2,y}\gamma_{2x+1,y}.
 \label{eq:app_detuning_raw_exponents}
 \end{aligned}
\end{equation}
These are reduced bulk identities; crossing a closing bond requires the same parity and twist factors as in the original transfer operator.

Both residual exponents act on the input slice $y$. 
Substituting $\gamma_{j,y}=u_{j,y}\Gamma_{j,y}$ from Eq.~\eqref{eq:app_moving_frame_definition} into Eq.~\eqref{eq:app_detuning_raw_exponents} gives their expressions in the clean moving frame. 
Define the clean-frame Majorana bilinears
\begin{equation}
\begin{aligned}
 B^{\mathrm h}_{x,y}=i\Gamma_{2x,y}\Gamma_{2x+1,y},\quad
 B^{\mathrm v}_{x,y}&=i\Gamma_{2x-2,y}\Gamma_{2x+1,y},
\end{aligned}
\label{eq:app_detuning_clean_vertices}
\end{equation}
and the signs $\xi^{\mathrm h,\mathrm v}$ which contain the accumulated trajectory signs 
\begin{equation}
    \begin{aligned}
\xi^{\mathrm h}_{x,y}&=\eta^{\mathrm h}_{x,y}u^R_{x,y}u^L_{x+1,y},\\
 \xi^{\mathrm v}_{x,y}&=\eta^{\mathrm h}_{x-1,y}\eta^{\mathrm h}_{x,y}u^R_{x-1,y}u^L_{x+1,y}.
    \end{aligned}
\end{equation}
These definitions agree with Eqs.~\eqref{eq:detuning_frame_signs} and \eqref{eq:detuning_bilinears}. 
The same residual exponents now read $A_{\mathrm h,y}=\beta_R\sum_x\xi^{\mathrm h}_{x,y}B^{\mathrm h}_{x,y}$ and $A_{\mathrm v,y}=i\phi(\beta_R)\sum_x\xi^{\mathrm v}_{x,y}B^{\mathrm v}_{x,y}$, so the exact ordered detuning operator is
\begin{equation}
\begin{gathered}
 e^{\mathcal P_y}=e^{A_{\mathrm v,y}}e^{A_{\mathrm h,y}}\\
=\exp\!\left[i\phi(\beta_R)\sum_x\xi^{\mathrm v}_{x,y}B^{\mathrm v}_{x,y}\right]
  \exp\!\left[\beta_R\sum_x\xi^{\mathrm h}_{x,y}B^{\mathrm h}_{x,y}\right].
 \label{eq:app_detuning_clean_product}
 \end{gathered}
\end{equation}
For real detuning, $A_{\mathrm h,y}^{\dagger}=A_{\mathrm h,y}$ and $A_{\mathrm v,y}^{\dagger}=-A_{\mathrm v,y}$. 
The horizontal residual is therefore generally nonunitary, whereas the vertical residual is unitary. 
The unitary frame change preserves this distinction; the detuned transfer row is generally nonunitary.

Near $\beta_R=0$, choose the logarithm continuous from $\mathcal P_y=0$. At fixed finite size, the Baker--Campbell--Hausdorff expansion and $\phi(\beta_R)=\beta_R+O(\beta_R^3)$ give
\begin{equation}
\begin{gathered}
 \mathcal P_y=\mathcal P_y^{(1)}+O(\beta_R^2),\\
 \mathcal P_y^{(1)}
 =\beta_R\sum_x\xi^{\mathrm h}_{x,y}B^{\mathrm h}_{x,y}
 +i\beta_R\sum_x\xi^{\mathrm v}_{x,y}B^{\mathrm v}_{x,y}.
 \label{eq:app_detuning_leading_vertices}
 \end{gathered}
\end{equation}
More precisely, $\mathcal P_y-\mathcal P_y^{(1)}=\tfrac12[A_{\mathrm v,y},A_{\mathrm h,y}]+O(\beta_R^3)$.
Commutators of Majorana bilinears remain bilinear, so the reduced transfer operator remains Gaussian at every order within each fixed fermion-parity sector. 

The substitution $\gamma_{j,y}=u_{j,y}\Gamma_{j,y}$ and the action of $G_y$ on auxiliary row states describe the same frame change: $\Gamma_{j,y}$ are the clean-frame Majoranas, while $G_y$ expresses the row states in that frame. 
For row-state products, we retain the symbols $\Gamma$, $B$, and $\mathcal P_y$ for their matrices in this common clean-frame representation. 
By Eq.~\eqref{eq:app_detuning_row_split}, $(\hat T_y^{(0)})^{-1}\hat T_y$ is the ordered residual factor in the original row coordinates. 
Its action on clean-frame row states is
\begin{equation}
 e^{\mathcal P_y}
 =G_y(\hat T_y^{(0)})^{-1}\hat T_yG_y^{-1}.
 \label{eq:app_detuning_frame_coordinates}
\end{equation}
This conjugation implements the substitution already used in Eq.~\eqref{eq:app_detuning_clean_product}, with the trajectory signs retained in $\xi^{\mathrm h,\mathrm v}$. 
Combining this identity with Eq.~\eqref{eq:app_moving_frame_row} gives the full detuned row:
\begin{equation}
 \hat T_y^{\mathrm{cl}}=G_{y+1}\hat T_yG_y^{-1}
 \doteq\hat T_0e^{\mathcal P_y}.
 \label{eq:app_detuning_full_clean_row}
\end{equation}
The subsequent products and sign averages use this representation. 

\subsection{Backward-layer alignment in the clean-SWAP frame}
\label{app:backward_clean_frame}
The syndrome-dependent signs $u_{j,y}$ of Appendix~\ref{app:signed_routing_frame} absorb the accumulated trajectory signs into the clean-frame Majoranas $\Gamma_{j,y}$. Here we introduce deterministic signs $\Omega_{j,y}$ only for the backward layer, aligning its leading detuning vertices with the forward layer while preserving clean routing. We suppress the replica index and use the clean-coordinate convention of Appendix~\ref{app:swap_perturbation}.

In the chosen Jordan--Wigner representation, odd-indexed Majorana matrices are real and even-indexed ones imaginary; the real signs $u_{j,y}$ preserve this property. Each bilinear in Eq.~\eqref{eq:app_detuning_clean_vertices} contains one index of each type, so
\begin{equation}
 \bigl(i\Gamma_{2p,y}^{B}\Gamma_{2q-1,y}^{B}\bigr)^*
 =(-i)(-\Gamma_{2p,y}^{B})\Gamma_{2q-1,y}^{B}
 =i\Gamma_{2p,y}^{B}\Gamma_{2q-1,y}^{B}.
 \label{eq:app_backward_real_bilinears}
\end{equation}
Both bilinears are therefore real, and conjugation reverses only the explicit $i$ in the leading vertical coefficient:
\begin{equation}
 \mathcal P_y^{(1);B}
 =\beta_R\sum_x\xi^{\mathrm h}_{x,y}B_{x,y}^{\mathrm h;B}
 -i\beta_R\sum_x\xi^{\mathrm v}_{x,y}B_{x,y}^{\mathrm v;B}.
 \label{eq:app_backward_raw_generator}
\end{equation}
Both layers retain the same syndrome signs $\xi^{\mathrm h,\mathrm v}$.

Conjugation preserves the layer order, $(\hat V_y^{(0)}\hat H_y^{(0)})^*=(\hat V_y^{(0)})^*(\hat H_y^{(0)})^*$, while reversing both endpoint SWAP angles. Each complete-row trajectory traverses one horizontal and one vertical node, so these two sign reversals cancel. The backward layer thus has the same clean bulk routes:
\begin{equation}
 \Gamma_{L,x,y}^{B}\longmapsto-\Gamma_{L,x-1,y+1}^{B},\quad
 \Gamma_{R,x,y}^{B}\longmapsto-\Gamma_{R,x+1,y+1}^{B},
 \label{eq:app_backward_clean_routes}
\end{equation}
where $\Gamma_{L,x,y}^{B}=\Gamma_{2x-1,y}^{B}$ and $\Gamma_{R,x,y}^{B}=\Gamma_{2x,y}^{B}$. Scalar factors and operator-lift phases remain separately tracked.

The spatial staggering $\omega_{2x}=\omega_{2x+1}=(-1)^x$ preserves the horizontal bilinear and reverses the transported vertical bilinear, as needed to align the coefficients. It also reverses each clean routing sign because $\omega_j\omega_{j\pm2}=-1$. Including $(-1)^y$ supplies a second minus sign between successive cuts, restoring the routing without changing any same-row bilinear. Thus Eq.~\eqref{eq:backward_flavor_relabeling} uses
\begin{equation}
 \widetilde\Gamma_{j,y}^{B}=\Omega_{j,y}\Gamma_{j,y}^{B},
 \qquad
 \Omega_{j,y}=(-1)^y\omega_j.
 \label{eq:app_backward_staggering}
\end{equation}
A route $\Gamma_{j,y}^{B}\mapsto s_j\Gamma_{k,y+1}^{B}$, with $s_j=\pm1$, acquires the coefficient $s_j\Omega_{j,y}\Omega_{k,y+1}$. For either bulk trajectory, $k=j\pm2$, and
\begin{equation}
 \Omega_{j,y}\Omega_{j\pm2,y+1}
 =(-1)^{2y+1}\omega_j\omega_{j\pm2}=1.
 \label{eq:app_backward_frame_route_identity}
\end{equation}
Thus both routes in Eq.~\eqref{eq:app_backward_clean_routes} are preserved. Within each perturbing bilinear, the two row factors cancel, leaving
\begin{equation}
\begin{gathered}
 \Omega_{2x,y}\Omega_{2x+1,y}=1,
 \qquad
 \Omega_{2x-2,y}\Omega_{2x+1,y}=-1,\\
 \widetilde B_{x,y}^{\mathrm h;B}=B_{x,y}^{\mathrm h;B},
 \qquad
 \widetilde B_{x,y}^{\mathrm v;B}=-B_{x,y}^{\mathrm v;B}.
 \label{eq:app_backward_bilinear_alignment}
 \end{gathered}
\end{equation}
Consequently, the backward generator becomes
\begin{equation}
 \mathcal P_y^{(1);B}
 =\beta_R\sum_x\xi^{\mathrm h}_{x,y}\widetilde B_{x,y}^{\mathrm h;B}
 +i\beta_R\sum_x\xi^{\mathrm v}_{x,y}\widetilde B_{x,y}^{\mathrm v;B}.
 \label{eq:app_backward_aligned_generator}
\end{equation}
Applying this redefinition to each backward replica gives the common $2n$-flavor vertices of Eq.~\eqref{eq:aligned_replicated_vertices}, with shared signs $\xi^{\mathrm h,\mathrm v}$ and unchanged clean bulk routing.

The alignment is local in the bulk. On a cylinder with $L_x$ cells around its circumference,
\begin{equation}
 \Omega_{j+2L_x,y}=(-1)^{L_x}\Omega_{j,y},
 \qquad
 \prod_{j=1}^{2L_x}\Omega_{j,y}=(-1)^{L_x}.
 \label{eq:app_backward_cylinder_signs}
\end{equation}
For odd $L_x$, the spatial mismatch enters the closing-bond or twist data. The second identity is the frame determinant: orientation is preserved for even $L_x$ and reversed for odd $L_x$. Hence odd $L_x$ permits no parity-preserving Gaussian implementation of the type $G_y$ in Appendix~\ref{app:signed_routing_frame}. With the same ordered-product convention, the parity operator formed from $\widetilde\Gamma^B$ is $(-1)^{L_x}$ times that formed from $\Gamma^B$; the physical sector is retained with its parity label adjusted accordingly.

\section{From the syndrome average to NL$\sigma$M}
\label{app:nlsm_derivation}
Starting from the aligned detuning vertices of Appendix~\ref{app:swap_basis}, we derive the local sign moments and their leading connected interaction, then state the continuum matching assumptions and construct the antisymmetric Hubbard--Stratonovich representation and class-D saddle manifold. 
The sign identities are exact within the specified bulk counting measure; the continuum projection, saddle selection, and stiffness matching require additional assumptions stated at those steps.

\subsection{Syndrome counting and local sign moments}
\label{app:random_average}
For fixed logical-sector labels in the replica expansions of Appendix~\ref{app:coherent_info}, let $F[\eta]$ be the complete gauge-invariant weight, including logical defects, scalar factors, and transformed boundary contractions. 
The layers share the reference field $\eta^{s,0}$, with their specified logical defects inserted separately. 
Vertex flips $\eta_{ij}\mapsto\epsilon_i\eta_{ij}\epsilon_j$ preserve the syndrome. 
Writing $\mathcal G$ for the effective gauge group and $\mathcal E_{\mathrm{ref}}$ for the union of the reference configurations' gauge orbits gives
\begin{equation}
 \sum_{s}F[\eta^{s,0}]
 =\sum_{[\eta]\in\mathcal E_{\mathrm{ref}}/\mathcal G}F[\eta]
 =\frac{1}{|\mathcal G|}\sum_{\eta\in\mathcal E_{\mathrm{ref}}}F[\eta].
 \label{eq:app_sign_orbit_counting}
\end{equation}
Each syndrome contributes one orbit of size $|\mathcal G|$. 
On a connected closed dual graph with $N_v$ vertices, $|\mathcal G|=2^{N_v-1}$ because the two uniform vertex flips act trivially; open boundaries require the allowed effective group. 
The replicated amplitudes in $F$ already contain the physical weights, so Eq.~\eqref{eq:app_sign_orbit_counting} introduces no additional Born factor. 
With conjugate branch choices, the vertical-link scalar $A_\eta(\beta)$, distinct from $A_\mu^s$, satisfies $A_\eta(\beta_p)A_\eta(\beta_p^*)=2\cosh(2\beta_R)$. 
Thus $n$ pairs supply a sign-independent factor $[2\cosh(2\beta_R)]^n$ per common bulk link away from defects; the remaining weight generally depends on the signs.

In the closed geometry, face products encode syndromes, while chosen noncontractible cycle products $H$ complete the gauge-orbit labels. 
Let $\mathcal E_H$ denote the bond configurations with these fixed cycle products. 
The specialization $\mathcal E_{\mathrm{ref}}=\mathcal E_H$ requires reference configurations with the same $H$ for every syndrome; fixed relative logical defects alone do not ensure this. 
Otherwise, retain $\mathcal E_{\mathrm{ref}}$, and accompany any reference change by the corresponding logical-sector relabeling in $F$.

For a given moment, let $E_{\mathrm{loc}}$ be the union of its full bond supports, including trajectory-frame histories. 
A uniform local counting measure applies when every possible sign assignment on $E_{\mathrm{loc}}$ occurs equally often among the globally allowed configurations. 
For binary product constraints at fixed $H$, this holds if no nontrivial combination of constraints is supported entirely on $E_{\mathrm{loc}}$. Then
\begin{equation}
 \langle f\rangle_{\mathrm{loc}}
 =\frac{1}{2^{|E_{\mathrm{loc}}|}}
 \sum_{\{\eta_e^{\mathrm{loc}}=\pm1\}}f,
 \qquad \langle1\rangle_{\mathrm{loc}}=1.
 \label{eq:app_local_counting_measure}
\end{equation}
Here $\eta_e^{\mathrm{loc}}$ is the restriction of the original bond field $\eta_e$ to $E_{\mathrm{loc}}$. 
The condition must be checked for each moment in the chosen reference ensemble, including any winding histories or boundary constraints. 
It does not require a uniform product measure over the whole cylinder, factorize $F$, or remove syndrome-dependent endpoint factors from the full average.

With $\ell=(x,y,\alpha)$ and $\alpha=\mathrm h,\mathrm v$, write $\xi_\ell=\xi^\alpha_{x,y}=\prod_{e\in S_\ell}\eta_e^{\mathrm{loc}}$. 
The support $S_\ell$ contains the bonds occurring an odd number of times in the vertex sign, including its frame history. 
The independent fundamental signs in Eq.~\eqref{eq:app_local_counting_measure} give
\begin{equation}
 \left\langle\prod_{r=1}^k\xi_{\ell_r}\right\rangle_{\mathrm{loc}}
 =\begin{cases}
 1,&S_{\ell_1}\oplus\cdots\oplus S_{\ell_k}=\varnothing,\\
 0,&\text{otherwise}.
 \end{cases}
 \label{eq:app_local_sign_moments}
\end{equation}
Thus a product survives precisely when every fundamental sign occurs an even number of times.

For Eq.~\eqref{eq:app_detuning_clean_vertices}, $u^{L/R}_{x,y}$ depends only on earlier rows. 
Holding that history fixed, the current-row supports are $\{(x,y,\mathrm h)\}$ for $\xi^{\mathrm h}_{x,y}$ and $\{(x-1,y,\mathrm h),(x,y,\mathrm h)\}$ for $\xi^{\mathrm v}_{x,y}$. 
Their nonempty, distinct supports give zero means and diagonal same-row covariance for the retained bulk nodes. 
For $y'>y$, average the later row first, conditioning on its preceding bond-sign history $\mathcal H_{y'}$:
\begin{equation}
 \left\langle\xi^\alpha_{x,y}\xi^{\alpha'}_{x',y'}\right\rangle_{\mathrm{loc}}
 =\left\langle\xi^\alpha_{x,y}
 \left\langle\xi^{\alpha'}_{x',y'}\mid\mathcal H_{y'}\right\rangle_{\mathrm{loc}}
 \right\rangle_{\mathrm{loc}}=0.
 \label{eq:app_sign_cross_row_covariance}
\end{equation}
Consequently, under this bulk product measure,
\begin{equation}
\begin{gathered}
     \langle\xi_\ell\rangle_{\mathrm{loc}}=0,\\
K_{\ell\ell'}\equiv\langle\xi_\ell\xi_{\ell'}\rangle_{\mathrm{loc}}
 -\langle\xi_\ell\rangle_{\mathrm{loc}}\langle\xi_{\ell'}\rangle_{\mathrm{loc}}
 =\delta_{\ell\ell'}.
\end{gathered}
 \label{eq:app_bulk_sign_covariance}
\end{equation}
Coincident supports or constraints on the retained histories require the full covariance kernel; zero means and diagonal covariance must then be checked in the constrained measure. 
Nor does diagonal covariance imply independent vertex signs: on the initial row, $u^{L/R}_{x,0}=1$ gives $\xi^{\mathrm h}_{x-1,0}\xi^{\mathrm h}_{x,0}\xi^{\mathrm v}_{x,0}=1$. 
Higher moments follow from Eq.~\eqref{eq:app_local_sign_moments}. The first two moments above supply the second-cumulant calculation in the next subsection.

\onecolumngrid
\subsection{Second cumulant and the leading Majorana interaction}
\label{app:sign_second_cumulant}

Use the aligned $2n$ flavors of Appendix~\ref{app:backward_clean_frame} and the node index $\ell=(x,y,\alpha)$. All node sums and residual row factors below refer to the retained bulk vertices, evaluated with the local counting law of Appendix~\ref{app:random_average}. Matching the second-order result to a constrained ensemble requires agreement of the relevant first and second sign moments, including the mixed moments in the BCH term. The flat-marginal criterion applied to each single or paired support is sufficient; a flat joint distribution over the whole cylinder is not required. The strictly linear row perturbation is
\begin{equation}
     \mathcal P_y^{(1)}=\sum_{\ell\in y}c_\ell\xi_\ell\mathcal B_\ell,
 \quad \mathcal B_\ell=\sum_{A=1}^{2n}B_\ell^A,\quad
 c_{\mathrm h}=\beta_R,\quad c_{\mathrm v}=i\beta_R.
 \label{eq:app_cumulant_linear_generator}
\end{equation}
All flavors share the same $\xi_\ell$. Since $\phi(\beta_R)=\beta_R+O(\beta_R^3)$, its nonlinear correction does not enter at second order.

The ordered row factorization $e^{\mathcal P_y}=e^{A_{\mathrm v,y}}e^{A_{\mathrm h,y}}$ in Eq.~\eqref{eq:app_detuning_clean_product} gives the Baker--Campbell--Hausdorff expansion
\begin{equation}
 \mathcal P_y=\mathcal P_y^{(1)}+\mathcal R_y^{(2)}+O(\beta_R^3),\qquad
 \mathcal R_y^{(2)}
 =\frac{i\beta_R^2}{2}\sum_{x,x'}
 \xi^{\mathrm v}_{x,y}\xi^{\mathrm h}_{x',y}
 [\mathcal B^{\mathrm v}_{x,y},\mathcal B^{\mathrm h}_{x',y}],
 \label{eq:app_cumulant_bch_remainder}
\end{equation}
where $\mathcal B^\alpha_{x,y}=\sum_A B^{\alpha;A}_{x,y}$. 
The commutator order follows from the vertical factor standing to the left of the horizontal factor. 
Distinct-flavor bilinears commute, while same-flavor commutators remain bilinear. Under the measure of Eq.~\eqref{eq:app_local_counting_measure}, $\langle\xi^{\mathrm v}_{x,y}\xi^{\mathrm h}_{x',y}\rangle_{\mathrm{loc}}=0$, so $\langle\mathcal R_y^{(2)}\rangle_{\mathrm{loc}}=0$: this BCH term produces no averaged bulk bilinear at second order.

For the ordered product of these bulk row factors, remove the deterministic clean propagation. Here $\hat T_0$ denotes the product of the clean row operators over the aligned flavors. 
Define $\mathcal P_{y,I}=\hat T_0^{-y}\mathcal P_y\hat T_0^y$, then
\begin{equation}
\begin{aligned}
 \prod_{y=N_y-1}^{0}(\hat T_0e^{\mathcal P_y})
 =\hat T_0^{N_y}e^{\mathcal P_{N_y-1,I}}\cdots e^{\mathcal P_{0,I}}=\hat T_0^{N_y}\mathcal U_I,
 \label{eq:app_cumulant_interaction_picture}
\end{aligned}
\end{equation}
with $\mathcal U_I\equiv e^{\mathcal P_{N_y-1,I}}\cdots e^{\mathcal P_{0,I}}$.
The subscript $I$ denotes the same conjugation for $\mathcal P_y^{(1)}$, $\mathcal R_y^{(2)}$, and each bilinear on row $y$. These are reduced bulk operators. 
The syndrome-dependent endpoint contractions, scalar factors, and logical defects remain in the full amplitude; Eq.~\eqref{eq:app_cumulant_interaction_picture} does not remove them from its average.

For a fixed finite set of retained vertices, $\langle\mathcal P_{y,I}^{(1)}\rangle_{\mathrm{loc}}=0$, and the ordered expansion is
\begin{equation}
     \langle\mathcal U_I\rangle_{\mathrm{loc}}=I
 +\sum_y\langle\mathcal R_{y,I}^{(2)}\rangle_{\mathrm{loc}}
 +\frac12\sum_y\left\langle(\mathcal P_{y,I}^{(1)})^2\right\rangle_{\mathrm{loc}}
 +\sum_{y>y'}\left\langle\mathcal P_{y,I}^{(1)}\mathcal P_{y',I}^{(1)}\right\rangle_{\mathrm{loc}}
 +O(\beta_R^3).
 \label{eq:app_cumulant_ordered_expansion}
\end{equation}
The mean BCH term vanishes as above. 
Since the averaged correction starts at second order, the logarithm continuous from the identity has the same second-order coefficient:
\begin{equation}
     \ln\langle\mathcal U_I\rangle_{\mathrm{loc}}
 =\Delta\mathcal L_I^{(2)}+O(\beta_R^3),\qquad
 \Delta\mathcal L_I^{(2)}
 =\frac12\sum_{\ell,\ell'}c_\ell c_{\ell'}K_{\ell\ell'}
 \mathcal T\!\left[\mathcal B_{\ell,I}\mathcal B_{\ell',I}\right].
 \label{eq:app_ordered_second_cumulant}
\end{equation}
The symbol $\mathcal T$ puts later rows to the left; equal-row products are symmetrized in this double sum. 
The factor $1/2$ retains the exponential combinatorics. 
Neither factorization of row averages nor vanishing higher sign cumulants is assumed.

For the bulk covariance $K_{\ell\ell'}=\delta_{\ell\ell'}$, only same-node squares remain. Since $(B_\ell^A)^2=I$ and the even bilinears commute between flavors,
\begin{equation}
 \frac12c_\ell^2\mathcal B_\ell^2
 =n c_\ell^2 I+c_\ell^2\sum_{A<B}B_\ell^AB_\ell^B.
 \label{eq:app_cumulant_flavor_square}
\end{equation}
The same identity holds after clean conjugation. 
To distinguish local row insertions from the interaction-picture logarithm, define
\begin{equation}
     \mathcal Q_y^{(2)}
 =\beta_R^2\sum_x\sum_{1\leq A<B\leq2n}
 \left[
 B_{x,y}^{\mathrm h;A}B_{x,y}^{\mathrm h;B}
 -B_{x,y}^{\mathrm v;A}B_{x,y}^{\mathrm v;B}
 \right],\qquad
 \Delta\mathcal L_I^{(2)}
 =\sum_y\hat T_0^{-y}\mathcal Q_y^{(2)}\hat T_0^y.
 \label{eq:app_bulk_majorana_quartic}
\end{equation}
This is the row contribution used in Eq.~\eqref{eq:local_hv_quartic}; each row has its own clean conjugation in the interaction-picture logarithm. 
The relative quartic minus sign follows from $c_{\mathrm h}^2=\beta_R^2$ and $c_{\mathrm v}^2=-\beta_R^2$ in the transfer exponent.

The quartic arises from averaging shared signs, while each fixed-syndrome transfer remains Gaussian within a fixed fermion-parity sector. Each $\mathcal B_\ell$ is invariant under a common global $O(2n)$ rotation, so the retained interaction has this bulk flavor symmetry. For supports subject to global or boundary constraints, recheck the first moments, the full covariance kernel, and the mean BCH term; same-flavor cross products need not be constants. Extending the result to the full cylinder requires these moment checks in its counting ensemble and retention of the syndrome-dependent boundary contraction.

The result fixes the order-$\beta_R^2$ lattice interaction under the stated bulk measure. Replacing $\beta_R$ by the exact $\phi(\beta_R)$ in the vertical coefficient changes its same-node second cumulant only at $O(\beta_R^4)$. This estimate concerns the retained node second cumulant; the full ordered expansion in Eq.~\eqref{eq:app_ordered_second_cumulant} retains an $O(\beta_R^3)$ remainder.

\subsection{Bond-resolved interaction and Hubbard--Stratonovich representation}
\label{app:eft_derivations}
Starting from the second-cumulant quartic in Eq.~\eqref{eq:app_bulk_majorana_quartic}, we retain the displaced endpoints, identify the zero-momentum cancellation, and construct the lattice Hubbard--Stratonovich (HS) representation. Integrating the fermions with the full lattice kernel gives the Pfaffian action used in Appendix~\ref{app:diffusive_saddle}.

\paragraph{Retaining the displaced lattice interaction.}
The horizontal and vertical quartics in Eq.~\eqref{eq:app_bulk_majorana_quartic} have opposite coefficients but different endpoints, which we retain before taking a continuum limit. 
Set the cell and row spacings to one, with $y$ the discrete transfer-row coordinate. 
For each chirality and flavor, an auxiliary Majorana partner permits a coherent-state path integral. An exact change of Grassmann variables absorbs the clean routing and row shift into $\phi$, making the physical insertion $\phi+\bar\phi$ local while retaining the shift in the free action below. The variables $\phi,\bar\phi$ are independent; define
\begin{equation}
 \chi=\frac{\phi+\bar\phi}{\sqrt2},\qquad
 \zeta=\frac{i(\bar\phi-\phi)}{\sqrt2}.
 \label{eq:app_lattice_coherent_components}
\end{equation}
The physical insertion is $\sqrt2\chi$; the complementary Grassmann component $\zeta$ completes the kinetic description. This auxiliary construction adds no replicas.

After separating the same-flavor constants, each retained inter-flavor quartic contains four distinct Majoranas. Their insertions supply the normalization factor $(\sqrt2)^4=4$. For $A<B$, Grassmann reordering at fixed bond endpoints gives $ (i\chi_R^A\chi_L^A)(i\chi_R^B\chi_L^B)
 =-\chi_R^A\chi_L^A\chi_R^B\chi_L^B
 =\chi_R^A\chi_R^B\chi_L^A\chi_L^B.$
Thus, the quartic is a product of the antisymmetric flavor currents
\begin{equation}
 M_\rho^{AB}(x,y)=\chi_\rho^A(x,y)\chi_\rho^B(x,y),
 \qquad \rho=R,L,\quad 1\leq A<B\leq2n.
 \label{eq:app_lattice_flavor_currents}
\end{equation}
With the common left endpoint labeled $(x,y)$, the horizontal bond samples $M_R^{AB}(x,y)$ and the vertical bond samples $M_R^{AB}(x-1,y)$. Their opposite coefficients therefore give
\begin{equation}
 \Delta_g=g\sum_{x,y}\sum_{A<B}
 \left[M_R^{AB}(x,y)-M_R^{AB}(x-1,y)\right]
 M_L^{AB}(x,y),\qquad g=4\beta_R^2.
 \label{eq:app_displaced_lattice_interaction}
\end{equation}
The matching $g=4\beta_R^2$ applies to the retained second cumulant. The weight is $e^{-S_0[\chi,\zeta]+\Delta_g}$, so the interaction contributes $-\Delta_g$ to the action. No continuum approximation has been made; higher-order interactions and physical boundary and sector factors require separate matching.

\paragraph{Cancellation at zero spatial current momentum.}
For periodic bulk sums, use the Fourier transformation $M_\rho^{AB}(x,y)=\tfrac{1}{\sqrt{N_{\mathrm{cell}}}}\sum_Q e^{i(Kx+\Omega y)}M_\rho^{AB}(Q)$, where $N_{\mathrm{cell}}$ counts the lattice cells. 
The current momentum $Q=(K,\Omega)$ is the sum of its constituent fermion momenta: $K=k_1+k_2$ and $\Omega=\omega_1+\omega_2$, modulo $2\pi$. The two components are conjugate to the cell and transfer-row coordinates. Equation~\eqref{eq:app_displaced_lattice_interaction} becomes
\begin{equation}
 \Delta_g=g\sum_Q\sum_{A<B}f(K)M_R^{AB}(Q)M_L^{AB}(-Q),\qquad
 f(K)=1-e^{-iK}.
 \label{eq:app_lattice_current_form_factor}
\end{equation}
At $K=0$, $f(0)=0$ cancels the interaction for every $\Omega$. Near this channel, $f(K)=iK+O(K^2)$ leaves derivative interactions but no derivative-free quartic. The finite-momentum lattice interaction survives, so its channel must be identified before dropping the bond displacement.

The same cancellation occurs when the local continuum action in Appendix~C of Ref.~\cite{YanBaoVijay2026} is read as two independent pointwise HS integrals of coincident Grassmann currents. 
Its displaced-bond HS identities remain valid, but a nonzero continuum interaction requires an explicit point-splitting or momentum-channel prescription. 
We therefore construct the HS theory directly from the retained lattice interaction.

\paragraph{Lattice Hubbard--Stratonovich fields and the full action.}
For $g=4\beta^2_R>0$, introduce independent real horizontal and vertical bond fields $H_{AB}(x,y)$ and $V_{AB}(x,y)$ for $A<B$, assembled into antisymmetric $2n\times2n$ matrices $H,V$. For one cell and flavor pair, suppress $A,B$ and write $X=M_R(x,y)$, $Z=M_R(x-1,y)$, $Y=M_L(x,y)$. In the following identity, $H,V$ denote the corresponding scalar entries. These even Grassmann currents commute and satisfy $X^2=Y^2=Z^2=0$, giving 
\begin{equation}
 \exp{\left[gXY-gZY\right]}
 =\frac1{2\pi g}\int_{\mathbb R^2}dH\,dV\,
 \exp{\left[-\frac{(H^2+V^2)}{2g}\right]}\cdot \exp{\left[iH(X-Y)+iV(Z+Y)\right]}.
 \label{eq:app_bond_hs_identity}
\end{equation}
The two Gaussian integrals generate $-g(X-Y)^2/2=gXY$ and $-g(Z+Y)^2/2=-gZY$. Thus real contours and imaginary source couplings reproduce both bond signs. Applying this identity throughout the lattice gives $e^{-S_0+\Delta_g}=\mathcal N_g\int\mathcal D H\,\mathcal D V\,e^{-S[\chi,\zeta,H,V]}$, with normalization $\mathcal N_g$ fixed by the Gaussian measures and extended action
\begin{equation}
 \begin{aligned}
 S[\chi,\zeta,H,V]={}S_0[\chi,\zeta]+S_G[H,V]
 -&i\sum_{x,y,A<B}H_{AB}(x,y)\left[M_R^{AB}(x,y)-M_L^{AB}(x,y)\right]\\
 -&i\sum_{x,y,A<B}V_{AB}(x,y)\left[M_R^{AB}(x-1,y)+M_L^{AB}(x,y)\right].
 \end{aligned}
 \label{eq:app_lattice_hs_action}
\end{equation}
The Gaussian cost of the HS fields is
\begin{equation}
 S_G[H,V]=\frac1{4g}\sum_{x,y}\operatorname{tr}(H^TH+V^TV)
 =\frac1{2g}\sum_{x,y,A<B}(H_{AB}^2+V_{AB}^2).
 \label{eq:app_bond_hs_gaussian_cost}
\end{equation}
The ordinary flavor trace counts each independent antisymmetric entry twice. The HS fields couple only to $\chi$ through the currents; $\zeta$ remains in the free kinetic action. All lattice momenta are retained, and no saddle has been selected.
Reindexing the displaced right-current term in Eq.~\eqref{eq:app_lattice_hs_action} collects the bond couplings into local fermion sources:
\begin{equation}
 \begin{gathered}
  S[\chi,\zeta,H,V]=S_0[\chi,\zeta]+S_G[H,V]-\frac i2\sum_{x,y,\rho}
 \chi_\rho(x,y)^T U_\rho(x,y)\chi_\rho(x,y),\\
 U_R(x,y)=H(x,y)+V(x+1,y),\qquad U_L(x,y)=-H(x,y)+V(x,y).
 \end{gathered}
 \label{eq:app_displaced_hs_sources}
\end{equation}
The antisymmetric matrices $U_R,U_L$ are fixed combinations of $H(x,y)$ and $V(x,y)$ with no independent measure; open-boundary sums retain only existing bonds. With the bulk Fourier convention above,
\begin{equation}
 U_R(Q)=H(Q)+e^{iK}V(Q),\qquad
 U_L(Q)=-H(Q)+V(Q),\qquad Q=(K,\Omega).
 \label{eq:app_hs_source_form_factors}
\end{equation}
A source at $Q$ couples to a current at $-Q$.
The displacement phase is exact, and the real HS fields obey $H(-Q)=H(Q)^*$ and $V(-Q)=V(Q)^*$.

\paragraph{Lattice propagation and fermion integration.}
In the coherent-state convention of Eq.~\eqref{eq:app_lattice_coherent_components}, $S_0=\bar\phi^{\,T}(T^{-1}-I)\phi$, where $T$ is the clean routing and row shift. For $\Phi_\rho=(\chi_\rho,\zeta_\rho)^\mathrm{T}$ and fermion momentum $\mathbf p=(k,\omega)$, define $p_{R/L}=\omega\pm k$. Using $T(p_\rho)=e^{-ip_\rho}$ gives
\begin{equation}
 S_0[\chi,\zeta]=\frac12\sum_{\rho,\mathbf p}
 \Phi_\rho(-\mathbf p)^\mathrm{T}
 [\mathsf K(p_\rho)\otimes I_{2n}]\Phi_\rho(\mathbf p),\qquad
 \mathsf K(p)=
 \begin{pmatrix}
 i\sin p&-i(1-\cos p)\\
 i(1-\cos p)&i\sin p
 \end{pmatrix}.
 \label{eq:app_lattice_fermion_kernel}
\end{equation}
The kernel acts on $(\chi,\zeta)$ and obeys $\mathsf K(p)^\mathrm{T}=-\mathsf K(-p)$. With free zero modes regulated, inverting the full kernel gives the physical propagator, while integrating out $\zeta$ gives its inverse:
\begin{equation}
 G_{\chi\chi}(p)=[\mathsf K(p)^{-1}]_{\chi\chi}
 =-\frac i2\cot\frac p2,\qquad
 D_{\chi,\mathrm{eff}}(p)
 =\mathsf K_{\chi\chi}
 -\mathsf K_{\chi\zeta}\mathsf K_{\zeta\zeta}^{-1}\mathsf K_{\zeta\chi}
 =2i\tan\frac p2
 =G_{\chi\chi}(p)^{-1}.
 \label{eq:app_lattice_physical_propagator}
\end{equation}
This Schur complement requires an invertible auxiliary block; otherwise, use the full kernel. Deleting $\zeta$ would instead give $1/(i\sin p)$ and a spurious pole at $p=\pi$. The full lattice dispersion is retained.

Hold $H,V$ fixed during fermion integration. The sources enter only the physical block:
\begin{equation}
 \begin{aligned}
 S[\chi,\zeta,H,V]&=S_G[H,V]+\frac12\sum_{\rho=R,L}\Phi_\rho^\mathrm{T}\mathcal A_\rho[H,V]\Phi_\rho,\qquad
 \mathcal A_\rho[H,V]
 =\mathsf K_\rho\otimes I_{2n}
 -i\begin{pmatrix}U_\rho&0\\0&0\end{pmatrix}.
 \end{aligned}
 \label{eq:app_lattice_hs_pfaffian_kernel}
\end{equation}
Here $\mathsf K_\rho$ has symbol $\mathsf K(p_\rho)$, and $\mathcal A_\rho^\mathrm{T}=-\mathcal A_\rho$ with lattice positions included in the transpose. Integrating out $\zeta$ first gives $D_{\chi,\mathrm{eff},\rho}I_{2n}-iU_\rho$ and a source-independent auxiliary Pfaffian. The two chiral Majorana integrals give
\begin{equation}
 \begin{aligned}
 \frac{Z_g}{Z_0}
 =\mathcal N_g\int_{\mathbb R}\mathcal D H\,\mathcal D V\,
 e^{-S_{\mathrm{eff}}[H,V]},\qquad
 e^{-S_{\mathrm{eff}}[H,V]}
 =e^{-S_G[H,V]}\prod_{\rho=R,L}
 \frac{\operatorname{Pf}\mathcal A_\rho[H,V]}
 {\operatorname{Pf}\mathcal A_\rho[0,0]}.
 \end{aligned}
 \label{eq:app_lattice_hs_effective_weight}
\end{equation}
The interacting and free integrals $Z_g,Z_0$ use identical regulator and boundary data, so source-independent factors cancel. The HS measures and normalization $\mathcal N_g$ are fixed by Eq.~\eqref{eq:app_bond_hs_identity}. Locally on the prescribed nonzero-Pfaffian branch,
\begin{equation}
 S_{\mathrm{eff}}[H,V]
 =S_G[H,V]-\frac12\sum_{\rho=R,L}\operatorname{Tr}
 \ln\!\left(\frac{\mathcal A_\rho[H,V]}{\mathcal A_\rho[0,0]}\right).
 \label{eq:app_hs_effective_action}
\end{equation}
Here $\operatorname{Tr}$ includes lattice, coherent-state-component, and flavor indices; the factor $1/2$ follows from Majorana integration. The Pfaffian phase must be retained: real HS contours ensure Gaussian convergence, not positivity of the fermion-integrated weight. This exact representation of the retained lattice theory supplies the action for the saddle analysis.

\subsection{Diffusive saddle, target manifold, and stiffness matching}
\label{app:diffusive_saddle}
Starting from the lattice HS action of Appendix~\ref{app:eft_derivations}, we identify the surviving spacetime channel $\Pi=(\pi,\pi)$ and calculate its fermionic potential to determine the saddle radius and test stationarity and radial stability. 
We then identify the degenerate orientations $SO(2n)/U(n)$ and integrate out the massive HS fluctuations to obtain the gradient action for the soft orientation field.
Within the retained saddle construction, this yields a class-D NL$\sigma$M with bare inverse stiffness $t_{\sigma,0}\propto\beta_R^2$.

\paragraph{Surviving spacetime channel.}
The bond form factor $f(K)=1-e^{-iK}$ contains no temporal momentum and is nonzero at many spatial momenta. 
To identify a candidate channel for the saddle construction, we combine it with clean propagation. 
A current with momentum $Q=(K,\Omega)$ contains two fermions whose momenta sum to $Q$. 
If both constituents of the right current lie on their clean shells $\omega_i+k_i=0$ modulo $2\pi$, then $\Omega+K=0$. 
The left current at $-Q$ similarly requires $\Omega-K=0$. The simultaneous exact-shell candidates are therefore
\begin{equation}
 \Omega+K=\Omega-K=0\pmod{2\pi}
 \quad\Longrightarrow\quad Q=(0,0)\ \text{or}\ Q=\Pi\equiv(\pi,\pi).
 \label{eq:app_surviving_current_carrier}
\end{equation}
The vertex vanishes at $K=0$ for every $\Omega$. 
Among these two candidates, it thus removes the uniform channel and retains the staggered one, with $f(\pi)=2$. 
This selection concerns the total current momentum; individual fermions can lie anywhere along their respective clean shell lines.

The temporal staggering makes this surviving channel compatible with both propagation directions. 
A source with momentum $Q=(K,\Omega)$ connects fermion modes
at $(k,\omega)$ and $(k+K,\omega+\Omega)$, changing their
trajectory momenta by $P_{R/L}=\Omega\pm K$.
At $\Pi=(\pi,\pi)$, both transfers vanish modulo $2\pi$, preserving $p_\rho$ and coupling near-shell modes to near-shell modes. 
In real space, the spacetime staggering factor $s_{x,y}=(-1)^{x+y}$ is constant along either clean step $(x,y)\mapsto(x\pm1,y+1)$: the row sign compensates the spatial sign change. 
Near $\Pi=(\pi,\pi)$, the transfers reduce to $P_{R/L}=\delta\Omega\pm\delta K$ which can be small modulo $2\pi$, motivating a slowly varying envelope about this staggered background. 
By contrast, $(\pi,0)$ alternates along each path and couples $p_\rho$ to $p_\rho+\pi$, changing the fermion response and hence the HS potential. 
This kinematic argument identifies a candidate channel, not a nonzero or dominant saddle; the subsequent Pfaffian calculation retains the full internal fermion Brillouin zone, including off-shell modes.

\paragraph{Staggered HS family.}
We test this channel using constant real antisymmetric $2n\times2n$ flavor matrices $H_0,V_0$:
\begin{equation}
 H(x,y)=s_{x,y}H_0,\qquad V(x,y)=s_{x,y}V_0,
 \qquad s_{x,y}=(-1)^{x+y}.
 \label{eq:app_staggered_hs_family}
\end{equation}
This trial family lies within the full HS integral and does not assume staggered microscopic error strengths. 
Define the real antisymmetric matrix combinations
\begin{equation}
 u=H_0-V_0,\qquad w=H_0+V_0,\qquad
 H_0=\frac{u+w}{2},\quad V_0=\frac{w-u}{2}.
 \label{eq:app_hs_sum_difference}
\end{equation}
Since $s_{x+1,y}=-s_{x,y}$, the displaced sources and Gaussian action density of HS fields become
\begin{equation}
 U_R(x,y)=s_{x,y}u,\quad U_L(x,y)=-s_{x,y}u,\qquad
 \frac{S_G}{N_{\mathrm{cell}}}
 =\frac1{8g}\operatorname{tr}(u^Tu+w^Tw),
 \label{eq:app_staggered_sources_cost}
\end{equation}
where $N_{\mathrm{cell}}$ is the number of bulk cells. 
Within this constant-envelope family, the fermion Pfaffian depends only on $u$. At fixed $u$ and $g>0$, the Gaussian cost is therefore minimized by
\begin{equation}
 w_\star=0\quad\Longrightarrow\quad H_0=\frac u2,\quad V_0=-\frac u2.
 \label{eq:app_hs_complement_minimum}
\end{equation}
Equal and opposite bond amplitudes thus produce the specified sources at the smallest Gaussian cost. 
These relations use spatial staggering alone and also hold for $s_{x,y}=(-1)^x$; the temporal carrier enters through the fermion Pfaffian. 
Nonuniform complementary fluctuations remain in the subsequent analysis, since neighboring-site differences can couple them to the fermions. 
The remaining question is whether the fermion contribution favors a nonzero $u$, which we determine next from its effective potential.

\paragraph{Fermion potential and saddle point.}
We first evaluate the equal-magnitude family under $w_*=0$
\begin{equation}
 u=qJ,\qquad
 H(x,y)=\frac q2s_{x,y}J,\quad
 V(x,y)=-\frac q2s_{x,y}J,
 \label{eq:app_hs_equal_radius_family}
\end{equation}
where $q\geq0$ is the saddle radius to be determined and $J^\mathrm{T}=-J$, $J^2=-I_{2n}$. 
The sources are $U_R=qsJ$, $U_L=-qsJ$, and $S_G/N_{\mathrm{cell}}=nq^2/(4g)$. 
The same calculation below tests whether unequal flavor-block magnitudes are favored.

At carrier $\Pi=(\pi,\pi)$, each clean trajectory sees a constant source $U=\sigma qJ_{(2)}$, where $\sigma=\pm1$ includes chirality and trajectory parity, and $J_{(2)}=\left(\begin{smallmatrix}0&1\\-1&0\end{smallmatrix}\right)$. 
The two flavor eigenkernels are $D(p)\pm q$, with $D(p)\equiv D_{\chi,\mathrm{eff}}(p)=2i\tan(p/2)$. The normalized determinant for one flavor block is therefore
\begin{equation}
 F_q(p)=\det\!\left[I_2-i\sigma qG_{\chi\chi}(p)J_{(2)}\right]
 =1+\frac{q^2}{4}\cot^2\frac p2.
 \label{eq:app_staggered_fermion_factor}
\end{equation}
Each chiral Pfaffian contributes one half of the full-zone logarithm; the two chiralities supply one copy, with no extra multiplicity from trajectory parity. 
With an even antiperiodic trajectory regulator, continue the normalized Pfaffian from $q=0$ along real $q$: the positive, nonzero determinant ratios fix this branch continuously. 
Define $\mathcal V_{\pi,\Omega}$ as the effective-action density per cell and per two-dimensional flavor block for the specified carrier, with $\mathcal V=\mathcal V_{\pi,\pi}$. 
In the long-trajectory limit,
\begin{equation}
 \mathcal V(q)-\mathcal V(0)
 =\frac{q^2}{4g}-\int_{-\pi}^{\pi}\frac{dp}{2\pi}\ln F_q(p)
 =\frac{q^2}{4g}-2\ln\left(1+\frac q2\right),\qquad q\geq0.
 \label{eq:app_lattice_saddle_potential}
\end{equation}
The near-shell fermions lower the potential by $-q+O(q^2)$, competing with the quadratic Gaussian cost. This linear term is a $-|q|$ cusp if the amplitude is extended to signed values: it requires the long-trajectory limit before the weak-coupling limit, with the full lattice dispersion retained. 
At fixed finite antiperiodic length, the momentum sum is analytic in $q^2$ near zero, and a nonzero saddle need not exist at arbitrarily small $g$. 
In the stated limit, balancing the two contributions gives
\begin{equation}
 \mathcal V'(q)=\frac q{2g}-\frac2{q+2}=0\longrightarrow
 q_\star=\sqrt{1+4g}-1=2g+O(g^2).
 \label{eq:app_lattice_saddle_radius}
\end{equation}
For unequal constant flavor-block magnitudes $q_\alpha$, the potential is $\sum_{\alpha=1}^n\mathcal V(q_\alpha)$. Each block is minimized at $q_\star$, selecting equal magnitudes and full rank within this family. 
With $g=4\beta_R^2$ at the retained order, $q_\star\sim8\beta_R^2$ is a collective HS self-energy amplitude that regulates the clean fermion singularity through $D(p)\pm q_\star$, rather than a deterministic microscopic mass. 
The radius vanishes at exact SWAP, $g=0$.

\paragraph{Stationarity and radial stability.}
To test stationarity beyond the scalar ansatz, let $G_\rho$ denote the physical $\chi\chi$ block of the inverse fermion kernel in this HS background. For $q>0$, its coincident values are $G_R(x,y;x,y)=-is_{x,y}J/(q+2)$ and $G_L(x,y;x,y)=+is_{x,y}J/(q+2)$. In the first variation of Eq.~\eqref{eq:app_hs_effective_action}, varying $V(x,y)$ changes $U_R(x-1,y)$, whose staggering sign is opposite to $s_{x,y}$. Including this displacement gives, for every independent entry $A<B$,
\begin{equation}
 \frac{\delta S_{\mathrm{eff}}}{\delta H_{AB}(x,y)}
 =s_{x,y}J_{AB}\,\mathcal V'(q),\qquad
 \frac{\delta S_{\mathrm{eff}}}{\delta V_{AB}(x,y)}
 =-s_{x,y}J_{AB}\,\mathcal V'(q),\qquad
\mathcal V''(q_\star)=\frac1{2g}+\frac2{(q_\star+2)^2}>0.
 \label{eq:app_lattice_saddle_stationarity}
\end{equation}
Both derivatives on HS fields vanish at $q=q_\star$, establishing stationarity against arbitrary bond-field variations in the retained long-trajectory theory. 
The positive curvature $  \mathcal V''(q_\star)$ per flavor block gives a restoring cost for independent constant block-radius fluctuations, while constant $w$ fluctuations retain their positive Gaussian cost.
Constant flavor orientations are degenerate by the symmetry discussed below; spatially varying orientations and other nonuniform fluctuations require the full HS fluctuation kernel. Stationarity and radial stability alone do not establish the saddle's global contribution to the physical integral.

\paragraph{Comparison with a purely spatial carrier.}
To see why the temporal choice matters dynamically, keep the same $H_0=u/2$, $V_0=-u/2$, and $u=qJ$, but replace $s_{x,y}$ by $(-1)^x$, corresponding to $(K,\Omega)=(\pi,0)$. 
The Gaussian cost and source relations are unchanged, but the source now transfers trajectory momentum $\pi$. 
In each flavor eigensector, the coupled modes $p,p+\pi$ have kernel
\begin{equation}
 \mathcal A_\pm(p)=
 \begin{pmatrix}D(p)&\pm q\\\pm q&D(p+\pi)\end{pmatrix},
 \qquad D(p)D(p+\pi)=4.
 \label{eq:app_spatial_carrier_pair_kernel}
\end{equation}
Its normalized determinant is $1-q^2/4$. 
Counting each momentum pair once and retaining both flavor sectors, both chiralities, and the Pfaffian factor gives, on the branch connected to $q=0$,
\begin{equation}
 \begin{aligned}
 \mathcal V_{\pi,0}(q)-\mathcal V_{\pi,0}(0)
 =\frac{q^2}{4g}-\ln\left(1-\frac{q^2}{4}\right)\longrightarrow
 \mathcal V_{\pi,0}'(q)
 =\frac q{2g}+\frac{2q}{4-q^2}>0,
 \quad 0<q<2.
 \end{aligned}
 \label{eq:app_spatial_carrier_potential}
\end{equation}
Thus this family has its minimum at $q=0$ and no nonzero radial saddle on the stated branch, whereas the $(\pi,\pi)$ carrier supports Eq.~\eqref{eq:app_lattice_saddle_radius}. 
This compares two backgrounds with different fermion momentum couplings.
The determinant vanishes at $q=2$, so the argument does not extend across that point or establish uniqueness among arbitrary HS backgrounds.

\paragraph{Saddle orientations and target space.}
At fixed radius, common flavor rotations change the saddle orientation without changing its potential. 
Using identical bulk boundary conditions and regularization for all flavors, $O\in SO(2n)$ rotates both chiralities and their auxiliary components, with $H\mapsto OHO^T$ and $V\mapsto OVO^T$. 
This preserves the lattice action and the Pfaffians, including their phases. The degenerate orientations are
\begin{equation}
 J=OJ_0O^T,\qquad
 J_0=\begin{pmatrix}0&I_n\\-I_n&0\end{pmatrix},
 \qquad O\in SO(2n).
 \label{eq:app_saddle_orbit}
\end{equation}
Rotations commuting with $J_0$ leave the same orientation unchanged. 
Their block form $R=\left(\begin{smallmatrix}a&-b\\b&a\end{smallmatrix}\right)$ identifies them with unitary matrices $a+ib$, so the stabilizer is $U(n)$. 
Quotienting out these redundant rotations gives the connected class-D saddle manifold
\begin{equation}
 \mathcal M=SO(2n)/U(n),\qquad
 \dim\mathcal M=n(n-1).
 \label{eq:app_classD_saddle_target}
\end{equation}
The normalized field obeys $J^T=-J$ and $J^2=-I_{2n}$. 
Allowing its orientation to vary slowly defines the candidate sigma-model field,
\begin{equation}
 J(x,y)=O(x,y)J_0O(x,y)^T\in\mathcal M,\qquad
 H(x,y)=\frac{q_\star}{2}s_{x,y}J(x,y),\quad
 V(x,y)=-\frac{q_\star}{2}s_{x,y}J(x,y).
 \label{eq:app_saddle_orientation}
\end{equation}
Equation~\eqref{eq:app_saddle_orientation} parametrizes the orientation family; gradients also induce complementary HS fluctuations, which must be included when calculating the stiffness. The orbit above is one connected component of $O(2n)/U(n)$.

\paragraph{Collective fluctuations and the nonlinear sigma model.}
We construct tangent fluctuations at integer $n\geq2$, before replica continuation. For fluctuations near lattice momenta $0$ and $\Pi$, write
\begin{equation}
 \delta H(x,y)=\delta H_0(x,y)+s_{x,y}\delta H_\Pi(x,y),\qquad
 \delta V(x,y)=\delta V_0(x,y)+s_{x,y}\delta V_\Pi(x,y).
\end{equation}
The four matrix envelopes vary slowly; their subscripts label the lattice momentum about which they fluctuate. 
Thus a constant $\delta H_\Pi$ has zero envelope momentum but staggered lattice momentum. 
Near $J_0$, orientation variations obey $\delta J^T=-\delta J$ and $\{J_0,\delta J\}=0$ and occupy the staggered difference $\delta H_\Pi-\delta V_\Pi$. 
They are soft because a constant orientation change costs no potential. 
In this tangent sector, the complementary sum $\delta H_\Pi+\delta V_\Pi$ and carrier-$0$ envelopes $\delta H_0,\delta V_0$ have positive restoring costs at zero envelope momentum. 
These and the radial restoring costs describe massive collective HS modes, not microscopic Majorana masses.

At envelope momenta $|\delta\Omega\pm\delta K|\ll q_\star$, the massive modes can be integrated out, leaving a theory for slowly varying orientations. 
They still respond to orientation gradients and thereby change the stiffness. 
Writing $\mathcal H$ for the Hessian of $S_{\mathrm{eff}}$ in the original real bond fluctuations, Gaussian elimination gives
\begin{equation}
 \mathcal H_{\mathrm{soft}}
 =\mathcal H_{ss}
 -\mathcal H_{sm}\mathcal H_{mm}^{-1}\mathcal H_{ms}.
 \label{eq:app_hs_soft_kernel}
\end{equation}
Here $s,m$ label soft and massive coordinates. 
Mixing between carriers $0,\Pi$ begins linearly in envelope momentum, so the Schur correction contributes at quadratic order. 
The full complex Hessian is required: imaginary mixing also changes the final real stiffness.

Set $q=q_\star$ and use unit cell and row spacings. Choose a tangent matrix $K_T$ with $K_T^T=-K_T$, $\{J_0,K_T\}=0$, and $\operatorname{tr}(K_T^TK_T)=4$. 
The displacement $\delta H=s\alpha K_T/2$, $\delta V=-s\alpha K_T/2$ obeys $\sum_{A<B}[(\delta H_{AB})^2+(\delta V_{AB})^2]=\alpha^2$ and corresponds to $\delta J=\alpha K_T/q$. Thus $\alpha$ is a normalized soft HS coordinate. 
After massive-mode elimination, its action is
\begin{equation}
 S_{\mathrm{soft}}^{(2)}=\frac12\int dx\,dy\,
 [\kappa_y(\partial_y\alpha)^2+\kappa_x(\partial_x\alpha)^2].
 \label{eq:app_soft_gradient_coefficients}
\end{equation}
The coefficients follow from the chiral fermion bubble
\begin{equation}
 \mathcal B(P)=-\int_{-\pi}^{\pi}\frac{dp}{2\pi}\,
 G_+(p)G_-(p+P)
 =B_0+ic_1P-b_2P^2+\cdots,\qquad
 G_\pm(p)=[D(p)\pm q]^{-1}.
 \label{eq:app_lattice_tangent_bubble}
\end{equation}
Here $p$ spans the full internal fermion Brillouin zone, while $P$ is the small transferred trajectory momentum. Evaluation of this response and the Schur complement gives
\begin{equation}
 B_0=\frac1{q(q+2)}=\frac1{4g},\quad
 c_1=\frac1{4q^2},\quad b_2=\frac{q^2+4}{32q^3}\longrightarrow
 \kappa_y=4b_2+\frac{4c_1^2}{B_0}
 =\frac{q^2+2q+8}{8q^3},\quad
 \kappa_x=\kappa_y+\frac{B_0}{4}.
 \label{eq:app_lattice_gradient_coefficients}
\end{equation}
The term $4b_2$ is the direct soft response, while $4c_1^2/B_0$ comes from imaginary mixing with massive carrier-$0$ modes; each contributes $1/(2q^3)$ at leading order. The residual $B_0/4$ in $\kappa_x$ retains the displaced bond phase after eliminating the massive staggered sum.

Since $\operatorname{tr}[(\partial_\mu J)^T(\partial_\mu J)]=4(\partial_\mu\alpha)^2/q^2$, matching to the ordinary-trace orientation action gives the directional inverse stiffnesses $t_x,t_y$:
\begin{equation}
 \frac1{t_\mu}=\frac{q^2\kappa_\mu}{4},\qquad
 t_{\sigma,0}=\sqrt{t_x t_y}
 =\frac4{q^2\sqrt{\kappa_x\kappa_y}}.
 \label{eq:app_orientation_stiffness_conversion}
\end{equation}
The geometric mean follows from rescaling the coordinates to remove the leading anisotropy. The coefficients already include the induced massive-mode response.

The positive gradient cost penalizes differences between neighboring orientations while leaving a uniform $J$ degenerate. Flavor symmetry extends the tangent action over the saddle manifold. To leading order in gradients, the bulk orientation action in rescaled coordinates is
\begin{equation}
 S_{\mathrm{bulk}}[J]
 =\frac1{2t_{\sigma,0}}\int d^2\mathbf r\,
 \sum_{\mu=1,2}\operatorname{tr}
 \left[(\partial_\mu J)^T(\partial_\mu J)\right],\qquad
 J(\mathbf r)\in SO(2n)/U(n),\quad
 J^T=-J,\quad J^2=-I_{2n}.
 \label{eq:app_nlsm_gradient_action}
\end{equation}
Here $\mathbf r$ denotes real network coordinates after rescaling; gradients act on the slow orientation envelope, with $s_{x,y}$ kept exact. The constraint to the orbit under finite rotations makes the theory nonlinear. With the ordinary flavor trace, $1/t_{\sigma,0}$ is the orientation stiffness. For the retained lattice action,
\begin{equation}
 t_{\sigma,0}=8g+O(g^2)
 =32\beta_R^2+o(\beta_R^2).
 \label{eq:app_conditional_stiffness_matching}
\end{equation}
The scaling follows from $\kappa_\mu\sim q_\star^{-3}$: converting HS displacements to normalized orientations supplies $q_\star^2$, so $1/t_{\sigma,0}\sim q_\star^{-1}$ and $t_{\sigma,0}\sim q_\star\sim g$. The numerical coefficient is the bare matching result of this retained saddle and Gaussian gradient calculation, before sigma-model renormalization.

As $g\to0$, the radius and the allowed envelope-momentum range vanish, so this construction does not extend to the ballistic SWAP endpoint. 
Local stationarity and the positive gradient action do not establish global saddle dominance. 
Application to the full physical model still requires establishing the saddle's contribution on the integration cycle, boundary and logical-sector matching, and control of omitted interactions and replica continuation. 
Higher-gradient, topological, and defect terms are not fixed by the displayed quadratic-gradient action. 
Subject to these qualifications, the result supplies the bare coupling for the following RG analysis.

\begin{figure}[h]
    \centering
    \includegraphics[width=0.7\linewidth]{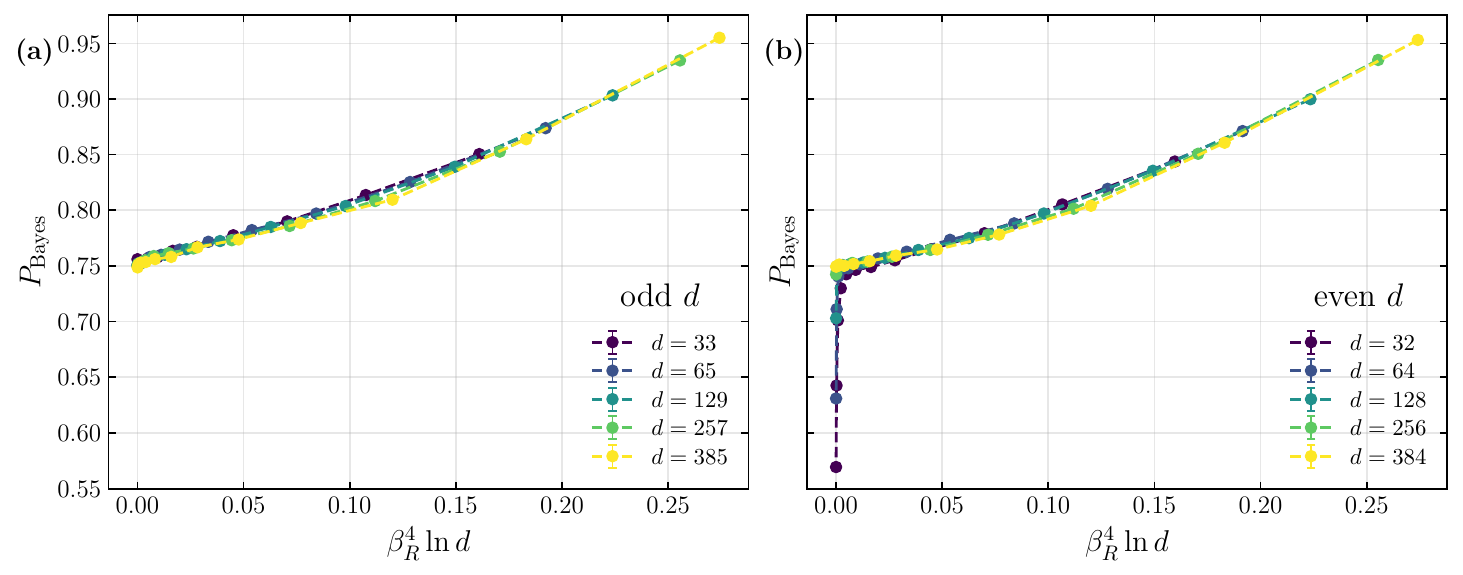}
    \caption{\textbf{Finite-size collapse of $P_\mathrm{Bayes}$}. 
     The left and right columns show $P_{\mathrm{Bayes}}$ against $X_4=\beta_R^4\ln d$ in $0.12\leq\theta/\pi\leq0.23$, for odd and even distances, respectively.    
Colors distinguish the distances listed in the legends. 
Error bars indicate one standard error; dashed lines connect points at fixed $d$ and are guides to the eye.}
    \label{fig:nlsm_fitting_ic_bayes}
\end{figure}

\subsection{Comparison of collapse powers}
\label{app:collapse_power}
\begin{figure}[t]
    \centering
    \includegraphics[width=1.0\linewidth]{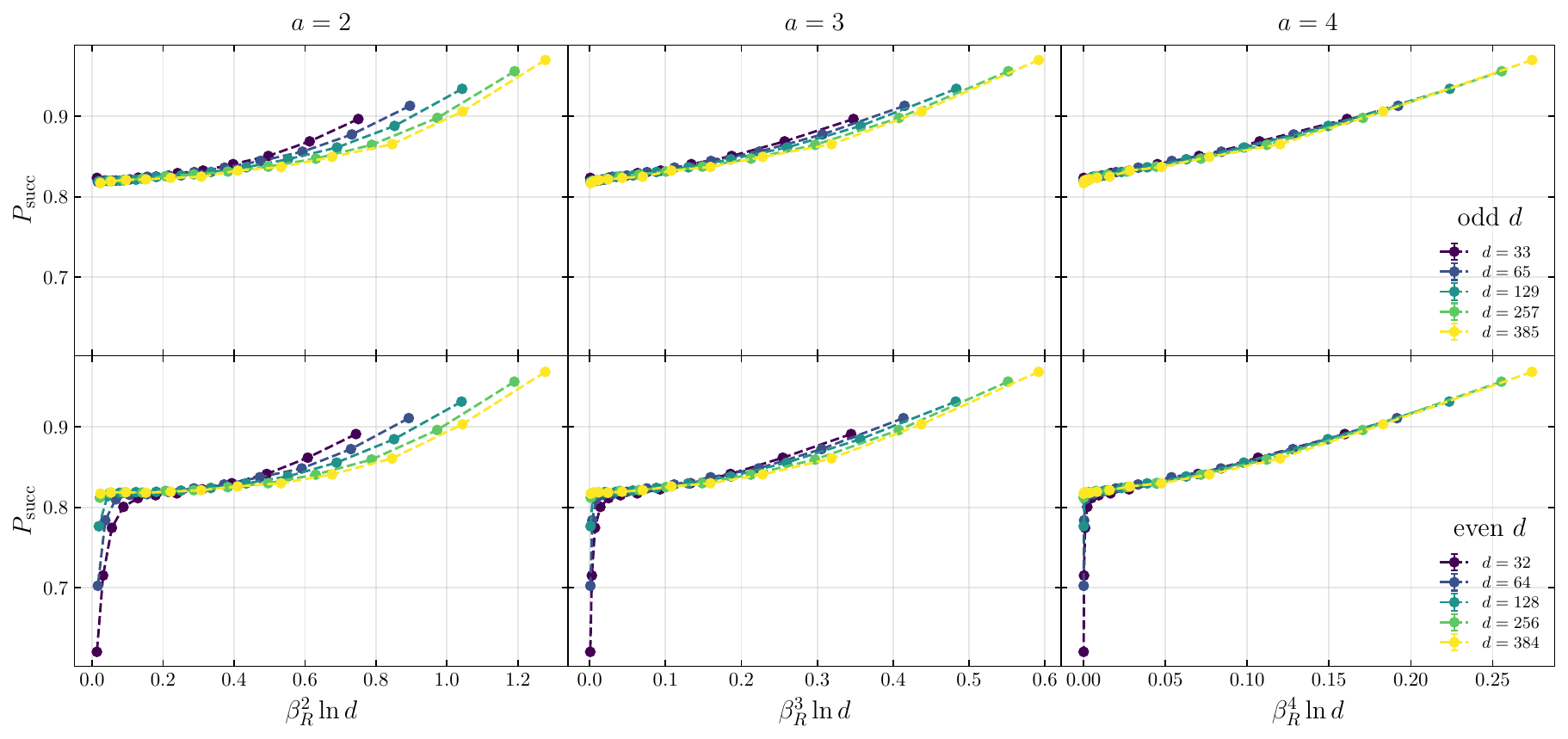}
    \caption{\textbf{Comparison of collapse powers.} The same $P_{\mathrm{succ}}$ data are plotted against $X_a=\beta_R^a\ln d$ with $a=2,3,4$ from left to right. The upper and lower rows show odd and even distances, respectively, for $0.12\leq\theta/\pi\leq0.23$. Colors distinguish the distances listed in the legends. The fourth-power variable reduces the separation between sizes along the rising branches, while deviations remain near the smallest collapse variables. Error bars indicate one standard error; dashed lines connect points at fixed $d$ and are guides to the eye.}
    \label{fig:nlsm_fitting_psucc_power}
\end{figure}
We compare alignment across code distances using
\begin{equation}
 X_a=\beta_R^a\ln d,
 \qquad \beta_R=-\frac12\ln\tan\theta.
 \label{eq:app_collapse_power}
\end{equation}
Each observable and parity is allowed a separate nonlinear scaling function. Section~\ref{sec:numerical_nlsm} and Appendix~\ref{app:diffusive_saddle} give the theoretical motivation for $a=4$.
Figs.~\ref{fig:nlsm_fitting}, \ref{fig:nlsm_fitting_ic_bayes}, and \ref{fig:nlsm_fitting_psucc_power} use $0.12\leq\theta/\pi\leq0.23$, with odd distances $d=33,65,129,257,385$ and even distances $d=32,64,128,256,384$. Each angle uses $50{,}000$ syndrome samples, except $d=384,385$, which use $10{,}000$. Odd $I_c/\ln2=1$ is fixed analytically and supplies no independent collapse test. The even measures share samples and are correlated.

At each measured point, we interpolate curves from other distances to the same $X_a$, linearly in $\beta_R$, and equally average their predictions. The score is the root-mean-square (RMS) difference between measurements and predictions. Each comparison uses identical measured points and contributing distances across the tested powers, without extrapolation. Reusing curves correlates the residuals, so these scores are descriptive and give no confidence interval for $a$. Error bars show sampling uncertainty, excluding interpolation bias and finite-size corrections.

Fig.~\ref{fig:nlsm_fitting_psucc_power} compares $a=2,3,4$. 
Restricting both measured points and prediction curves to $0.12\leq\theta/\pi\leq0.20$ gives 43 points for both even and odd cases, and the lowest RMS for $a=4$ in both parities. 
Over the full plotted interval, odd distances still favor $a=4$, while near-SWAP deviations dominate the even score and favor lower powers. 
The fourth-power advantage therefore concerns the resolved rising branches.

In exploratory comparisons, $a=5$ gives closer alignment in the selected middle-response band $0.83\leq P_{\mathrm{succ}}\leq0.89$, but its ranking relative to $a=4$ depends on the comparison window, interpolation, and retained sizes.

We retain $a=4$ as the theory-motivated reference scaling variable without assigning a unique asymptotic power. 
The omitted $L_0(\theta)$ dependence and endpoint response are possible sources of the ambiguity.
The logarithmic size dependence limits long-distance tests: at $\theta/\pi=0.22$, a tenfold increase in $d$ changes $X_4$ by only $1.86\times10^{-4}$. A change of $10^{-3}$ requires a size ratio of about $2.4\times10^5$, motivating a wider size range and denser angle sampling.

\subsection{Bulk Lyapunov rate and the scattering-length proxy}
\label{app:bulk_lyapunov}
We evaluate the scalar-free, single-layer Majorana transfer matrix on a long strip of $W$ transverse cells with periodic transverse boundaries. 
In axial gauge, the horizontal signs are independent and uniform, and the vertical layer is deterministic. 
This quenched diagnostic supports the transfer-growth scale used in Sec.~\ref{subsec:swap_variables_parity}; its relation to transport in the Born-weighted problem requires matching.

\paragraph{Reduced row and Lyapunov spectrum.}
Label the $2W$ Majoranas by $a_x=\gamma_{2x}$ and $b_x=\gamma_{2x+1}$, with $x=0,\ldots,W-1$. For $O=\sum_j c_j\gamma_j$, conjugation by $e^{\alpha\gamma_i\gamma_j}$ acts on the coefficient column $(c_i,c_j)^T$ through
\begin{equation}
 q(\alpha)=
 \begin{pmatrix}
 \cos(2\alpha)&\sin(2\alpha)\\
 -\sin(2\alpha)&\cos(2\alpha)
 \end{pmatrix}.
 \label{eq:app_lyapunov_pair}
\end{equation}
The coefficient matrix is the transpose of the operator-action matrix. A row applies $H_y$ on $(b_x,a_{x+1})$, then $V$ on $(a_x,b_x)$, with indices modulo $W$ and angles
\begin{equation}
 \alpha_{{\rm h},x,y}=\eta_{x,y}\omega_x
 \left(\frac{\pi}{4}+i\beta_R\right),\qquad
 \alpha_{\rm v}=-\frac{\pi}{4}+\phi(\beta_R),\qquad
 \phi(\beta_R)=\arctan(\tanh\beta_R),
 \label{eq:app_lyapunov_angles}
\end{equation}
where $\eta_{x,y}=\pm1$ are the independent local signs, $\omega_x=1$ except on the closing bond, and $\omega_{W-1}=\omega=\pm1$ is a fixed seam twist. The row is $M_y^{(\omega)}=VH_y^{(\omega)}\in SO(2W,\mathbb C)$. Scalar Ising prefactors are excluded from the growth rate. At $\beta_R=0$, the rows are real orthogonal signed-SWAP matrices and all Lyapunov exponents vanish.

For the product $\mathsf M_L^{(\omega)}=M_L^{(\omega)}\cdots M_1^{(\omega)}$, let $\sigma_j$ denote the singular values in decreasing order. The Lyapunov exponents, in units of inverse transfer rows, are
\begin{equation}
 \gamma_j^{(\omega)}(W,\beta_R)
 =\lim_{L\to\infty}\frac{1}{L}\ln\sigma_j\!\left(\mathsf M_L^{(\omega)}\right),
 \qquad j=1,\ldots,2W.
 \label{eq:app_lyapunov_exponents}
\end{equation}
Since $\mathsf M_L^{-1}=\mathsf M_L^T$ and transposition preserves singular values, complex orthogonality gives reciprocal singular-value pairs and hence $\gamma_j^{(\omega)}=-\gamma_{2W+1-j}^{(\omega)}$. The expanding $W$-dimensional volume therefore grows at a rate $\sum_{j=1}^W\gamma_j^{(\omega)}$. We define its rate per transverse cell by
\begin{equation}
 \Gamma_W^{(\omega)}(\beta_R)
 =\frac{1}{W}\sum_{j=1}^{W}\gamma_j^{(\omega)}(W,\beta_R),
 \qquad
 \overline\Gamma_W=\frac{\Gamma_W^{(+)}+\Gamma_W^{(-)}}{2}.
 \label{eq:app_lyapunov_rate}
\end{equation}
Thus $\Gamma_W$ averages the nonnegative half of the spectrum; it does not define the strip localization length. After discarding a warmup, repeated complex QR decompositions give $\widehat\gamma_j=L_{\rm meas}^{-1}\sum_k\ln|(R_k)_{jj}|$, where $L_{\rm meas}$ counts rows. Sorting these estimates and applying Eq.~\eqref{eq:app_lyapunov_rate} yields $\widehat{\overline\Gamma}_W$.

The twists are evaluated separately using the same local-sign stream. Flipping the independently sampled seam sign preserves the disorder distribution, so their mean is a bulk estimator, not a physical logical-sector average. The external logical defect remains in the physical channel and replica construction.

\paragraph{Numerical results.}
We used $W=8,12,16,24,32$, $\beta_R=0,0.01,0.015,0.02,0.03,0.04$, and four disorder seeds: 120 twist-paired cases, with QR orthogonalization every eight rows. Nonzero-detuning runs satisfy $\beta_R^2L_{\rm warm}\geq20$ and $\beta_R^2L_{\rm meas}\geq200$; at $\beta_R=0.01$, this gives $200{,}000$ warmup and $2{,}000{,}000$ measurement rows. Endpoint runs use 4096 warmup and $20{,}000$ measurement rows and return zero rates. The maximum exponent-sum and pair-sum residuals are $8.9\times10^{-15}$ and $4.8\times10^{-5}$. Length convergence remains untested: a $1/L$ bias would also scale as $\beta_R^2$ under this schedule. No independent length-doubling or QR-interval study was performed.

An unweighted common-slope fit of the 100 nonzero-detuning seed estimates,
\begin{equation}
 \ln\widehat{\overline\Gamma}_W\simeq c_W+p\ln\beta_R,
 \label{eq:app_lyapunov_power_fit}
\end{equation}
with width-dependent intercepts gives $p=2.00004$. A delete-one-seed-cluster jackknife with a Student-$t$ multiplier (three degrees of freedom) gives the fit-conditional 95\% interval $[1.9973,2.0028]$; it excludes finite-length and fitting-model uncertainty.

Averaging $\widehat{\overline\Gamma}_W/\beta_R^2$ over the five positive detunings for each seed gives the four-seed means $a_W$ and sampling standard errors in Table~\ref{tab:bulk_lyapunov_rate}.
\begin{table}[t]
 \centering
 \caption{Quadratic Lyapunov-rate coefficients, averaged over detunings and then four seeds. Standard errors (SE) include sampling uncertainty only; detuning, finite-length, and width-extrapolation effects are excluded.}
 \label{tab:bulk_lyapunov_rate}
 \begin{minipage}{0.56\linewidth}
 \centering
 \begin{ruledtabular}
 \begin{tabular}{ccc}
 $W$ & $a_W$ & SE\\
 \hline
 8  & 1.720334 & 0.001788\\
 12 & 1.813807 & 0.001919\\
 16 & 1.859701 & 0.001556\\
 24 & 1.905743 & 0.001105\\
 32 & 1.928770 & 0.000987
 \end{tabular}
 \end{ruledtabular}
 \end{minipage}
\end{table}
The unweighted extrapolation ansatz $a_W=a_\infty+b/W$ gives $a_\infty=1.99845$. Restricting it to $W\geq12$ or $W\geq16$ gives $1.99773$ and $1.99784$, while adding a $1/W^2$ term using all widths gives $1.99688$. These checks support a coefficient near two, without establishing the extrapolation form or an exact coefficient. Separate linear extrapolations at each detuning give intercepts from $1.99194$ to $2.00201$, with sample standard deviation $0.00382$; this spread is not a confidence interval or a width-extrapolation uncertainty.

These finite-strip results support the numerical summary
\begin{equation}
 \Gamma(\beta_R)\equiv\Gamma_\infty(\beta_R)\simeq2\beta_R^2,
 \qquad
 \ell_\Gamma\equiv\Gamma^{-1}\simeq\frac{1}{2\beta_R^2},
 \qquad 0.01\leq\beta_R\leq0.04,
 \label{eq:app_lyapunov_quadratic}
\end{equation}
where $\Gamma_\infty$ is the width-extrapolated rate. In the main-text convention $u=\beta_R^2d$, accumulated transfer growth is $\Gamma d\simeq2u$. Identifying $\ell_\Gamma$ with the physical mean free path requires transport and geometry matching, as assumed in Sec.~\ref{subsec:swap_variables_parity}. It is distinct from the odd-response scale $v=\beta_Rd$ and the conditional RG crossover. The measured coefficient fixes neither the replicated coupling nor the NL$\sigma$M stiffness, and these data establish neither a diffusive saddle nor a decoding threshold.

\twocolumngrid
\bibliography{references}

\end{document}